\documentclass[12pt]{article}

\pdfoutput=1
 
\usepackage[top=80pt,bottom=85pt,left=85pt,right=85pt]{geometry}
\usepackage{amssymb}
\usepackage{amsmath}
\usepackage{amsfonts,amsxtra, mathrsfs,amsthm}
\usepackage{setspace}
\usepackage{sectsty}
\usepackage{accents}
\usepackage{comment}
\usepackage[english]{babel}

\usepackage[usenames]{xcolor}
\usepackage{graphicx,subcaption}
\usepackage{float}
\usepackage{array,multirow,booktabs,longtable}
\usepackage{bbm}
\usepackage[vcentermath]{youngtab}

\def\node#1#2{\overset{#1}{\underset{#2}{{\color{gray} \bullet}}}}
\def\Node#1#2{\overset{#1}{\underset{#2}{{ \bullet}}}}
\def\Nodeblue#1#2{\overset{#1}{\underset{#2}{{\color{blue} \bullet}}}}
\def\Nodeblueleft#1#2#3{\overset{#1}{\llap{$\textstyle#3$}\underset{#2}{{\color{blue} \bullet}}}}
\def\NNode#1#2{\overset{#1}{\underset{#2}{{\color{blue} \bullet}}}}

\def\ver#1#2{\overset{{\llap{$\scriptstyle#1$}\displaystyle{\color{gray} \blacksquare}{\rlap{$\scriptstyle#2$}}}}{\scriptstyle\vert}}
\def\verbl#1#2{\overset{{\llap{$\scriptstyle#1$}\displaystyle{\color{black} \blacksquare}{\rlap{$\scriptstyle#2$}}}}{\scriptstyle\vert}}
\def\vver#1#2{\overset{{\llap{$\scriptstyle#1$}\displaystyle{\color{red} \blacksquare}{\rlap{$\scriptstyle#2$}}}}{\scriptstyle\vert}}
\def\Ver#1#2{\overset{{\llap{$\scriptstyle#1$}\displaystyle\blacksquare{\rlap{$\scriptstyle#2$}}}}{\scriptstyle\vert}}

\definecolor{link}{rgb}{.8,.15,.1}

\usepackage[debug,pageanchor=false,bookmarks=false]{hyperref}
\hypersetup{colorlinks=true,linkcolor=link,citecolor=link,urlcolor=link,linktocpage}

\newcommand{\rr}{\mathbb{R}}
\newcommand{\cc}{\mathbb{C}}
\newcommand{\zz}{\mathbb{Z}} 
\newcommand{\pp}{\mathbb{P}}

\DeclareMathOperator{\tr}{tr}
\DeclareMathOperator{\Tr}{Tr}
\DeclareMathOperator{\rk}{rk}

\DeclareMathOperator{\ord}{ord}
\DeclareMathOperator{\Ind}{Ind}
\DeclareMathOperator{\SU}{SU}
\DeclareMathOperator{\U}{U}
\DeclareMathOperator{\SO}{SO}
\DeclareMathOperator{\Sp}{Sp}
\DeclareMathOperator{\USp}{USp}

\makeatletter
\@addtoreset{equation}{section}
\makeatother

\begin{document}

% title page
\begin{titlepage}

\begin{center}

\vskip .3in \noindent

{\Large \bf{AdS$_7$/CFT$_6$ with orientifolds}}

\bigskip

Fabio Apruzzi$^{a,b}$ and Marco Fazzi$^{c,d}$

%\bigskip

%Version of \DTMNow~(w.r.t. GMT)

\bigskip
{\small 

$^a$ Department of Physics, University of North Carolina, Chapel Hill, NC 27599, USA \\	
\vspace{.25cm}
$^b$ Department of Physics and Astronomy, University of Pennsylvania, \\ Philadelphia PA, 19104-6396, USA \\	
\vspace{.25cm}
$^c$ Department of Physics, Technion, 32000 Haifa, Israel\\
\vspace{.25cm}
$^d$ Department of Mathematics and Haifa Research Center for Theoretical Physics and Astrophysics, University of Haifa, 31905 Haifa, Israel
	
}

\vskip .3cm
{\small \tt \href{mailto:fabio.apruzzi@unc.edu}{fabio.apruzzi@unc.edu} \hspace{.5cm} \href{mailto:mfazzi@physics.technion.ac.il}{mfazzi@physics.technion.ac.il}}
\vskip .6cm
     	{\bf Abstract }
\vskip .1in
\end{center}

AdS$_7$ solutions of massive type IIA have been classified, and are dual to a large class of six-dimensional $(1,0)$ SCFT's whose tensor branch deformations are described by linear quivers of $\SU$ groups. Quivers and AdS vacua depend solely on the group theory data of the NS5-D6-D8 brane configurations engineering the field theories. This has allowed for a direct holographic match of their $a$ conformal anomaly. In this paper we extend the match to cases where O6 and O8-planes are present, thereby introducing $\SO$ and $\USp$ groups in the quivers. In all of them we show that the $a$ anomaly computed in supergravity agrees with the holographic limit of the exact field theory result, which we extract from the anomaly polynomial. As a byproduct we construct special AdS$_7$ vacua dual to nonperturbative F-theory configurations. Finally, we propose a holographic $a$-theorem for six-dimensional Higgs branch RG flows.

\noindent

\vfill
\eject

\end{titlepage}

% end titlepage

\tableofcontents

\newpage 
%%%%%%%%%%%%%%%%%%%%%%%%%%%
\section{Introduction} % sec (intro)
\label{sec:intro}
%%%%%%%%%%%%%%%%%%%%%%%%%%%

Six-dimensional superconformal field theories (SCFT's henceforth) have received a great deal of attention in recent years. The reasons for such a renewed interest are numerous, and arguably well-justified.

First of all, their existence is ascertained only through an embedding into string \cite{witten-comments,hanany-zaffaroni,brunner-karch,intriligator-6d,intriligator-6d-II} or M/F-theory \cite{strominger-6d,heckman-morrison-vafa}, but a rigorous and purely field-theoretic definition is still lacking. Most notably, a lagrangian description (in terms of fundamental, microscopic fields) of the quantum theory is not available at the moment. (Classical Lagrangians for $N$ $(2,0)$ tensor multiplets coupled to $(1,0)$ vector multiplets have been constructed in \cite{samtleben-sezgin-wimmer-1,samtleben-sezgin-wimmer-2}.)

For $(2,0)$ SCFT's of type $A_{N-1}$, i.e. the theory on $N$ coincident M5's, it has been known for a long time \cite{gubser-klebanov,henningson-skenderis,henningson-skenderis-talk,deser-schwimmer,tseytlin,beem-rastelli-vanrees} that the number of degrees of freedom grows like $N^3$, which is more than (naively) expected for a theory of $N$ tensors in six dimensions. This number can be estimated by computing the so-called conformal anomaly of the theory, an observation that we will heavily exploit. %\footnote{\label{foot:bootstrapA}\color{red}Actually, exact formulae are known for all $\Gamma_G=ADE$ types. For type $A_{N-1}$, \cite{deser-schwimmer,tseytlin} computed explicitly the so-called $a$ and $c_{1,2,3}$ conformal anomalies in the trace of the stress-energy tensor of one such $(2,0)$ theory. At large $N$, they indeed all scale like $N^3$, matching the holographic result of \cite{henningson-skenderis}. \cite{beem-rastelli-vanrees} later gave closed formulae for the $c_i$ (for all types) that depend only on the $G$ group theory data ($G$ being associated with $\Gamma$ through the $\SU(2)$ McKay correspondence). E.g. $c_i(A_{N-1}) =(4N^3-3N-1) c_i^\text{tens}$, where $c_i^\text{tens}$ is a normalization constant associated with the abelian (i.e. $N=1$) theory.} 
The $N^3$ growth suggests that these theories are interacting, and follow the rough scaling pattern for an SCFT in $d$ dimensions given by $N^{d/2}$.\footnote{Examples of theories evading this ``paradigm'' are well-known in odd dimensions, where the free energy $F:=- \log |Z_{S^d} |$ of the theory ($Z_{S^d}$ being its $d$-sphere partition function) can be used to estimate the number of degrees of freedom (see e.g. \cite{drukker-marino-putrov} for the ABJM \cite{abjm} case, and \cite{jafferis-pufu} for five-dimensional theories with AdS$_6$ massive IIA duals \cite{brandhuber-oz,bergman-rodriguezgomez}).  For instance, the three-dimensional $\mathcal{N}=3$ Chern--Simons-matter necklace quivers of \cite{gaiotto-t,gaiotto-t2} exhibit an $N^{5/2}$ scaling \cite{aharony-jafferis-t-zaffaroni}, and five-dimensional $\mathcal{N}=1$ SCFT's of ``long quiver'' type \cite{dhoker-gutperle-uhlemann-II} engineered by simple $N$ D5, $M$ NS5 brane webs exhibit an $N^2 M^2$ scaling \cite{gutperle-marasinou-trivella-uhlemann} (i.e. $N^4$ when $M\sim N \to \infty$).} However their less-supersymmetric counterparts -- $(1,0)$ theories -- with only eight Poincar\'e supercharges and which make up a much richer class of theories \cite{gaiotto-t-6d,delzotto-heckman-tomasiello-vafa,heckman-morrison-rudelius-vafa}, are characterized by an even more surprising scaling. The number of degrees of freedom depends in this case on multiple parameters (a fact first discovered in \cite{gaiotto-t-6d,afrt,ads7prl}). Even when the latter are taken to scale in the same way (like $N$) we get an $N^5$ growth, which clearly does not exhibit the expected dimension-dependent exponent. This behavior can be explained by looking at the M-theory origin of the field theories. 

In M-theory a large class of ``orbifold'' $(1,0)$ SCFT's can be constructed by having a stack of $N$ coincident M5-branes probe a line of singularities $\rr \times \cc^2 / \Gamma$, with $\Gamma$ a discrete subgroup of $\SU(2)$, i.e. one in the $ADE$ list. In particular, in the $A_{k-1}$ ($k \geq 2$) and $D_k$ ($k \geq 4$) cases, the extra parameter is provided by the order of the finite group -- $k$ and $2k$ respectively -- and it can be shown that the number of degrees of freedom scales like $| \Gamma |^2 N^3$, explaining the $N^5$ growth when $k \sim N$ as $N \to \infty$.\footnote{Notice that for $\Gamma=E_n$ the limit $n \to \infty$ is not meaningful, nor is $N\to \infty$ given the lack of a weakly-coupled (eleven-dimensional) supergravity description that could produce an $nN^3$ growth. The $a$ conformal anomaly has been computed exactly at finite $N$ in \cite{mekareeya-rudelius-tomasiello}.} %\footnote{One might wonder whether the scaling such a scaling pattern also holds for the three exceptional groups $E_n$, $n=6,7,8$. Notice however that in this case the order is a finite number fixed by $n$, and moreover the growth cannot be reproduced via a (type IIA) supergravity computation, given that a weakly-coupled (type IIA) string theory construction is not available. One can (and indeed must) use F-theory to compute the $a$ conformal anomaly \cite{mekareeya-rudelius-tomasiello} of the theory of M5's probing $ \cc^2 / E_n$ \cite{delzotto-heckman-tomasiello-vafa}. This number will depend on $N$; however its large $N$ limit (in the sense of supergravity, which would entail having a weakly-coupled description) is not meaningful.} 
Although the theories we consider in this paper do not have a realization in M-theory as simple orbifolds (because of the presence of D8's in their brane engineering), we will see that such a scaling behavior carries through nonetheless. %\footnote{The ultimate reason for this fact may be the following. The $\cc^2 / A_{k-1}$ orbifold $(1,0)$ theories with the addition of T-brane data \cite{delzotto-heckman-tomasiello-vafa} can also be constructed as F-theory configurations where, in the base of the elliptic fibration, D7-branes wrap a collection of compact mutually-intersecting curves ($\pp^1$'s). At the tails of the configuration we have instead noncompact curves, carrying the T-brane data. Upon T-duality, this turns into an NS5-D6 suspended brane configuration with two D8-brane stacks at the tails. See \cite[Sec. 3.3]{delzotto-heckman-tomasiello-vafa} and \cite[Sec. 2.2]{mekareeya-rudelius-tomasiello} for cases without orientifolds.}

Second, despite the abundance of nonperturbative constructions and embeddings into string or M/F-theory, very few exact results in field theory are known for these SCFT's. For instance, the conformal bootstrap program has not been applied to constrain the space of $(1,0)$ theories and check the classification efforts of \cite{heckman-morrison-vafa,heckman-morrison-rudelius-vafa} (however see \cite{chang-lin} for attempts in this direction), nor has been localization to compute their $S^6$ partition function, given the lack of a lagrangian description.\footnote{We thank B.~Van Rees and F.~Yagi for discussion on this point. Exact results for compactifications on $S^1$ or $T^2$ are known for some $(1,0)$ SCFT's. See e.g. \cite{kim-lee,gadde-haghighat-kim-kim-lockhart-vafa,kim-kim-lee-vafa,hayashi-kim-lee-yagi,yun} and references therein.} (It is true however that such embeddings have been very fruitful. For instance, they allow us to classify six-dimensional theories \cite{heckman-morrison-rudelius-vafa} and, partially, their compactifications \cite{delzotto-vafa-xie,razamat-vafa-zafrir,bah-hanany-maruyoshi-razamat-tachikawa-zafrir,kim-razamat-vafa-zafrir,morrison-vafa-N1,apruzzi-hassler-heckman-melnikov-dglsm}; compute quantities such as dimensions of moduli spaces \cite{mekareeya-rudelius-tomasiello}, defect and autmorphism groups \cite{delzotto-heckman-park-rudelius,apruzzi-heckman-rudelius}; determine RG flows and their hierarchy \cite{heckman-morrison-rudelius-vafa-flow,heckman-rudelius-tomasiello} and the global symmetries \cite{bertolini-merkx-morrison,merkx,delzotto-heckman-tomasiello-vafa}; compute anomalies from the six-dimensional anomaly polynomial \cite{ohmori-shimizu-tachikawa-yonekura}.)

Therefore it appears particularly important to check the stringy constructions against properties of the field theories they supposedly give rise to. Focusing on the (massive) type IIA string theory embeddings of $(1,0)$ SCFT's (dating back to \cite{brunner-karch,hanany-zaffaroni}), an independent and explicit check of their soundness can in principle be obtained through the AdS/CFT correspondence. Indeed one expects that the holographic limit of quantities that can be computed purely in terms of the brane configuration data match those computed in the AdS supergravity duals. Very few tests of the AdS/CFT duality in this higher-dimensional setting have been attempted to date, starting with \cite{gaiotto-t-6d} and culminating in the ``precision test'' of \cite{cremonesi-tomasiello}. There it was shown that the $a$ conformal anomaly of $(1,0)$ theories engineered by NS5-D6-D8 brane configurations in type IIA perfectly agrees with the supergravity result computed using the massive AdS$_7$ vacua of \cite{afrt,gaiotto-t-6d}. 

Emboldened by this nontrivial result, we extend the six-dimensional holographic $a$ anomaly match to cases where orientifolds are present.\footnote{\label{foot:oplanes}We use conventions whereby an O$p^\pm$-plane has $\pm 2^{p-4}$ D$p$ charge.} We may in fact add O6 and O8-planes to the aforementioned suspended brane configurations in order to engineer $\SO$ and $\USp$ gauge and flavor groups. The supergravity data associated with such setups change, but we will show that the holographic match holds true in all of these cases just as in \cite{cremonesi-tomasiello}.  The leading order of the $a$ anomaly takes the simple form
\begin{equation}\label{eq:a-intro}
a \sim \frac{192}{7} (\eta^{-1})_{ij} \, h^\vee_{G_i} h^\vee_{G_j}\ , \nonumber
\end{equation}
where $h^\vee_{G_i}$ are the dual Coxeter numbers of gauge groups $G_i$ in a linear quiver description of the SCFT tensor branch, and $\eta$ its so-called Dirac pairing. In \cite{cremonesi-tomasiello} all gauge groups are $\SU(r_i)$, and $h^\vee_{G_i} = r_i$. Here the groups will be allowed to be $\SU$, $\SO$ and $\USp$ according to the theory at hand. We thus provide further compelling evidence for the advocated duality between the AdS$_7$ vacua of \cite{afrt,gaiotto-t-6d}, the brane configurations of \cite{brunner-karch,hanany-zaffaroni}, and a vast class of $(1,0)$ SCFT's.

To obtain such a result we had to generalize the simple combinatorial formalism of \cite{cremonesi-tomasiello} in order to construct more general AdS$_7$ vacua featuring orientifold sources. (The possiblity of having vacua with an O8-plane source was suggested in \cite{afrt} but left unexplored. \cite{bpt} recently constructed a first concrete example which is dual to the so-called massive E-string theory.) As an interesting byproduct of this, we exhibit for the first time the supergravity duals to some of the ``formal'' massive IIA brane setups of \cite{mekareeya-rudelius-tomasiello}, which are characterized by the same $a$ conformal anomaly as certain nonperturbative F-theory configurations. We argue that these type IIA AdS$_7$ solutions can be understood as gravity duals to the F-theory quivers, thus complementing a very scarce class of AdS vacua of type IIB with varying and monodromic axiodilaton.

Finally, we propose a version of the holographic $a$-theorem for six-dimensional RG flows induced by Higgs branch deformations of the quiver theories. We identify a monotonic function in the supergravity duals which decreases along the flow. The function is extremely simple, and controls the position of D8-brane sources in the supergravity vacua.\\

This paper is organized as follows. In section \ref{sec:solz} we explain how massive type IIA brane configurations can be used to construct $(1,0)$ SCFT's on the tensor branch, and how very general dual supergravity solutions can be constructed by relying on the same combinatorial data. (This data also determines the various integration constants the supergravity solutions depend on. The relevant computations are carried out in appendices \ref{app:var} and \ref{app:int-const}.) In section \ref{sec:aexamples} we compute exactly in field theory the $a$ conformal anomaly of general $(1,0)$ SCFT's, whose tensor branch is characterized by a linear quiver of $\SU,\SO,\USp$ gauge and flavor groups, and matter in various representations. (This is done by exploiting the six-dimensional anomaly polynomial, whose derivation we carry out in appendix \ref{app:aFT}.) We then take the holographic limit of the exact field theory result. In section \ref{sec:match} we match this limit to the supergravity result, which can be obtained as an internal space integral (carried out for general AdS$_7$ solutions in appendix \ref{app:agrav}). Section \ref{sec:newex} contains several new examples, obtained by specializing the formulae of section \ref{sec:aexamples} to concrete linear quivers. We show how the formalism put forward in this paper can be used to check the AdS$_7$/CFT$_6$ correspondence in particularly interesting cases, such as when the dual SCFT can be engineered nonperturbatively in F-theory or when both O6 and O8 sources are present in supergravity. In section \ref{sec:athm} we provide evidence for the existence of a holographic $a$-theorem. We present an outlook and our conclusions in section \ref{sec:conc}.

\begin{comment}
\subsection{Summary of results}

\begin{itemize}
\item We have generalized the formalism introduced in \cite{cremonesi-tomasiello} in order to construct AdS$_7$ vacua of massive type IIA featuring not only D-brane but also O-plane sources, i.e. corresponding to NS5-D6-O6-D8-O8 brane setups. The construction only depends on combinatorial data that can be read off of the dual six-dimensional quiver.

\item We have used this formalism to compute in supergravity the holographic $a$ conformal anomaly of several classes of AdS$_7$ vacua.

\item On the field theory side we have computed $a$ exactly, by leveraging the six-dimensional anomaly polynomial, for all the above classes of theories.

\item Finally, we have specialized the holographic match to a few cases of interest. For all of them we have also constructed for the first time the corresponding AdS$_7$ vacuum. These include: a formal massive type IIA brane setup among those in \cite{mekareeya-rudelius-tomasiello} that engineers an alternating $\SO$-$\USp$ quiver labeled by the principal orbit of $\mathfrak{so}(2k)$ (and which can also be engineered in F-theory); a brane configuration featuring a single O8$^-$-plane; a brane configuration featuring a combined O6$^+$-O8$^-$ orientifold projection.

\end{itemize}
\end{comment}

% fold sec (intro)

%%%%%%%%%%%%%%%%%%%%%%%%%%%%%%%%%%
\section{Brane configurations in massive IIA and supergravity solutions} % sec (solz)
\label{sec:solz}
%%%%%%%%%%%%%%%%%%%%%%%%%%%%%%%%%%

\subsection{The dictionary between branes, quivers, and vacua}
\label{sub:dictionary}

We shall now summarize the proposed correspondence between NS5-D6(-O6)-D8(-O8) suspended brane configurations of \cite{brunner-karch,hanany-zaffaroni}, $(1,0)$ linear quivers, and the massive type IIA AdS$_7$ vacua of \cite{afrt,ads7prl,gaiotto-t-6d,cremonesi-tomasiello}.

\subsubsection{\texorpdfstring{Only $\SU(k)$ groups: M5's on $\cc^2/\zz_{k}$}{Only SU(k) groups: M5's on C2/Zk}}
\label{subsub:pureSUk}

Consider $N$ M5-branes probing the $\cc^2/\zz_{k}$ singularity, i.e. the quotient of the transverse space $\rr^4 \subset \rr^5$ by the discrete subgroup $A_{k-1}$ of $\SU(2)$. Resolving the singularity produces the so-called $k$-center Taub-NUT space, which gives rise to $k$ D6-branes upon reduction to type IIA \cite{gubser-tasi}, together with $N$ NS5-branes. The situation is summarized in table \ref{tab:M5}. 
\begin{table}
\centering
{\renewcommand{\arraystretch}{1.2}%
\begin{tabular}{@{}c c c c c c c  @{}}
& & $\rr$ & \multicolumn{4}{c}{$\cc^2/\Gamma$} \\
& $x^{0\ldots 5}$ & $x^6$ & \multicolumn{1}{|c}{$x^7$} & $x^8$ & $x^9$ & $S^1_\text{M}$\\
\toprule\toprule 
$i$-th M5 & $\times$ & $\phi_i$ & $\cdot$ & $\cdot$ & $\cdot$ & $\cdot$ \\
$i$-th NS5 & $\times$ & $\phi_i$ & 0 & 0 & 0 \\
$r_i$ D6's & $\times$ & $\left[\phi_i,\phi_{i+1}\right]$ & 0 & 0 & 0 \\
O6$^\pm$ & $\times$ & $\left[\phi_i,\phi_{i+1}\right]$ & 0 & 0 & 0 \\
$f_i$ D8's & $\times$ & $\cdot \in \left[\phi_i,\phi_{i+1}\right]$ & $\times$ & $\times$ & $\times$ \\
O8$^\pm$ & $\times$ & 0 & $\times$ & $\times$ & $\times$ \\
\toprule
\end{tabular}}
\caption{NS5-D6(-O6)-D8(-O8) brane scan. A $\cdot$ means the brane is sitting at a point along that direction; $\times$ means it is infinitely extended along that noncompact direction. When $\Gamma=D_k$ O6-planes are present, and are overlaid onto the D6-branes. The O8-plane can either sit at 0, between the first NS5 at $\phi_1$ and its image at $-\phi_1$, or be stuck on the first NS5 say at $\phi_1$, which we choose to put at 0.}
\label{tab:M5}
\end{table}

\begin{comment}
\begin{table}
\centering
{\renewcommand{\arraystretch}{1.2}%
\begin{tabular}{@{}c c c c c c c c c c c c  @{}}
& & & & & & & $\rr$ & \multicolumn{4}{c}{$\cc^2/\Gamma$} \\
& $x^0$ & $x^1$ & $x^2$ & $x^3$ & $x^4$ & $x^5$ & $x^6$ & \multicolumn{1}{|c}{$x^7$} & $x^8$ & $x^9$ & $S^1_\text{M}$\\
\toprule\toprule 
$i$-th M5 & $\times$ & $\times$ & $\times$ & $\times$ & $\times$ & $\times$ & $\phi_i$ & $\cdot$ & $\cdot$ & $\cdot$ & $\cdot$ \\
$i$-th NS5 & $\times$ & $\times$ & $\times$ & $\times$ & $\times$ & $\times$ & $\phi_i$ & 0 & 0 & 0 \\
$r_i$ D6's & $\times$ & $\times$ & $\times$ & $\times$ & $\times$& $\times$ & $\left[\phi_i,\phi_{i+1}\right]$ & 0 & 0 & 0 \\
O6$^\pm$ & $\times$ & $\times$ & $\times$ & $\times$ & $\times$& $\times$ & $\left[\phi_i,\phi_{i+1}\right]$ & 0 & 0 & 0 \\
$f_i$ D8's & $\times$ & $\times$ & $\times$ & $\times$ & $\times$& $\times$ & $\cdot \in \left[\phi_i,\phi_{i+1}\right]$ & $\times$ & $\times$ & $\times$ \\
O8$^\pm$ & $\times$ & $\times$ & $\times$ & $\times$ & $\times$& $\times$ & 0 & $\times$ & $\times$ & $\times$ \\
\toprule
\end{tabular}}
\caption{NS5-D6-D8 brane scan. A $\cdot$ means the brane is sitting at a point along that direction; $\times$ means it is infinitely extended along that noncompact direction. When $\Gamma=D_k$ O6-planes are present, and are overlaid onto the D6-branes. The O8-plane can either be sitting at 0, between the first NS5 say at $\phi_1$ and its image at $-\phi_1$, or be stuck on the first NS5 say at $\phi_1$, which we choose to put at 0.}
\label{tab:M5}
\end{table}
\end{comment}

Supersymmetry is halved due to the orbifold, and the $(2,0)$ tensor multiplets from the M5's each reduce to a $(1,0)$ tensor multiplet plus a $(1,0)$ hypermultiplet. We then have $N-1$ finite-length stacks each containing $k$ D6-branes, giving rise to a chain of $\SU(k)$ gauge groups, as well as two semi-infinite D6 stacks.\footnote{In principle we would have $\U(k)$ gauge groups. The $\U(1)$ centers are all anomalous, but the Green--Schwarz--West--Sagnotti mechanism involved in the anomaly cancellation renders them massive \cite{hanany-zaffaroni}. They are therefore decoupled from the low-energy dynamics of the linear quiver description.} The $N$ NS5's contribute $N_\text{T}=N-1$ tensor multiplets as well as $N_\text{T}-1=N-2$ bifundamental hypermultiplets. This is the type IIA description of this orbifold $(1,0)$. 

The real scalars $\phi_i$ inside the tensor multiplets are related to the positions of the NS5's along direction $x^6$: We say we are on the tensor branch of the $(1,0)$ SCFT when we give (nonzero) vevs to these scalars. This corresponds to having finite-coupling Yang--Mills terms in the Lagrangian of the quiver, and separates all NS5's.\footnote{The numbers $N_\text{T}$ of dynamical tensor multiplets and $N_\text{T}-1$ of bifundamental hypermultiplets are now explained. One tensor multiplet scalar, corresponding to the center-of-mass motion of the quiver along $x^6$, decouples from the dynamics. Supersymmetry tells us the whole multiplet is lost. Then, only $N-1-1$ bifundamental hypermultiplets coming from the NS5's will be charged under two neighboring gauge groups engineered by the D6-branes.} In particular, we see from figure \ref{fig:NS5D6} that the left- and rightmost $\SU(k)$'s are actually flavor groups, since they are associated with (two stacks of) semi-infinite D6-branes. Through a Hanany--Witten move we can trade each of the two for a stack of D8-branes sourcing a nonzero Romans mass $F_0$ (although the latter has to vanish globally), where each D6 ends on a different D8. The $k$ D8's contribute $k$ fundamental hypermultiplets of the left- and rightmost gauge groups (see figure \ref{fig:quiverNS5D6}). We can now activate vevs for the former (much as in \cite{hanany-witten,gaiotto-witten-1}), and slide finite segments of D6-branes trapped between two D8's off to infinity. We have modified the tail structure of the linear quiver by moving onto the Higgs branch of the SCFT.

In particular, its quiver will be characterized by two ``massive tails'' (of ``lengths'' $i=1,\ldots,L$ and $N-i=N-1,\ldots,N-R$), where D8's cross D6-branes, and a central ``massless plateau'' (of length $N-L-R$) where there are no D8-branes and the Romans mass is identically zero. (Clearly, there can be nongeneric situations where the plateau disappears or we only have one massive region.) This engineers a situation (depicted in figure \ref{fig:quiverNS5D6D8}) where we can have $f_i$ fundamental flavors of the $i$-th gauge group, for $i\neq0,N$. The ranks $r_i -1$ of the $\SU(r_i)$ gauge groups need not equal $k-1$ anymore (but $\underset{1\leq i \leq N-1}{\text{max}} r_i = k$), due to the various Higgsings we have performed. However, in the massless plateau $r_i = k$ for $i=L,\ldots,N-R$: We will dub this number ``height'' of the plateau. To all this data one can easily associate combinatorial objects, in the form of two Young tableaux $\rho_\text{L},\rho_\text{R}$ (one for each tail). They are associated to a (ordered) partition of the maximal rank $k$ (and therefore to a nilpotent orbit of $\mathfrak{su}(k)$ \cite{collingwood-mcgovern}) as follows.\footnote{See \cite{mekareeya-rudelius-tomasiello} for a full exploitation of this observation in the more general context of $(1,0)$ quivers engineered through F-theory.} Define the depth $(\rho^\text{t})_i$ of the rows of the transposed tableau $\rho^\text{t}$ by $(\rho^\text{t})_i := r_i-r_{i-1} =: s_i$, $i=1,\ldots,L$ (for $\rho_\text{L}$) or $i=N-R,\ldots,N-1$ (for $\rho_\text{R}$). Then $\rho_\text{L} = \left[ \rho_1, \rho_2, \ldots,  \rho_{l} \right]$ and $-\rho_\text{R} = -\left[ \rho_1, \rho_2, \ldots,  \rho_{r} \right]$ are partitions of $k$:
\begin{equation}
\sum_{i=1}^l (\rho_\text{L})_i = \sum_{i=1}^L (\rho^\text{t}_\text{L})_i = r_L = k = r_{N-R} = -\sum_{i=1}^R (\rho^\text{t}_\text{R})_{N-i} = -\sum_{i=1}^r (\rho_\text{R})_{N-i} \ .
\end{equation}
(The numbers $l,r$ depend on the specifics of the tableaux at hand, and can easily be found by transposing $\rho_\text{L,R}^\text{t}$.) In the above equation we have crucially assumed $r_0=r_N=0$; this assumption will be relaxed momentarily. The theory at the origin of the Higgs branch -- the ``unHiggsed'' theory in figure \ref{fig:NS5D6} -- will be labeled by two trivial partitions $\rho^\text{t}_\text{L}=-\rho^\text{t}_\text{R}=[1,1,\ldots,1]=:[1^k]$ (both corresponding to the trivial nilpotent orbit $\{ 0\}$ of dimension zero), since $\rho_\text{L}=-\rho_\text{R}=[k]=r_{1}=-(-r_{N-1})$. The Higgsed quiver of figure \ref{fig:NS5D6D8} has instead
\begin{equation}\label{eq:young-ex-suk}
	\rho^\text{t}_\text{L}={\tiny \yng(4,3,1,1,1)}\  \quad \rho_\text{L}={\tiny \yng(5,2,2,1)}\quad \quad ; \quad \quad -\rho^\text{t}_\text{R}={\tiny \yng(2,2,2,2,1,1)}\  \quad  -\rho_\text{R}={\tiny \yng(6,4)}
\end{equation}
corresponding to the nilpotent orbits $\mathcal{O}^\text{L}_{[5,2^2,1]}$ and $\mathcal{O}^\text{R}_{[6,4]}$ of $\mathfrak{su}(10)$.

Finally, gauge-anomaly cancellation implies \cite{hanany-zaffaroni}
\begin{equation}\label{eq:flavors}
f_i=2r_i-r_{i+1}-r_{i-1} = - (r_{i+1} - r_i) + (r_i - r_{i-1}) = -s_{i+1} + s_i > 0\ .
\end{equation}
The positivity of the $f_i$ implies that the function $i \mapsto r_i$ be convex. A simple consequence of this is that for $i = 1,\ldots, L$ the numbers $r_i$ have to grow, and to decrease for $i=N-R,\ldots,N$.\footnote{The minus in front of $\rho_\text{R}$ accounts for the fact that in the right Young tableau the columns have ``negative depth'', given that $r_i > r_{i+1}$ (for $i=N-R,\ldots,N$) implies $s_{i+1} < 0$. However the $r_i$ themselves are obviously positive for all $i$.} Given that the $f_i$ are the numbers of D8-branes sourcing a nonzero Romans mass $F_0$, the latter will be monotonous and decreasing along $x^6$, crossing a region where it is zero (the massless plateau) and eventually becoming negative (the right massive tail), so that we always have D8-branes instead of anti-D8's.

As already mentioned, we can further generalize this situation by slightly modifying the quiver in figure \ref{fig:quiverNS5D6D8}. In fact, as long as relation \eqref{eq:flavors} is satisfied at each node, we can have nonzero numbers $r_0,r_N$ of flavor D6-branes escaping off to infinity at the left and right of the quiver. For $i=0,N$ the left hand side of \eqref{eq:flavors} then reads $r_0+f_1$ and $r_N+f_{N-1}$ respectively. The left, right Young tableau will give a partition of $k_\text{L}:= r_L-r_0$, $k_\text{R}:= r_{N-R}-r_N$ respectively, with $k=r_L=r_{N-R}$ the height of the plateau. As we will see, although we are simply adding some flavors of the first and last gauge groups, this has the effect of modifying the ``poles'' of the internal space of the dual supergravity AdS vacuum (topologically, an $S^3$).\\

We now move on to describe how the AdS$_7$ vacua of \cite{afrt,ads7prl,gaiotto-t-6d} are related to the above constructions. A possible interpretation of these vacua as near-horizon limits of the brane configurations first appeared in \cite{gaiotto-t-6d}. (See also \cite{imamura-D8,janssen-meessen-ortin,bobev-dibitetto-gautason-truijen,bpt,macpherson-t} for more general Ansatze of localized intersecting brane metrics with AdS$_7$ near-horizon.) Bringing all NS5's on top of each other (the origin of the tensor branch, where the SCFT sits), we can imagine zooming in close to the NS5-D6 intersection, say at $x^6=0$. This limit cannot forget the information about the D6-D8 intersection though, which labels the particular SCFT and is collected in the Young tableaux $\rho_\text{L,R}$ of the linear quiver. In fact, the D6-D8's transform into magnetized D8 sources in the supergravity solution (with D6 charge smeared over their common worldvolume); the $N$ NS5's turn into $N$ units of quantized $H$ flux. Intuitively, two among directions $x^6$ and $x^{7,8,9}$ mix, and parameterize the base of the three-dimensional (compact) internal space $M_3$ of the AdS vacuum, plus its radial direction. In fact unbroken supersymmetry dictates that this space be a fibration $S^2 \hookrightarrow M_3 \to I=[0,N]$, where the (finite-length) base interval is now parameterized by a coordinate we call $z$. The remaining seven directions parameterize AdS$_7$ and are filled by the D8 sources, which also ``wrap'' an $S^2$ fiber inside $M_3$.\footnote{The D6 charge of the magnetized D8's is equivalent to turning on a (nontrivial) gauge bundle on the $S^2$.} Their position along $z$ is fixed by supersymmetry \cite{afrt,cremonesi-tomasiello}. Moreover the Romans mass $F_0$ that is sourced by the branes will be a step function supported on $I$: Its value decreases whenever we cross a D8 stack starting from $z=0$. 

The supergravity vacuum (metric, dilaton, warping factor, fluxes) can be defined in terms of a single cubic polynomial of $z$ that we call $\alpha(z)$; it is defined piecewise in the subintervals $[i,i+1]$ we decide to divide $I$ into ($i=0,\ldots,N-1$). In the coordinate $z$, the position of the $i$-th D8 stack is conveniently fixed to be at $z=i$ (i.e. the lower endpoint of $[i,i+1]$) by the second derivative of $\alpha(z)$, namely $\ddot{\alpha}(z)|_{z=i}=-(9\pi)^2 r_i$. The  number of D8's in the $i$-th stack at $z=i$ defines the value of the Romans mass $F_0$ in $[i,i+1]$ (which is related to the third derivative of $\alpha(z)$ via \eqref{eq:third-der-alpha} and \eqref{eq:F0subint}, given below). This way, the supergravity vacuum depends on the quiver data only through $F_0$. The data associated with the tails of the quiver (i.e. the Young tableaux, when D8's are present, or simply the groups $\SU(f_1)$, $\SU(f_{N-1})$ when we have semi-infinite flavor D6's) dictates what the coefficients of the polynomial $\alpha(z)$ are for $i \in [0,L]$ (where $F_0 >0$) and $i \in [N-R,N]$ (where $F_0<0$). In particular, for $i=0,N$, such coefficients will be called ``boundary data'', and can be related to what kind of brane sources are located in the vicinity of the ``poles'' of $M_3$ at the extrema $z=0,N$ of the base interval $I$. 

We remark that, in case $r_0,r_N \neq 0$, the left impression in figure \ref{fig:solNS5D6D8} will be slightly modified (as depicted in the right one): The smooth poles of the internal space $M_3$ will now be singular points for the metric due to the presence of (flavor) D6-brane sources. The creases representing magnetized D8 sources will be displaced along $z$ so as to satisfy \eqref{eq:flavors}. The correspondence that we have just sketched will be made much more precise in section \ref{sec:solz}.

\subsubsection{\texorpdfstring{Alternating $\SO$-$\USp$ groups: M5's on $\cc^2/D_k$}{Alternating SO-USp groups: M5's in C2/Dk}}
\label{subsub:alternatingSOUSp}

In case $N$ M5-branes probe the $\cc^2/ D_k$ singularity, there are two interesting effects. In M-theory, the M5's ``fractionate'' (i.e. we have $N=2M$ half-M5-branes) \cite{witten-toroidal}; in the IIA reduction, we have O6-planes on top of D6-branes (intuitively, the former ``lift'' to the extra generator of $D_k$ which is not present in $A_{k-1}$) suspended bewteen NS5-branes. The NS5's also fractionate, producing a sequence of alternating $\SO(2k)$ and $\USp(2(k-4))$ gauge groups \cite{delzotto-heckman-tomasiello-vafa}.\footnote{For real compact symplectic groups we use the following notation: $\USp(2k)=\Sp(k)$, of real dimension $k(2k+1)$ and rank $2k$. The notation implies that the real compact symplectic group is isomorphic to the one of unitary $2k \times 2k$ symplectic matrices. Indeed we could also write $\Sp(k)=\U(k,\mathbb{H})$, unitary $k\times k$ matrices on the quaternions. The real compact special orthogonal group $\SO(2k)$ has real dimension $k(2k-1)$ and rank $2k$.} (See figure \ref{fig:NS5D6O6} for the brane setup and \ref{fig:quiverNS5D6O6} for the unHiggsed quiver.) They will contribute $N_\text{T} = N-1=2M-1$ $(1,0)$ tensor multiplets. The first gauge group is engineered through an O6$^-$ projection on $\SU(k)$, the second through an O6$^+$ one (the O6 charge changes sign whenever the plane crosses an NS5, an effect first discussed in \cite{evans-johnson-shapere} for O4's). The rank of both gauge groups is always even, a fact that is related to the number of $k$ mirror pairs of D6's under the O6 projection. Moreover, the ``jump'' in the ranks of consecutive gauge groups ($2k \to 2k-8$ or $2k-8 \to 2k$) can be explained in field theory as a consequence of condition \eqref{eq:flavors}, and in string theory as the fact that an O6$^\pm$-plane has $\pm4$ D6 charge (thereby modifying the effective number of D6's in a finite-length stack).

As before, we can modify the tail structure of the orbifold SCFT by adding D8-branes, as depicted in figure \ref{fig:NS5D6O6D8}. Gauge-anomaly cancellation at each node enforces the following condition \cite[Eqs. (4.11), (4.12)]{mekareeya-rudelius-tomasiello} (which can be derived from \eqref{eq:ti-gaugeanom}):
\begin{equation}\label{eq:flavors-sok}
f_i+16 = 2p_i -q_i -q_{i-1}\ , \quad g_i-16 = 2q_i -p_i -p_{i-1}\ ; \quad i=1,\ldots,M\ .
\end{equation}
$f_i$ ($g_i$) is the number of half-hypermultiplets in the (real) vector representation of a gauge $\SO(p_i)$ ($\USp(q_i)$) group; $N-1=2M-1$ when $f_1=p_1=0$ (i.e. the quiver starts off with a $\USp(q_1)$ gauge group), or $N-1=2M$ if $f_1, p_1 \neq 0$.

As usual, adding D8-branes corresponds to a Higgsing of the theory which will be specified by two nilpotent orbits of $\mathfrak{so}(2k)$ (one for each tail), defined by two (transposed) ``even'' partitions $\lambda^\text{t}$ of $2k$ \cite[Thm. 5.1.4]{collingwood-mcgovern}. The quiver can now be written as in (the left frame of) figure \ref{fig:quiverNS5D6O6D8}. (Notice that in a Higgsed quiver we might encounter odd-rank $\SO$ groups as well, see e.g. \cite[Fig. 6]{heckman-rudelius-tomasiello}. This corresponds to a so-called $\widetilde{\text{O6}}^-$, whereby a half-D6 is stuck on the O-plane. K-theory then requires having an odd quantum $n_0$ of Romans mass $F_0$ \cite{bergman-gimon-sugimoto}, i.e. an odd number of D8-branes crossing the $\text{D6}$-$\widetilde{\text{O6}}^-$ stack.)

For each massive tail, suppose we define $\rho_i := \lambda^\text{t}_i - \lambda^\text{t}_{i+1}$ for $i=1,\ldots,n-1$ and $\rho_n := \lambda^\text{t}_n$ using $\lambda^\text{t} = [\lambda^\text{t}_1, \lambda^\text{t}_2,\ldots,\lambda^\text{t}_n]$. (The number $n$ can be found by transposing the chosen partition.) The $\rho_i$ give the numbers of D8-branes crossing the $i$-th D6-O6 stack, whereas the ``ranks'' are defined as sums of parts of $\lambda^\text{t}$ as follows \cite[Eq. (4.34)]{mekareeya-rudelius-tomasiello}:
\begin{equation}\label{eq:ranks-sok}
\varrho_i := -8 + \sum_{k=1}^i \lambda^\text{t}_k\ , \ \text{for $i$ odd}\ ; \quad \varrho_i :=  \sum_{k=1}^i \lambda^\text{t}_k\ , \ \text{for $i$ even}\ .
\end{equation}
Here $\varrho_i$ is the rank of a gauge $\USp$ ($\SO$) group for $i$ odd (even). In figure \ref{fig:quiverNS5D6O6D8}, the former is represented by a black circle, the latter by a gray one. The number of D6-branes in each color stack is given by $2r_i:=\sum_{k=1}^i \lambda^\text{t}_k$, and this is what we will call rank of the gauge group in the large $N$ computation of sections \ref{sub:O6poles} and \ref{sub:O6polesgrav}. (The factor of 2 comes from counting physical branes and their images, in our conventions.) Using this partition-inspired notation, \eqref{eq:flavors-sok} reads (for each massive tail) \cite[Eq. (4.36)]{mekareeya-rudelius-tomasiello}:
\begin{equation}\label{eq:flavors-sok-part}
\rho_{2i-1}+16 = 2\varrho_{2i-1} - \varrho_{2i-2} -\varrho_{2i}\ , \quad \rho_{2i}-16 = 2\varrho_{2i} - \varrho_{2i-1} -\varrho_{2i+1}\ ; \quad i=1,\ldots,n\ .
\end{equation}
If there is a massless plateau, the maximum rank is given by $2k$, and in this region the quiver looks like that in figure \ref{fig:quiverNS5D6O6}.

We also remark that, in the $\SO$-$\USp$ case, the perturbative IIA picture may break down due to the appearance of (hypermultiplet) spinor representations which cannot be engineered in string theory. One must then turn to the F-theory description of the $(1,0)$ theory \cite{delzotto-heckman-tomasiello-vafa,heckman-rudelius-tomasiello,mekareeya-rudelius-tomasiello}. However one may still use a ``formal'' massive IIA brane configuration (where we formally allow for a non-positive number of D6-branes in some finite-length stacks) to compute the $a$ conformal anomaly of the $(1,0)$ quiver engineered in F-theory \cite{mekareeya-rudelius-tomasiello}. As it turns out, the result agrees with the nonperturbative F-theory calculation. \cite{mekareeya-rudelius-tomasiello} also found a necessary and sufficient condition to engineer one such formal massive IIA quiver: The largest part $\lambda^\text{t}_1$ of an ordered transposed even partition $\lambda^\text{t}$ of $2k$ is less or equal to 8. One immediately notices that the principal (or regular) orbit $\mathcal{O}_{[2k-1,1]}$ of $\mathfrak{so}(2k)$ satisfies this condition (since $\lambda^\text{t}=[2,1^{2k-2}]$). In section \ref{sub:formal} we will construct for the first time the AdS$_7$ vacuum dual to this quiver (depicted in figure \ref{fig:formalquiv55}), and we will extract its $a$ conformal anomaly at large $N$.\\

The corresponding supergravity vacua will differ from those dual to NS5-D6-D8-engineered quivers only for the presence of a nonzero constant term of the cubic polynomial $\alpha(z)$ in the intervals $[0,1]$ and $[N-1,N]$. We call them $\alpha_0$ (when $\alpha(z)$ is supported on $[0,1]$) and $\alpha_N$ (when $\alpha(z)$ is supported on $[N-1,N]$). These constants are vanishing in the pure $\SU(k)$ case, but are nonvanishing when O6-planes of negative charge are present at the end of the quiver. As we will explain in greater detail in section \ref{sub:boundary}, $\alpha_0,\alpha_N$ can indeed be related to the effective D6 charge of a D6-O6$^-$ source localized at the poles of the internal space $M_3$. This charge is given by $\tilde{r}_{0,N}:=-4+2 n_2=-4,-2,0$, with $n_2=0,1,2$ pairs of D6-branes, or by $\tilde{r}_{0,N}:=-4+2 n_2 +2\tfrac{1}{2}=-3,-1$ with $n_2=0,1$ if also a half-D6 is present. (For $n_2=2$ the total D6 charge is zero, and the pole is regular.) A nonvanishing D6 charge $n_2$ (or $n_2+\tfrac{1}{2}$) can then be associated with flavor symmetries $\SO(g_{1,M}=2n_2)$ (or $\SO(g_{1,M}=2n_2+1)$) of positive rank in the unHiggsed quiver of figure \ref{fig:quiverNS5D6O6} or the left Higgsed quiver of figure \ref{fig:quiverNS5D6O6D8} (when $f_1=p_1=0$).

In both cases, the metric will be singular at the poles, and the dilaton divergent. The $S^2$ fibers of $M_3$ will be replaced by $\rr\pp^2$ ones due to the antipodal action of the O6-planes.

\subsubsection{\texorpdfstring{$\SO$ or $\USp$ flavor, $\USp, \SO, \SU$ gauge groups: O8$^\pm$ onto D8's}{SO or USp flavor, USp, SO, SU gauge groups: O8+- onto D8's}}
\label{subsub:O8pureSUk}

Finally, let us discuss what happens when we overlay an O8-plane onto a stack of flavor D8-branes in the pure $\SU(k)$ case, and then in the alternating $\SO(2k)$-$\USp(2(k-4))$ case. In the first case we have to distinguish two possibilities: When the O8 sits between two consecutive NS5's along $x^6$ (i.e. an NS5 at $\phi_i>0$ and its image at $-\phi_i$) and when the O8 is stuck at the location $\phi_i$ of an NS5.\footnote{Given that $x^6$ parameterizes $\rr$, we need not worry about D8 charge cancellation, but we still need to impose gauge-anomaly freedom, \eqref{eq:ti-gaugeanom}, i.e. an appropriately modified version of \eqref{eq:flavors}.} In the second case there exists only the first possibility. Given that the O8-plane acts as a mirror along direction $x^6$, we decide to put the origin of $x^6$ at its position. We will describe the linear quiver that ensues by considering only the ``physical half'' (say the one supported on $[0,+\infty)$) of the gauge and flavor groups.\\

Let us first discuss the case without O6-planes. All consistent brane configurations are depicted in figure \ref{fig:NS5D6D8O8unstuck-stuck}. In the first situation the O8$^\pm$-plane (carrying $\pm 16$ units of D8 charge) will cross the $(i=0)$-th segment of $r_0=k$ D6-branes, projecting the gauge group to $\SO(2k)$ or $\USp(2k)$ respectively.\footnote{Notice that, because of the projection around $x^6=0$, the 0-th stack of D6-branes engineers a gauge group, rather than a flavor one as in the pure $\SU(k)$ case. That stack is connected to its image at finite distance on the other side of the O8.} (The first NS5 at $\phi_1 >0$ contributes a single $(1,0)$ tensor multiplet and a decoupled hypermultiplet which is neutral under the 0-th gauge group.) The next finite-length D6 stacks are not affected by the O8 projection, and their gauge group will be $\SU(r_i)$ with even ranks $r_i=2k\mp i(8 \pm n_0)$ for $i=1,\ldots,N-1$. As we will see, this is obtained by repeatedly applying condition \eqref{eq:ti-gaugeanom} (which is a generalization of \eqref{eq:flavors}) at each node, starting with $r_0=2k$ and $f_0=2n_0$ half-hypermultiplets (i.e. $\epsilon^\text{flv}_0=\tfrac{1}{2}$ in \eqref{eq:ti-gaugeanom}). For $i=N$ the semi-infinite D6-branes engineer an $\SU(r_N=2k\mp N(8-n_0))$ flavor group.

We can now add more D8-branes (as in the pure $\SU(k)$ case), and engineer a left massive region, followed by a massless plateau, followed by a right massive region as long as condition \eqref{eq:ti-gaugeanom} is satisfied. (Notice that, without extra D8's, the number $k$ is constrained by $N$ and $n_0$ upon requiring $2k\mp N(8\pm n_0)\geq0$, in order to have a meaningful $\SU(r_N)$ flavor group). Moreover, the $n_0$ D8 pairs overlaid onto the O8$^\pm$ engineer an extra $\USp(2n_0)$ ($\SO(2n_0)$) flavor symmetry. Finally, on top of the $f_{N-1}$ fundamental (of $\SU(r_{N-1})$) hypermultiplets contributed by D8-branes, we can have $r_N$ flavor D6-branes escaping off to infinity. The quivers are depicted in figure \ref{fig:quiverNS5D6D8O8-unstuck}.

A very interesting subcase arises when $k=0 \Rightarrow r_0=2k=0$, i.e. the first gauge group is empty, and we stick say $8-n_0$ D8 pairs on the O8$^-$ (see figure \ref{fig:massiveE}). There is no orientifold projection on any of the gauge groups, and condition \eqref{eq:flavors} simply imposes $r_i=n_0 i$ for $i=1,\ldots,N-1$ (with a flavor symmetry $\SU(r_N=Nn_0)$ on the right). The $(1,0)$ SCFT corresponding to this quiver is very similar to the so-called (rank-$N$) E-string theory \cite{heckman-morrison-vafa}, with the crucial difference that it cannot be engineered in M-theory because of the D8's (sourcing a nonzero Romans mass $F_0=\frac{n_0}{2\pi}$). For this reason it was dubbed (rank-$N$) ``massive E-string theory'' in \cite{bpt}. (In particular when $n_0=1$ we have an extra $E_8$ flavor symmetry on the left, whose presence can be argued for by lifting the particular D8-O8$^-$ system to M-theory as in \cite{polchinski-witten};\footnote{The nonperturbative enhancement $\mathrm{SO}(2(8-n_0)) \to E_{1+(8-n_0)}$ is due to D0-branes \cite{ganor-hanany,seiberg-5d}, which become tensionless as $g_\mathrm{s} \to \infty$ since $T_\text{D0} \sim g_\mathrm{s}^{-1}$.} more generally, for $1\leq n_0 \leq 8$ we have an $E_{1+(8-n_0)}$ symmetry, using the definitions of \cite{seiberg-5d}.) The quiver was constructed in \cite[Fig. 6]{bpt} (which we reproduce in figure \ref{fig:massiveE}), the dual AdS$_7$ vacuum is given by \cite[Eq. (5.2)]{bpt} and its $a$ conformal anomaly at large $N$ by \cite[Eq. (5.13)]{bpt}.\footnote{In that formula $M$ is the number of D6-branes in the rightmost semi-infinite flavor stack, i.e. $M:=Nn_0$, which also diverges as $N \to \infty$.} Given that this case has already been discussed at length in \cite{bpt}, in the remainder we will only treat the generic case where $k \neq 0$, i.e. $r_0 \neq 0$. \\

In the second situation, the O8$^\pm$ sits on top of a half-NS5-brane stuck on the plane, at $x^6=0$. The orientifold projects out the tensor multiplet contributed by that NS5, does not act on the gauge group $\SU(r_0$), but acts on the bifundamental matter coming from strings D6$_0$-image D6$_0$: We have a hypermultiplet in the (anti)symmetric of $\SU(r_0=k)$. If we also overlay $n_0$ pairs of image D8's onto the O8$^\pm$, all gauge groups will be $\SU(r_i= k \mp i(8\pm2n_0))$ for $i=1,\ldots,N-1$, and we have a flavor symmetry $\SU(r_{N}=k\mp N(8\pm2n_0))$ (again, this is due to \eqref{eq:ti-gaugeanom}). We can also add D8-branes as usual. The quivers are depicted in figure \ref{fig:quiverNS5D6D8O8-stuck}.\\

The corresponding supergravity vacua will be defined in terms of a polynomial $\alpha(z)$ that, in the first subinterval $z\in [0,1]$, is characterized by a nonzero constant term $\alpha_0$ as well as coefficient $r_0$ of the quadratic term, and in the last subinterval $z\in[N-1,N]$ by a nonzero quadratic coefficient $r_N$ (this is a singular pole), but vanishing constant term $\alpha_N$, as the right tail of the quiver is the same as for the pure $\SU(k)$ case. The correspondence will be made more precise in section \ref{sec:solz}.\\

Let us now discuss the case with O6-planes overlaid onto D6-branes. The brane configurations are depicted in figure \ref{fig:NS5D6O6D8O8unstuck}. Given that the O6 charge changes sign whenever the former crosses an NS5, the situation where an O8 is stuck on the latter at $x^6=0$, reflecting an O6$^\pm$ into itself, is not consistent. Thus we only need to consider the first situation (O8-plane between an NS5 at $\phi_1>0$ and its image at $-\phi_1$). The combined O6$^+$-O8$^-$ projection produces an $\SU(r_0=k)$ gauge group (i.e. only half of the $k$ D6 pairs under the O6 projection count), followed by a sequence of gauge groups $\SO(p_i) \times \USp(q_i)$ for $i=1,\ldots,N-1$, and an $\SO(p_N)$ or $\USp(q_N)$ flavor symmetry, according to the particular theory at hand (i.e. $q_N=0$ or $p_N=0$ respectively). Notice that by simultaneously flipping the O6 and O8 charge we can exchange the gauge factors as to produce a sequence $\USp(q_i) \times \SO(p_i)$. Once again, we can add flavors of each gauge group by inserting D8-branes across the finite-length D6 stacks (e.g. $n_0$ D8 pairs overlaid onto the O8$^\pm$-plane will produce a $\USp(2n_0)$, respectively $\SO(2n_0)$, flavor group) as long as conditions \eqref{eq:ti-gaugeanom} are satisfied at each node. The quivers are depicted in figure \ref{fig:quiverNS5D6O6D8O8-unstuck}.

The supergravity vacua will be characterized by a smooth pole of $M_3$ at $z=0$, i.e. the cubic polynomial $\alpha(z)$ supported on $[0,1]$ will have nonvanishing constant term $\alpha_0$ and quadratic coefficient $r_0=k$, and by a singular pole at $z=N$, that is the cubic polynomial will have nonvanishing constant term $\alpha_N$ but vanishing quadratic coefficient $r_N$ (due to the rightmost D6-O6$^-$ stack of total negative charge), or vanishing $\alpha_N$ but nonvanishing $r_N$ (due to the rightmost D6-O6$^\pm$ with total positive charge). The fibers of $M_3$ are $\rr\pp^2$'s due to the antipodal action of the O6-planes.

% PURE SU CASE
\begin{figure}[!ht]
\vspace{-.75cm}
\centering	
\begin{minipage}{1\textwidth}
\captionsetup{type=figure}
\centering

\subcaptionbox{The reduction to IIA of $N$ M5's probing $\cc^2/A_{k-1}$. Circles represent NS5's spread out along $x^6$ (the horizontal direction); solid black lines represent finite-length D6-branes. $r_0=r_N=0$, given that the 0-th and $N$-th stacks are made of (semi-infinite) flavor D6-branes. \label{fig:NS5D6}}[1\columnwidth]{\centering \includegraphics[scale=1]{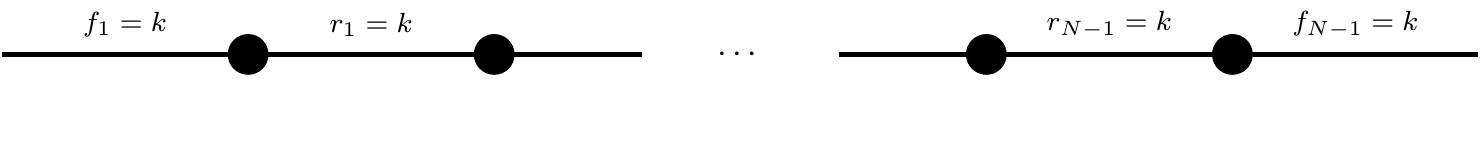}}

\vspace{.5cm}

\subcaptionbox{The fully unHiggsed quiver engineered by the brane configuration in figure \ref{fig:NS5D6}. Blue circles represent $\SU(r_i)$ gauge nodes (vector multiplets, D6$_i$-D6$_i$ strings), connected by bifundamental hypermultiplets (D6$_i$-D6$_{i+1}$ strings) and tensor multiplets (D6$_i$-NS5$_i$ strings). Red boxes represent $\SU(f_i)$ flavor nodes, connected to the former by fundamental hypermultiplets (flavor D6$_i$ - color D6$_i$ strings). The $f_1,f_{N-1}=k$ fundamental flavors of the $\SU(r_1=k)$ and $\SU(r_{N-1}=k)$ gauge groups respectively are equivalently engineered by $k$ semi-infinite D6-branes or $k$ D8-branes via a simple Hanany--Witten move (that does not modify the rest of the configuration). \label{fig:quiverNS5D6}}[1\columnwidth]{
$\NNode{\vver{}{f_{1}=k}}{r_1=k} - \Nodeblue{}{r_2=k} - \cdots - \hspace{.1cm} \Nodeblue{}{r_{N-2}=k} \hspace{.1cm}  - \NNode{\vver{}{f_{N-1}=k}}{r_{N-1}=k}$
}

\vspace{.5cm}

\subcaptionbox{Adding D8-branes to the setup in figure \ref{fig:NS5D6}, and Higgsing the theory as in \cite{gaiotto-witten-1}. Vertical lines (extending along directions $x^{7,8,9}$) represent flavor D8-branes crossing the D6-branes. \label{fig:NS5D6D8}}[1\columnwidth]{\centering \includegraphics[scale=1]{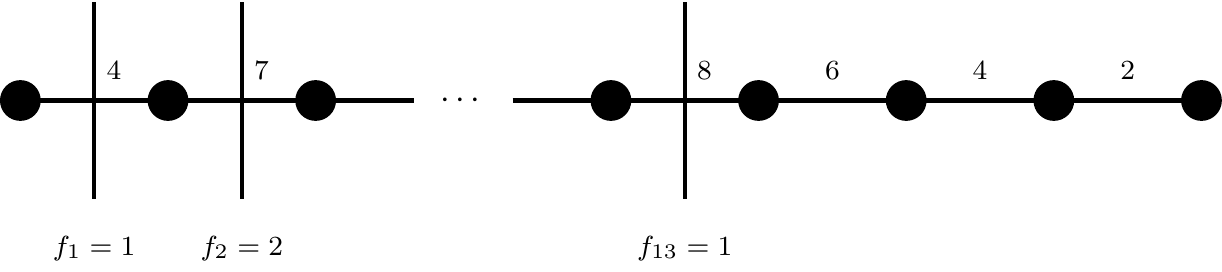}}

\vspace{.5cm}

\subcaptionbox{A more general quiver corresponding to the brane setup of figure \ref{fig:NS5D6D8}: $N=17$, $L=5$, $R=6$, the partitions of $10$ are given by \eqref{eq:young-ex-suk}, and $r_0=r_N=0$. The SCFT is at a generic point of its tensor and Higgs branch. \label{fig:quiverNS5D6D8}}[1\columnwidth]{
$\NNode{\vver{}{1}}{4}-\NNode{\vver{}{2}}{7}-\Nodeblue{}{8}-\Nodeblue{}{9}-\NNode{\vver{}{1}}{10}-\Nodeblue{}{10}-\Nodeblue{}{10}-\Nodeblue{}{10}-\Nodeblue{}{10}-\Nodeblue{}{10}-\NNode{\vver{}{1}}{10}-\Nodeblue{}{9}-\NNode{\vver{}{1}}{8}-\Nodeblue{}{6}-\Nodeblue{}{4}-\Nodeblue{}{2}$
}

\vspace{.5cm}

\subcaptionbox{The internal space $M_3 \cong S^3$ of the AdS$_7$ vacuum which is dual to the quiver in figure \ref{fig:quiverNS5D6D8}. The impression depicts the $S^2$ fibers over the base interval $I=[0,N]$ parameterized by $z$ (related to $x^6$ by the near-horizon limit). Notice that the poles of $S^3$, at the extrema of the base interval, are smooth points for the metric (the $S^2$ fiber smoothly shrinks to zero size). A black crease represents a stack of $f_i$ magnetized D8-branes (we will call stack even a single D8-brane) wrapping a particular $S^2$ fiber over the point $z=i$. \label{fig:solNS5D6D8}}[1\columnwidth]{\includegraphics[scale=.55]{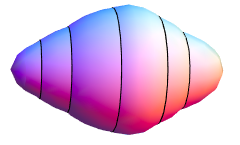}}
%The right impression depicts the singular poles of $M_3$ due to $r_0,r_N$ D6-brane sources localized there.

\end{minipage}
\caption{Brane configurations and quivers for $\SU$ gauge and flavor groups.}
\label{fig:SUk}
\end{figure}

% ALTERNATING  SO-USp CASE
\begin{figure}[!ht]
\centering	
\begin{minipage}{1\textwidth}
\captionsetup{type=figure}
\centering

\subcaptionbox{The reduction to IIA of $M$ M5's probing $\cc^2/D_k$. A $\frac12$ superposed on a black dot indicates a half-NS5-brane. A red dashed line represents an O6$^+$ overlaid onto $k$ pairs of D6-branes (the stack has an effective $2(k-4)$ D6 charge), a blue one an O6$^-$ overlaid onto $k$ pairs. \label{fig:NS5D6O6}}[1\columnwidth]{\centering \includegraphics[scale=1.1]{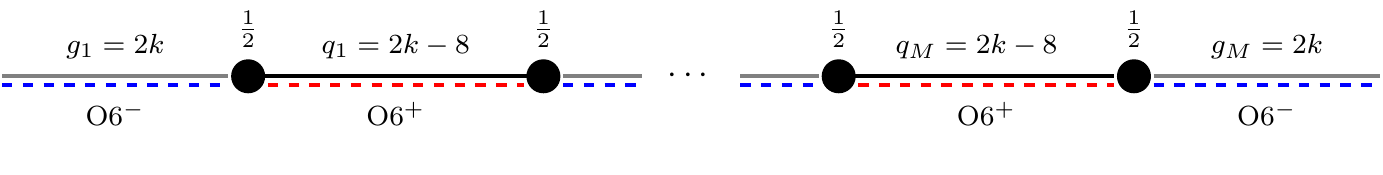}}

\vspace{.5cm}

\subcaptionbox{The quiver engineered by the brane configuration in figure \ref{fig:NS5D6O6}. Black (gray) circles represent $\USp(q_i)$ ($\,\SO(p_i)$) gauge groups, whereas black (gray) squares represent $\USp(f_i)$ ($\,\SO(g_i)$) flavor groups.  There are $M-1$ $\SO(2k)$ gauge groups and $M$ $\USp(2(k-4))$ gauge groups, $N_\text{T}=N-1=2M-1$ tensor multiplets and $N-2=2M-2$ hypermultiplets (both represented by a $-$). \label{fig:quiverNS5D6O6}}[1\columnwidth]{
$\Node{\ver{}{g_1=2k}}{q_1=2k-8} -\node{}{p_1=2k}-\Node{}{q_2=2k-8}-\cdots -\node{}{p_{M-1}=2k} -\Node{\ver{}{g_M=2k}}{q_M=2k-8}$
}

\vspace{.5cm}

\subcaptionbox{Stacks of $\rho_i$ D8-branes crossing the D6-O6 stacks in a massive tail. The O6$^\pm$ projects the $\SU$ flavor group engineered by the $i$-th D8 stack to $\SO(\rho_i)$ ( $\USp(\rho_i)$), which is represented by a vertical gray (black) line. All NS5's are half-branes.\label{fig:NS5D6O6D8}}[1\columnwidth]{\centering \includegraphics[scale=1.1]{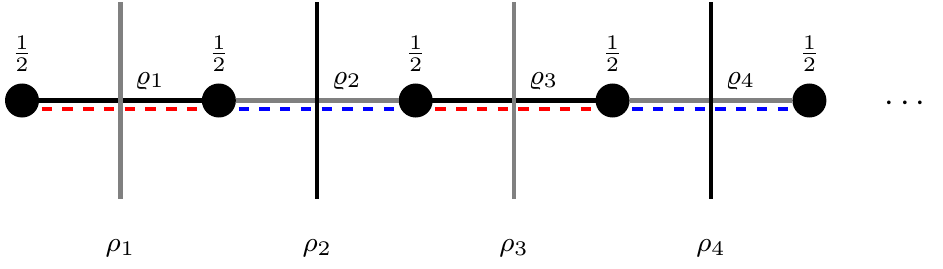}}

\vspace{.5cm}

\subcaptionbox{A more general quiver corresponding to the brane setup of figure \ref{fig:NS5D6O6D8}. On the left we use the same convention as in figure \ref{fig:quiverNS5D6O6} (there are $N=2M$ or $2M-1$ gauge groups, the latter when $p_1=f_1=0$); on the right we use the partition-inspired convention \eqref{eq:ranks-sok}, \eqref{eq:flavors-sok-part}.\label{fig:quiverNS5D6O6D8}}[1\columnwidth]{
$\node{\Ver{}{f_1}}{p_1} -\Node{\ver{}{g_1}}{q_1}-\cdots-\node{\Ver{}{f_{i}}}{p_{i}}-\Node{\ver{}{g_{i}}}{q_{i}}-\cdots-\node{\Ver{}{f_{M}}}{p_M} -\Node{\ver{}{g_{M}}}{q_{M}} \quad\quad\quad \cong \quad\quad\quad  \Node{\ver{}{\rho_{1}}}{\varrho_{1}}-\node{\Ver{}{\rho_{2}}}{\varrho_{2}}-\Node{\ver{}{\rho_3}}{\varrho_3}-\node{\Ver{}{\rho_{4}}}{\varrho_{4}}-\cdots$
}

%\vspace{.5cm}

%\subcaptionbox{The internal space $M_3$ with $\rr\pp^2$ fibers of the AdS$_7$ vacuum which is dual to the quiver in figure \ref{fig:quiverNS5D6O6D8}. The singular poles of $M_3$ denote localized O6-sources at $z=0,N$. \label{fig:solNS5D6O6D8}}[1\columnwidth]{{\color{red}ADD PICTURE HERE OR REMOVE SUBFIGURE ENTIRELY}}%\includegraphics[scale=.55]{}}

\end{minipage}
\caption{Brane configurations and quivers for alternating $\SO$-$\USp$ gauge and flavor groups.}
\label{fig:SOk}
\end{figure}

% O8 CASE NO O6
\begin{figure}[!ht]
\centering	
\begin{minipage}{1\textwidth}
\captionsetup{type=figure}
\centering

\subcaptionbox{NS5-D6-D8-O8 configuration with an O8$^+$-plane between an NS5 and its image, and crossing the 0-th D6 stack. A red dashed line and two vertical black lines represent an O8$^+$ overlaid onto $n_0$ pairs of D8-branes, engineering a $\USp(2n_0)$ flavor group and contributing $2n_0$ half-hypermultiplets. A blue (gray) horizonatal line represents a D6 stack engineering an $\SU$ ($\,\SO$) gauge group.\label{fig:NS5D6D8O8+unstuck}}[1\columnwidth]{\centering \includegraphics[scale=1]{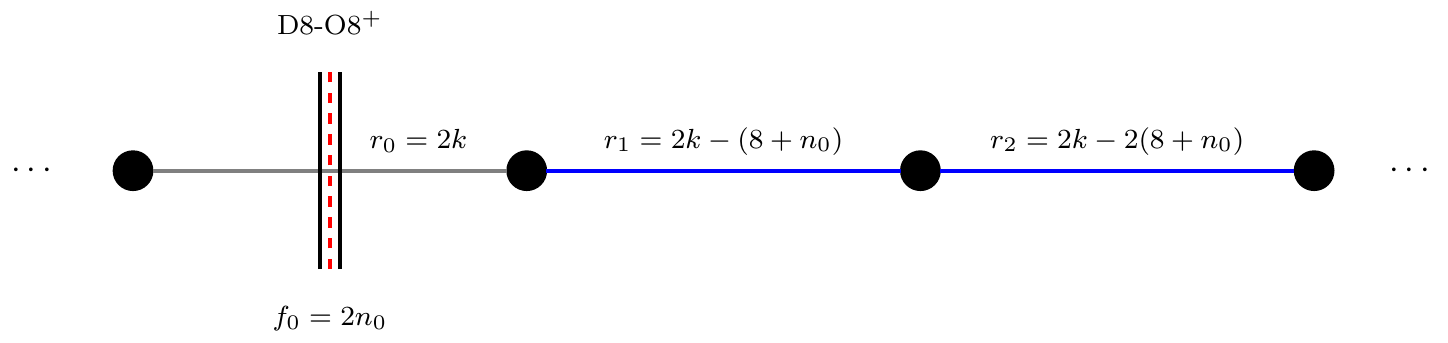}}

\subcaptionbox{NS5-D6-D8-O8 configuration with an O8$^-$-plane between an NS5 and its image, and crossing the 0-th D6 stack. A blue dashed line and two vertical gray lines represent an O8$^-$ overlaid onto $n_0$ pairs of D8-branes, engineering an $\SO(2n_0)$ flavor group and contributing $2n_0$ half-hypermultiplets. A blue (black) horizonatal line represents a D6 stack engineering an $\SU$ ($\,\USp$) gauge group. \label{fig:NS5D6D8O8-unstuck}}[1\columnwidth]{\centering \includegraphics[scale=1]{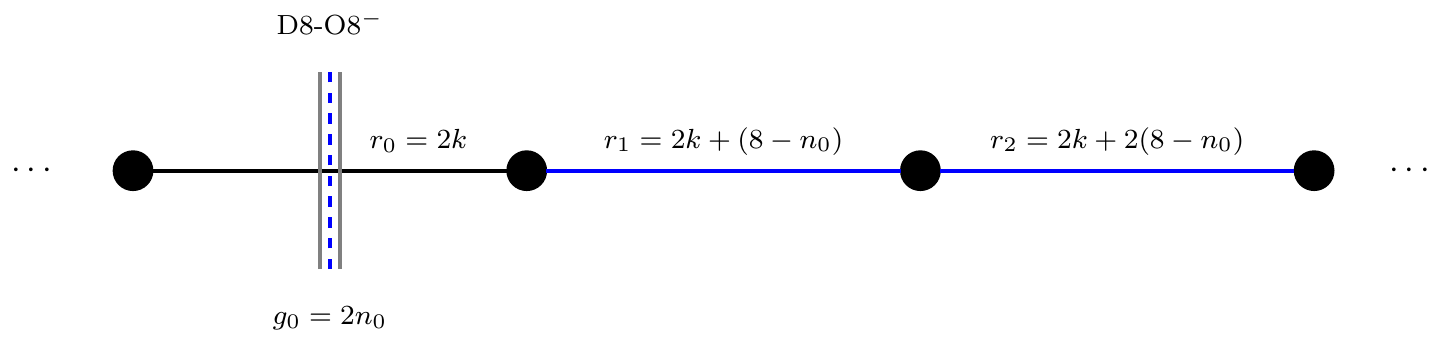}}

\subcaptionbox{NS5-D6-D8-O8 configuration with O8$^+$-plane overlaid onto $n_0$ pairs of D8's (engineering a flavor $\USp(f_0=2n_0)$ flavor symmetry and contributing $2n_0$ full hypermultiplets), and a stuck half-NS5. \label{fig:NS5D6D8O8+stuck}}[1\columnwidth]{\centering \includegraphics[scale=1]{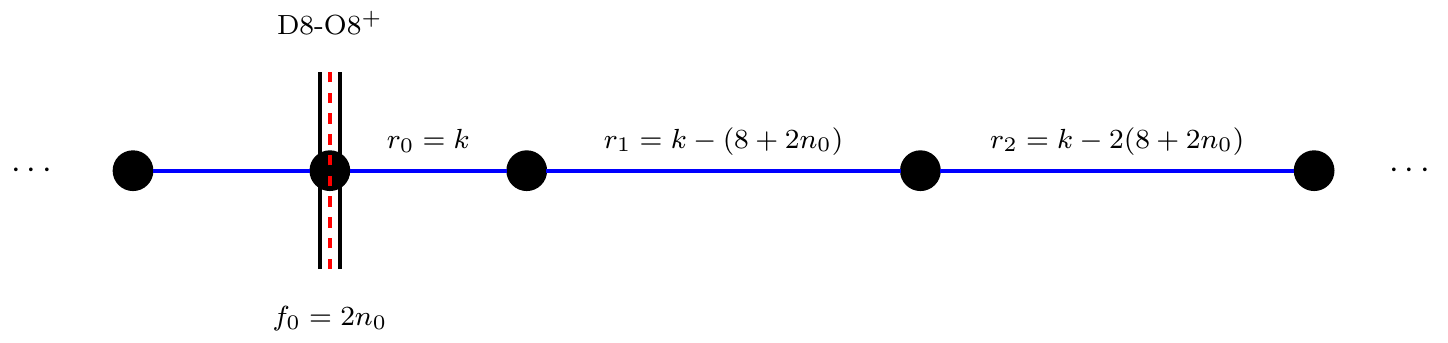}}

\subcaptionbox{NS5-D6-D8-O8 configuration with O8$^-$-plane overlaid onto $n_0$ pairs of D8's (engineering a flavor $\SO(g_0=2n_0)$ flavor symmetry and contributing $2n_0$ full hypermultiplets), and a stuck half-NS5. \label{fig:NS5D6D8O8-stuck}}[1\columnwidth]{\centering \includegraphics[scale=1]{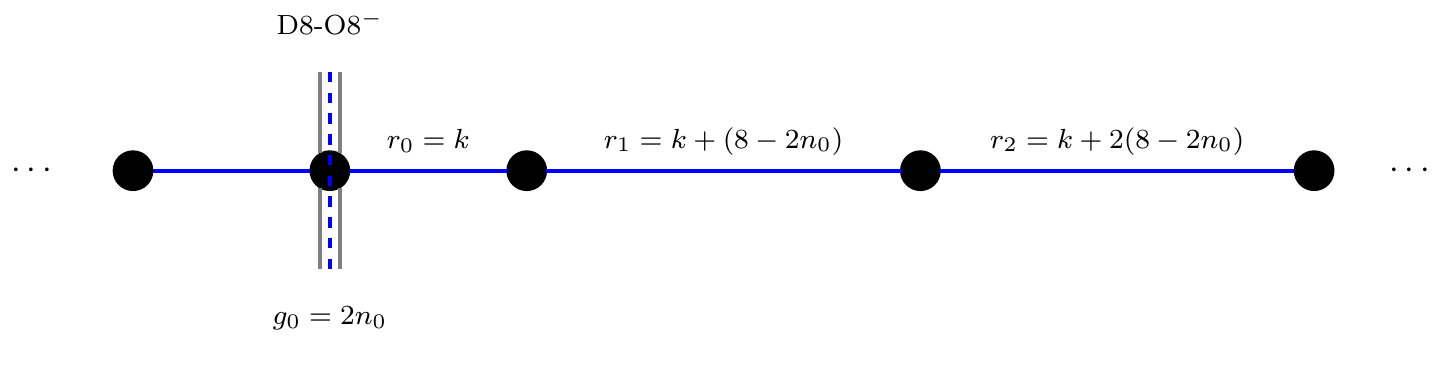}}

\end{minipage}
\caption{Possible NS5-D6-D8-O8 brane configurations without O6-planes or extra D8-branes.}
\label{fig:NS5D6D8O8unstuck-stuck}
\end{figure}

% O8 W/O O6: QUIVERS
\begin{figure}[!ht]
\centering	
\begin{minipage}{1\textwidth}
\captionsetup{type=figure}
\centering

\subcaptionbox{The left quiver corresponds to the brane setup of figure \ref{fig:NS5D6D8O8+unstuck} (O8$^+$ between NS5 and its image), the right one to that of figure \ref{fig:NS5D6D8O8-unstuck} (O8$^-$ between NS5 and its image). \label{fig:quiverNS5D6D8O8-unstuck}}[1\columnwidth]{
$\node{\Ver{}{f_0}}{r_{0}=2k}-\NNode{\vver{}{f_1}}{r_1}-\NNode{\vver{}{f_2}}{r_2}-\NNode{\vver{}{f_3}}{r_3}-\cdots \quad \quad \quad \quad \Node{\ver{}{g_0}}{r_0=2k}-\NNode{\vver{}{f_1}}{r_1}-\NNode{\vver{}{f_2}}{r_2}-\NNode{\vver{}{f_3}}{r_3}-\cdots$
}

\vspace{1cm}

\subcaptionbox{The left quiver corresponds to the brane setup of figure \ref{fig:NS5D6D8O8+stuck} (O8$^+$ stuck on a half-NS5); there is a hypermultiplet in the symmetric representation of $\SU(r_0=k)$, which we represent by $\circlearrowright$. The right one to that of figure \ref{fig:NS5D6D8O8-stuck} (O8$^-$ stuck on a half-NS5); there is a hypermultiplet in the antisymmetric representation of $\SU(r_0=k)$, which we represent by $\circlearrowleft$.  \label{fig:quiverNS5D6D8O8-stuck}}[1\columnwidth]{
$\Nodeblueleft{\Ver{}{f_0}}{r_{0}=k}{\circlearrowright} -\NNode{\vver{}{f_1}}{r_1}-\NNode{\vver{}{f_2}}{r_2}-\NNode{\vver{}{f_3}}{r_3}-\cdots \quad \quad \quad \quad \Nodeblueleft{\ver{}{g_0}}{r_{0}=k}{\circlearrowleft}-\NNode{\vver{}{f_1}}{r_1}-\NNode{\vver{}{f_2}}{r_2}-\NNode{\vver{}{f_3}}{r_3}-\cdots$
}

\vspace{1cm}

\subcaptionbox{NS5-D6-D8-O8$^-$ brane configuration engineering a rank-$N$ massive $E_{1+(8-n_0)}$-string theory on the tensor branch (simply called massive E-string theory when $n_0=1$). There are $8-n_0$ pairs of D8-branes overlaind onto the O8$^-$. The flavor group $\SO(2(8-n_0))$ can be argued to enhance to $E_{1+(8-n_0)}$ at strong coupling \cite{seiberg-5d}. \label{fig:massiveE}}[1\columnwidth]{\centering \includegraphics[scale=1]{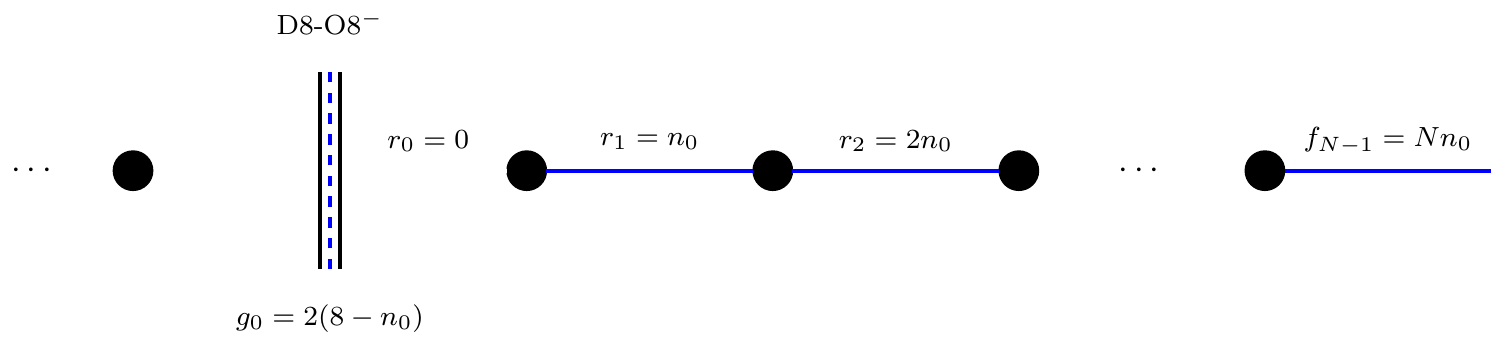}}

\vspace{1cm}

\subcaptionbox{The rank-$N$ massive $E_{1+(8-n_0)}$-string theory quiver. The first gauge group is empty, but there is a tensor multiplet (with $E_{1+(8-n_0)}$ enhanced flavor symmetry) which we represent by the leftmost $-$. \label{fig:quiver-massiveE}}[1\columnwidth]{
$\NNode{\ver{}{g_0=2(8-n_0)}}{r_{0}=0}-\Nodeblue{}{r_1=n_0}-\Nodeblue{}{r_2=2n_0}-\Nodeblue{}{r_3=3n_0}- \cdots - \NNode{\vver{}{f_{N-1}=Nn_0}}{r_{N-1}=(N-1)n_0}$
}

\end{minipage}
\caption{Quivers engineered by NS5-D6-D8-O8$^\pm$ brane configurations. Notice that we have added possible flavors for each gauge node for $i=1,\ldots,N-1$. ($f_0=g_0=2n_0$ is the rank of the leftmost flavor $\USp / \SO$ group respectively.) This can be done as long as condition \eqref{eq:ti-gaugeanom} is satisfied at each node. We use same colors and names as those in figures \ref{fig:quiverNS5D6D8} and \ref{fig:quiverNS5D6O6D8}. In figures \ref{fig:massiveE} and \ref{fig:quiver-massiveE} we see the brane engineering and the quiver describing the rank-$N$ massive E-string theory.}
\label{fig:quiverNS5D6D8O8stuck-unstuck}
\end{figure}

% O8 CASE WITH O6
\begin{figure}[!ht]
\centering	
\begin{minipage}{1\textwidth}
\captionsetup{type=figure}
\centering

\subcaptionbox{NS5-D6-O6-D8-O8 configuration with O8$^+$-plane between an NS5 and its image, and crossing the 0-th D6-O6$^-$ stack. A red vertical dashed line paired up with two black lines represent an O8$^+$ overlaid onto $n_0$ pairs of D8-branes, engineering a $\USp(2n_0)$ flavor symmetry and contributing $2n_0$ half-hypermultiplets. A blue (black/gray) horizonatal line represents a D6 stack engineering an $\SU$ ($\,\USp$/$\,\SO$) gauge group.\label{fig:NS5D6O6D8O8+unstuck}}[1\columnwidth]{\centering \includegraphics[scale=1.1]{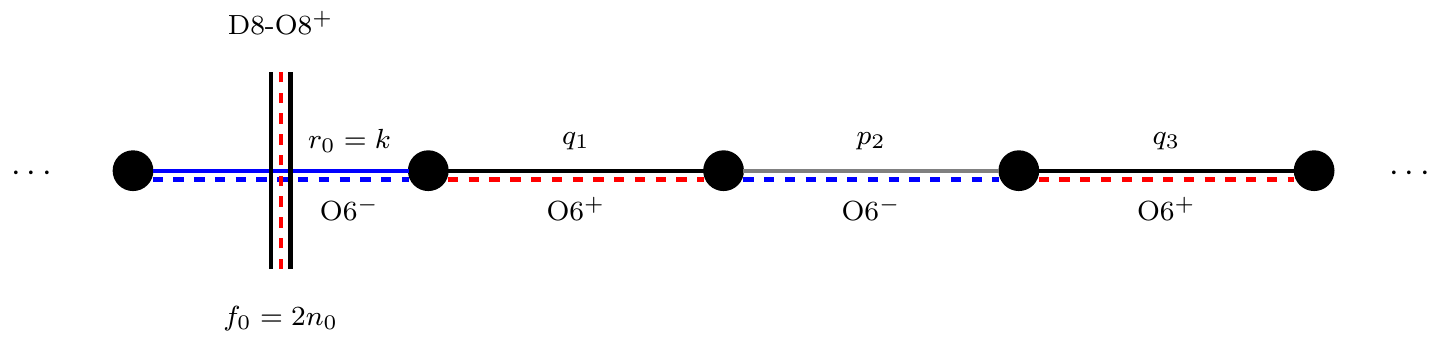}}

\vspace{.5cm}

\subcaptionbox{NS5-D6-O6-D8-O8 configuration with O8$^-$-plane between an NS5 and its image, and crossing the 0-th D6-O6$^+$ stack. A blue vertical dashed line paired up with two black lines represent an O8$^-$ overlaid onto $n_0$ pairs of D8-branes, engineering an $\SO(2n_0)$ flavor symmetry and contributing $2n_0$ half-hypermultiplets. A blue (gray/black) horizonatal line represents a D6 stack engineering an $\SU$ ($\,\SO$/$\,\USp$) gauge group.\label{fig:NS5D6O6D8O8-unstuck}}[1\columnwidth]{\centering \includegraphics[scale=1.1]{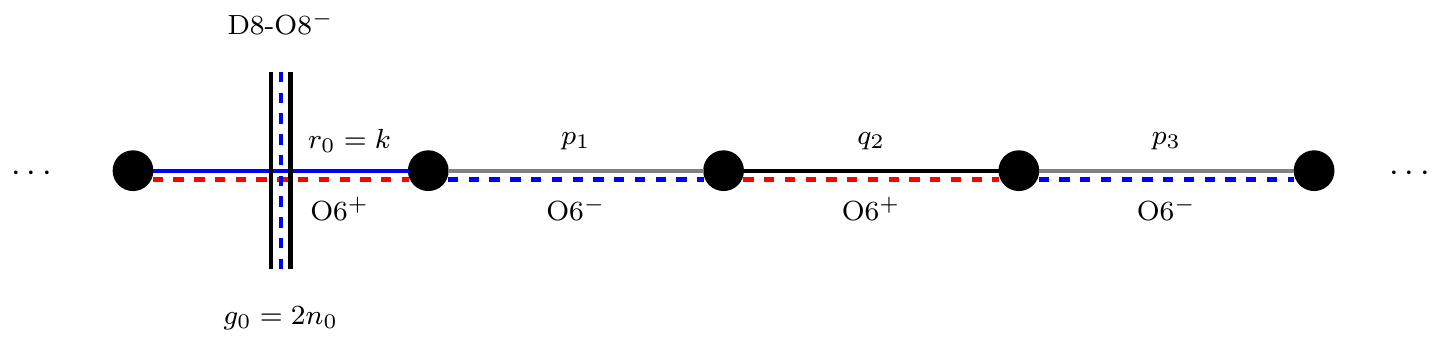}}

\vspace{.5cm}

\subcaptionbox{The left quiver corresponds to the brane setup of figure \ref{fig:NS5D6O6D8O8+unstuck} (O8$^+$-O6$^-$ combined projection), the right one to that of figure \ref{fig:NS5D6O6D8O8-unstuck} (O8$^-$-O6$^+$ combined projection). Notice that we have added possible flavors for each gauge node. This can be done as long as conditions \eqref{eq:ti-gaugeanom} are satisfied. Notice the difference with figure \ref{fig:quiverNS5D6O6D8} (we use same colors and names for gauge and flavor nodes). \label{fig:quiverNS5D6O6D8O8-unstuck}}[1\columnwidth]{
$\NNode{\verbl{}{f_0}}{r_{0}=k}-\Node{\ver{}{g_1}}{q_1}-\node{\Ver{}{f_2}}{p_2}-\Node{\ver{}{g_3}}{q_3}-\cdots \quad \quad \quad \quad \NNode{\ver{}{g_0}}{r_{0}=k}-\node{\Ver{}{f_1}}{p_1}-\Node{\ver{}{g_2}}{q_2}-\node{\Ver{}{f_3}}{p_3}-\cdots$
}

\end{minipage}
\caption{Possible NS5-D6-O6-D8-O8 brane configurations and the linear quivers they engineer.}
\label{fig:NS5D6O6D8O8unstuck}
\end{figure}

\clearpage

\subsubsection{The holographic limit}
\label{subsub:hololim}

Having heuristically explained how the near-horizon limit of the various brane configurations might work, we now set out to find the correct ``holographic limit''. By this we mean the limit that suppress curvature and $g_\text{s}$ corrections to the closed string spectrum sourced by the brane setup, allowing us to reliably use the classical AdS$_7$ supergravity vacua.  Usually this also turns out to be a so-called large $N$ limit in the dual field theory. 

For the NS5-D6(-O6)-D8(-O8) configurations that engineer six-dimensional $(1,0)$  linear quivers, \cite{cremonesi-tomasiello} identified the correct limit to achieve the aforementioned suppression and at the same time keep track of the nontrivial information contained in the Young tableaux $\rho_\text{L,R}$. (We know that this information labels the Higgsed theory and is associated with the massive tails of the quiver, so it should not be washed away in the limit.) The limit is the following:
\begin{equation}\label{eq:hololimit}
N,L,R,k,r_i \to \infty\ , \quad \frac{L}{N},	\frac{R}{N},\frac{k}{N},	\frac{r_i}{N} \quad \text{finite}\ .
\end{equation}
In particular we see that, in six dimensions, ``large $N$'' means infinite number of gauge groups. $k,r_i \to \infty$ also tells us that the ranks of the various gauge and flavor groups are infinite. In light of table \ref{tab:groupconst} this means that their dual Coxeter numbers (which will play an important role in the holographic match of $a$) are infinite, and approximate the ranks: $h^\vee_G \sim \rk G \to \infty$. We will write $\sim$ to indicate the holographic limit of all relevant quantities.

\subsection{Constructing generic solutions with $z$}
\label{sub:gensol}

We shall now describe in greater detail how to construct the supergravity AdS$_7$ vacua dual to the quivers just introduced, by relying on the same combinatorial data.\footnote{Notice that, once we construct a general AdS$_7$ vacuum of massive IIA, we can easily obtain AdS$_5$ and AdS$_4$ ones from it by applying the one-to-correspondences in \cite{afpt,rota-t}. It is also possible to construct a nonsupersymmetric AdS$_7$ vacuum following the construction in \cite{apruzzi-dibitetto-tizzano}.}\\

The generic AdS$_7$ supergravity vacuum of massive type IIA can be described in terms of a single function $\alpha(z)$ on which all physical fields (metric, dilaton, warping factor, fluxes) depend. The coordinate $z$ parameterizes the base interval $I$ of the three-dimensional internal space $M_3$, which is a fibration of two-spheres over $I$. The total space of the fibration can be made compact by requiring that the fiber shrink at the extrema of $I$. (Thus, topologically, $M_3 \cong S^3$.) This in turn imposes boundary conditions for the internal metric at the poles of $S^3$. Different boundary conditions correspond to different physical sources, such as branes and orientifolds. The existence of these global solutions was first established numerically in \cite{afrt}.\footnote{See \cite{blaback-danielsson-junghans-vanriet-wrase-zagermann,gautason-junghans-zagermann} for an earlier AdS$_7 \times S^3$ Ansatz with smeared sources, and \cite{blaback-danielsson-junghans-vanriet-wrase-zagermann-2} for a local construction.} The solutions were later given a fully-analytic expression in \cite{afpt}, where the function $\alpha(z)$ we just introduced was called $\beta(y)$. Here, we will present the solutions as in \cite{cremonesi-tomasiello}, which further generalizes the formalism of \cite{afpt}.\footnote{A change of variables is needed to go from the presentation of \cite{afpt} to that of \cite{cremonesi-tomasiello} and the present paper. It is shown in appendix \ref{app:var}.}

In \cite{cremonesi-tomasiello} it was imposed that at the poles of $S^3$ the metric be either regular or have the asymptotics corresponding to D6-brane sources. Under the correspondence between NS5-D6(-O6)-D8(-O8) brane configurations and AdS$_7$ vacua we explained in the previous section, a regular asymptotics corresponds to having a stack of D8-branes (with D6 charge smeared on the common worldvolume) wrap an $S^2$ fiber in the vicinity of the pole, whereas the second case to a stack of D6-branes localized at the pole. Both D6 and D8 sources are spacetime filling. (These statements are to be understood after having taken the near-horizon limit of the localized closed string spectrum -- sourced by the brane configuration -- which produces the AdS vacuum.)

In this work we generalize this situation, and allow for several new boundary conditions. For instance, we will construct the most general solution with D6-brane poles and an O8-D8 wall along the equator of $S^3$. A version of this solution -- dual to the so-called massive E-string theory -- has already appeared in \cite{bpt}; here we will generalize it further. We will also see how to introduce O6-planes on top of D6-branes, and show what the boundary conditions for such a combined object look like.\\

Explicitly, the ten-dimensional metric reads
\begin{equation}\label{eq:10dmetricz}
	\frac1{\pi \sqrt2} ds^2_{10}= 8\sqrt{-\frac \alpha{\ddot \alpha}}ds^2_{{\rm AdS}_7}+ \sqrt{-\frac {\ddot \alpha}\alpha} \left(dz^2 + \frac{\alpha^2}{\dot \alpha^2 - 2 \alpha \ddot \alpha} ds^2_{S^2}\right)\ ,
\end{equation}
whereas the dilaton
\begin{equation}\label{eq:dilatonz}
	e^\phi(z)=2^{5/4}\pi^{5/2} 3^4 \frac{(-\alpha/\ddot \alpha)^{3/4}}{\sqrt{\dot \alpha^2-2 \alpha \ddot \alpha}}\ .
\end{equation}
(Morever, as reviewed in appendix \ref{app:var}, $\ddot \alpha<0$.) We also have
\begin{equation}\label{eq:fluxesBF}
	B=\pi \left( -z+\frac{\alpha \dot \alpha}{\dot \alpha^2-2 \alpha \ddot \alpha}\right) {\rm vol}_{S^2}\ ,\qquad F_2 =  \left(\frac{\ddot \alpha}{162 \pi^2}+ \frac{\pi F_0\, \alpha \dot \alpha}{\dot \alpha^2-2 \alpha \ddot \alpha}\right) {\rm vol}_{S^2}\ .
\end{equation}
The (continuous) coordinate $z$ parameterizes the base interval $I=\left[0, N\right]$, which will be divided into subintervals $\left[i, i+1\right]$, $i=0,\ldots,N-1$. The integer $N$ is precisely the number of NS5-branes in the IIA configuration (and is related to the quantized flux of $H$ via $N=-\frac{1}{4\pi^2} \int_{M_3} H$, see e.g. \cite[Eq. (5.42)]{afpt}). The Romans mass $F_0$ is a step function with different values for different subintervals $[i,i+1]$, namely
\begin{equation}\label{eq:F0subint}
F_0=\{F_{0,1},\ldots, F_{0,N}\}\ , \quad F_{0,i+1}=2 \pi n_{0,i+1}=2\pi s_{i+1} :=2 \pi( r_{i+1}-r_i)\ ,
\end{equation}
where $n_{0,i} \in \mathbb{Z}$ (due to flux quantization) and $r_i-1$ are the ranks of the gauge groups $\SU(r_i)$ in a linear quiver description of the dual SCFT's (in the pure $\SU$ case). The above combinatorial relation between Romans mass and ranks was derived in \cite[Eq. (2.15)]{cremonesi-tomasiello}

As discovered in \cite{afpt}, the supergravity equations that each vacuum is a solution to reduce to a single ODE, which in the language of the present paper can be recast in the following form:\footnote{For its derivation see \eqref{eq:PDE} and explanations around it.}
\begin{equation}\label{eq:third-der-alpha}
\dddot{\alpha}(z) =-(9\pi)^2 s_{i+1}\ , \quad z \in \left[i,i+1\right]\ .
\end{equation}
This allows us to determine $\alpha$ as well as its first and second derivative (which will be very useful in the following) by successive integration. Calling
\begin{equation}\label{eq:yzdot}
y(z):=-\frac{1}{18\pi} \dot{\alpha}(z)\ , \quad q(z):= \frac{1}{9\pi} \dot{y}(z) = -\frac{1}{2(9\pi)^2} \ddot{\alpha}(z) \geq 0 \ ,
\end{equation} 
in each of the intervals $z \in \left[i,i+1\right]$ we have
\begin{subequations}\label{eq:newalphanew}
\begin{align}
\alpha(z) &= \alpha_i -2(9\pi) y_i (z-i) -\frac{(9\pi)^2}{2}r_i (z-i)^2 -\frac{(9\pi)^2}{6} s_{i+1} (z-i)^3\ , \label{eq:alphaz}\\
 y(z) &=y_i + \frac{9\pi}{2} r_i (z-i) +\frac{9\pi}{4} s_{i+1}(z-i)^2\ , \label{eq:yz} \\
q(z) &= \frac{1}{2} r_i + \frac{1}{2} s_{i+1}(z-i)\ , \label{eq:2qz}
\end{align}
\end{subequations}
where $y_i, \alpha_i$ are integration constants. To determine the latter it suffices to impose continuity of $\alpha(z), y(z)$ at every interval upper endpoint $z=i+1$, for $i=0,\ldots,L-1$ and then $N-R,\ldots,N-2$. The results depend on the ``boundary data'' $y_0, y_N, \alpha_0, \alpha_N$ (which will be determined shortly) and the physical ranks $r_i$, and read
\begin{subequations}
\begin{equation}\label{eq:yialphai}
\frac{2}{9\pi} y_i = \frac{2}{9\pi} y_0 + \frac{1}{2}(r_0 +r_i) +\sum_{k=1}^{i-1}r_k\ ,
\end{equation}
and
\begin{equation} \label{eq:alphai}
\alpha_i = \alpha_0 - (9\pi) (2 i) y_0 -\frac{(9\pi)^2}{6} ((3i-1)r_0+r_i)- (9\pi)^2\sum_{k=1}^{i-1}(i-k) r_k
\end{equation}
for $i\in\left[1,L\right]$ and 
\begin{align} \label{eq:yialphaN-i}
&\frac{2}{9\pi} y_{N-i} = \frac{2}{9\pi} y_N - \frac{1}{2}(r_N+r_{N-i}) -\sum_{k=1}^{i-1}r_{N-k} \ , \\
& \alpha_{N-i} = \alpha_N +(9\pi) (2 i) y_N - \frac{(9\pi)^2}{6} ((3i-1)r_N+r_{N-i})-(9\pi)^2\sum_{k=1}^{i-1} (i-k) r_{N-k}
\end{align}
\end{subequations}
for $i\in\left[1,R \right]$. (The derivation is carried out in appendix \ref{app:int-const}.) 

We now have to determine the boundary data themselves. This can be done by imposing continuity at $z=L$ and $z=N-R$, which in turn implies the useful constraints \cite[Eq. (2.20) and above (A.5)]{cremonesi-tomasiello}:\footnote{\label{foot:typo-constr}Notice that there is a typo on the right-hand side of \cite[Eq. (2.20)]{cremonesi-tomasiello}: $\frac{9\pi}{4}$ should be replaced by $\frac{9\pi}{2}$.}
\begin{equation}\label{eq:constraints}
\begin{split}
&y_{N-R} - y_L = \frac{9\pi}{2} k(N-R-L) \ ,\\
&\alpha_{N-R} - \alpha_L =\frac{2}{k}(y_L^2-y_{N-R}^2)\ ;
\end{split}\quad  \Leftrightarrow \quad
\begin{split}
&y_{N-R} - y_L = \frac{9\pi}{2}  k(N-R-L) \ ,\\
&\alpha_{N-R} - \alpha_L =-9\pi (N-R-L)(y_L+y_{N-R})\ ,
\end{split}
\end{equation}
where $r_L= r_{N-R}=k$ is the height of the massless plateau, which is equivalent to maximum rank in each of the two massive regions. \eqref{eq:constraints} are two equations, and in general cannot determine four independent parameters ($y_0, y_N, \alpha_0, \alpha_N$). However in all practical situations we will encounter we only need to determine a subset of them, as some may be identically vanishing. This is because different brane or orientifold sources impose different boundary conditions on the internal part of the metric \eqref{eq:10dmetricz}, telling us which boundary data are vanishing, and which are not. 

A remark is in order here. In presence of O6-planes, finite-length D6 stacks will actually comprise $r_i \rightarrow 2 r_i$ branes (i.e. we count both the physical ones and their images) due to the orientifold projection, and the height of the plateau (the maximum rank) becomes $k\rightarrow 2k$ (see the discussion in section \ref{subsub:alternatingSOUSp}). (Moreover an $\SO$ gauge group engineered can have odd rank if there is a stuck half-D6 on top of the O6$^-$, in which case the latter is known as $\widetilde{\text{O6}}^-$.)

\subsection{Boundary conditions}
\label{sub:boundary}

Let us describe the possible boundary conditions of the internal metric around $z=0$. (Those around $z=N$ can be found analogously.) Let us define the quantity $\sigma(z):= \dot{\alpha}(z)^2-2\alpha(z) \ddot{\alpha}(z)$ for convenience. Given \eqref{eq:newalphanew}, at the lower endpoint of each subinterval $\left[i,i+1\right]$ we have:
\begin{equation}
\alpha(z)|_{z=i} = \alpha_i\ , \quad \dot{\alpha}(z)|_{z=i} = -2(9\pi)\, y_i\ , \quad \ddot{\alpha}(z)|_{z=i} = -(9\pi)^2\, r_i\ ,
\end{equation}
and
\begin{equation}
\sigma(z)|_{z=i} =: \sigma_i =  2(9\pi)^2(r_i\alpha_i +2 y_i^2)\ .
\end{equation}
\begin{comment}
\begin{subequations}
\begin{align}
\alpha(z)|_{z=i} =&\ \alpha_i \ , & \alpha(z)|_{z=i+1} =&\ \alpha_i -2(9\pi) y_i -\frac{(9\pi)^2}{2}r_i  -\frac{(9\pi)^2}{6} s_{i+1} \ ; \\
\dot{\alpha}(z)|_{z=i} =& -2(9\pi)\, y_i\ , & \dot{\alpha}(z)|_{z=i+1} =& -2(9\pi)\, y_i - \frac{(9\pi)^2}{2}(r_i +r_{i+1})\ ;\\
\ddot{\alpha}(z)|_{z=i} =& -(9\pi)^2\, r_i\ , &  \ddot{\alpha}(z)|_{z=i+1} =& -(9\pi)^2\, r_{i+1}\ ;
\end{align}
and
\begin{align}
&\sigma(z)|_{z=i} =  2(9\pi)^2(r_i\alpha_i +2 y_i^2) =: \sigma_i \ , \\
&\sigma(z)|_{z=i+1} =  2(9\pi)^2(r_{i+1} \alpha_i +2 y_i^2) - (9\pi)^4 s_{i+1} \left( \frac{1}{3}r_i +\frac{4}{3}s_{i+1} + \frac{2}{9\pi}y_i \right) =: \sigma_{i+1}\ .
\end{align}
\end{subequations}
\end{comment}
In terms of $\alpha(z), \sigma(z)$ the metric of the internal space $M_3$ reads (see \eqref{eq:10dmetricz}):
\begin{equation}\label{eq:radius}
\frac{1}{\pi \sqrt{2}}\, ds^2_{M_3} = \left(-\frac{\ddot{\alpha}(z)}{\alpha(z)}\right)^{1/2} dz^2 + R^2(z)\,ds^2_{S^2}\ ,\quad R^2(z):=  \left(-\frac{\ddot{\alpha}(z)}{\alpha(z)}\right)^{1/2}\frac{\alpha(z)^2}{\sigma(z)}\ .
\end{equation}
$R^2(z)$ is the squared radius of the $S^2$ fiber over the generic point $z \in \left[0, N\right]$. To have a compact $M_3$ we should impose $R^2(0)=R^2(N)=0$. Focusing on the first condition we see that this is equivalent to requiring
\begin{equation}
R^2(0) = \frac{r_0^{1/2}\alpha_0^{3/2}}{18 \pi(\alpha_0r_0+2y_0^2)} = 0\ \Leftrightarrow\ r_0 = 0\cup \alpha_0 = 0\ . 
\end{equation}
Moreover, recalling \eqref{eq:dilatonz}, the boundary value of the dilaton is found to be
\begin{equation}\label{eq:indil}
e^{\phi(0)}=2^{3/4}\pi^{4} 3^7 \frac{\alpha_0^{3/4}}{r_0^{3/4}(r_0 \alpha_0 + 2 y_0^2)^{1/2}}\ .
\end{equation}
The criteria of \cite[Sec. 5.1]{bpt} to determine which kind of physical object we have at the endpoint $z=0$ can now be phrased as follows:
\begin{itemize}
\item regular pole (the metric is finite and the space approximates $\mathbb{R}^3$): $\alpha_0 = r_0 = 0$, $\sigma_0 \neq 0 \Rightarrow y_0 \neq 0$. These are the boundary conditions considered in \cite{cremonesi-tomasiello}, and correspond to having magnetized D8-branes wrapping an $S^2$ fiber close to the pole;
\item D6 pole: $\alpha_0 =0$, $r_0 \neq 0$, $\sigma_0 \neq 0 \Rightarrow y_0 \neq 0$. We will call  D6 pole even one produced by a D6-O6$^\pm$ stack whose total effective D6 charge is positive; however in this case the fibers of the internal space are $\rr \pp^2$ rather than $S^2$ (due to the antipodal action of the O6-plane around the $z$ direction);

\item O6 pole: $\alpha_0 \propto \tilde{r}_0 \neq 0$, $r_0 = 0$, $\sigma_0 \neq 0 \Rightarrow y_0 \neq 0$. Also in this case the $S^2$ fiber is replaced by $\mathbb{RP}^2$. The total D6 charge $\tilde{r}_0$ of the D6-O6$^-$ source is negative;

\item O8 pole without D6 charge: $r_0 =\sigma_0 = 0 \Rightarrow y_0 =0$. In this case $\phi(z), R^2(z) \to \infty$ as $z \to 0$, as is appropriate for a D8-O8 source of divergent dilaton type.\footnote{See \cite{brandhuber-oz} for another well-known example.} ($\alpha_0$ may not be zero, for otherwise $R^2(z)$ tends to a constant as $z \to 0$.) These boundary conditions are appropriate for the AdS$_7$ vacuum constructed in \cite[Eq. (5.2)]{bpt}, dual to the massive E-string theory (described in section \ref{subsub:O8pureSUk}). Therefore we will neglect this case in the following;
\item O8-D8 pole with D6 charge: $r_0 \neq 0$, $y_0 = 0 \Rightarrow \sigma_0 = 2 (9\pi)^2 r_0 \alpha_0$ (and as before $\alpha_0 \neq 0$). In this case $\phi(z), R^2(z)$ are finite and nonvanishing at $z=0$, which corresponds to the equator of $M_3 \cong S^3$. (As already explained, the physical half of the internal space lies in $\left[0,N\right]$.) In other words, the D6-brane charge $r_0$ resolves the dilaton and metric singularity at $z=0$. (For $r_0 \rightarrow 0$ this case reduces to the previous one.)
\end{itemize}
One can indeed check \cite{afrt} that the metric $ds^2_{M_3}$, dilaton, and the relevant bulk fluxes close to $z=0$ have the correct asymptotics to justify the presence of the above brane and orientifold sources. We summarize all possible requirements in table \ref{tab:configurations}.\footnote{The analysis of the boundary conditions at the other endpoint, $z=N$, is greatly simplified if one labels the subintervals starting from the latter rather than $z=0$, i.e. $z\in \left[N-(i+1),N-i\right]$ with $i=0,\ldots,R-1$.} We now realize that there exist only two cases with a nonvanishing subset of boundary data:
\begin{itemize}
\item if regular or D6 poles occur at $z=0$ and $z=N$ then $\alpha_0=\alpha_N=0$ automatically, and we simply need to determine $y_0, y_N$. These are two parameters, and can be determined by \eqref{eq:constraints}. The result is given in \eqref{eq:y0N-asym-gen} (plugging in $\alpha_0=\alpha_N=0$). If O6 poles occur instead, $\alpha_0$ and $\alpha_N$ do not vanish but can be determined via an independent physical argument (namely by expanding the bulk $F_2$ flux in the vicinity of $z=0,N$, respectively)\footnote{This is shown in section \ref{sub:F2alpha0N}.} which suggests the definitions
%\begin{equation}\label{eq:F20}
%\frac{1}{2} \tilde r_0 \sim F_2|_0 \sim -\frac{\pi  \alpha_0 r_1}{9 y_0}\ ,
%\end{equation}
%which implies
\begin{equation}\label{eq:effectivealphas}
\alpha_0 := \frac{9}{2\pi} \frac{\tilde{r}_0 y_0}{r_1}\ , \quad \alpha_N := -\frac{9}{2\pi} \frac{\tilde{r}_N y_N}{r_{N-1}}\ ,
\end{equation}
where $\tilde{r}_0, \tilde{r}_N=-4,\ldots,-1$ can be interpreted as the effective D6 charge of a flavor D6-O6$^-$ stack.\footnote{\label{foot:alpha-signs}Notice that $\alpha_{0,N}>0$. E.g. by expanding the $dz^2$ component of the metric \eqref{eq:radius} around $z=0$, with $\alpha(z)$ in $[0,1]$ given by \eqref{eq:alphaz} with $i=0$, the constant term is found to be proportional to $\sqrt{\frac{r_1^2}{\tilde{r}_0 y_0}}$, which requires $\tilde{r}_0 y_0>0$. Given that $\tilde{r}_0<0$, we must also have $y_0<0$. At large $k,N$, this can be proven by directly inspecting \eqref{eq:y0limit} (since $\sum_i r_i > \sum_i \frac{i}{N}r_i$). A similar argument holds for $\alpha_N$.} Once again we can use \eqref{eq:constraints} to determine $y_0,y_N$. The result is given in \eqref{eq:y0N-asym-generic-eff}.

\item The second case corresponds to having an O8 pole at $z=0$. (The orientifold acts as a wall around the origin of the $z$ direction, and we choose to parameterize the physical half of the space by $z\in\left[0,N\right]$.) In this case $y_0=0$, hence we only need to determine $\alpha_0, y_N, \alpha_N$. Moreover $\alpha_N$ either vanishes (in case of a regular or D6 pole at $z=N$) or can be defined as in \eqref{eq:effectivealphas} in terms of $y_N$ and the effective charge of a D6-O6$^-$ stack, if the latter is present. We can then use \eqref{eq:constraints} to find expressions for $\alpha_0$ and $y_N$, which are given in \eqref{eq:alpha0O8} and \eqref{eq:yNO8} respectively.
\end{itemize}

\begin{table}[!htb]
\centering
{\renewcommand{\arraystretch}{1.2}%
\begin{tabular}{@{} c  c  c  c c  @{}}
asymptotics of $ds_{M_3}^2$ at  $z=0,N$ resp. &  $\alpha_{0,N}$ & $y_{0,N}$ & $r_{0,N}$  \\
\toprule \toprule
regular point: D8-branes ($\dot{\alpha}^2-2\alpha\ddot{\alpha} \neq 0$)  & \multicolumn{1}{!{\vrule width 1pt}c}{$0$} & $\neq 0$ & $0$  \\

D6 pole ($\dot{\alpha}^2-2\alpha\ddot{\alpha} \neq 0$)  & \multicolumn{1}{!{\vrule width 1pt}c}{$0$} & $\neq 0$ &  $>0$  \\

O6 pole ($\dot{\alpha}^2-2\alpha\ddot{\alpha} \neq 0$) & \multicolumn{1}{!{\vrule width 1pt}c}{$> 0$ } & $\neq 0$ &  $0$ \\

O8 pole with D6 charge at $z=0$: $\dot{\alpha}^2-2\alpha\ddot{\alpha} \propto r_{0} \alpha_{0}$  & \multicolumn{1}{!{\vrule width 1pt}c}{$\neq 0$} & $0$ & $\neq 0$  \\
\toprule
\end{tabular}}
\caption{The requirements for having a regular point, a D6 or D6-O6$^\pm$ source, a D8-O8 source with (smeared) D6 charge at the poles of the supergravity solution, characterized by an internal space $S^2 \hookrightarrow M_3 \to I=\left[0,N\right]$ (or $\rr\pp^2 \hookrightarrow M_3 \to I=\left[0,N\right]$). Different sources impose different boundary conditions on the metric of $M_3$ at the extrema of the base interval $I$. Notice that we can have an O8-plane only at $z=0$. The former acts as a mirror along direction $z$, and we choose to parameterize the physical half of the space by $z\in \left[0,N\right]$.}
\label{tab:configurations}
\end{table}

% fold sec (solz)

%%%%%%%%%%%%%%%%%%%%%%%%%%%%%%%%%%%%%%%%%%%%%%
\section{Computation of $a$ in field theory} % sec (aexamples)
\label{sec:aexamples}
%%%%%%%%%%%%%%%%%%%%%%%%%%%%%%%%%%%%%%%%%%%%%%

After having explained how to engineer $(1,0)$ theories with massive IIA AdS$_7$ duals, we now explain how to extract a very important observable of the SCFT, namely its $a$ conformal anomaly. We will then take the holographic limit of the latter, and compare it to the result obtained in supergravity.\\

The (eight-form) anomaly polynomial $\mathcal{I}$ of a six-dimensional $(1,0)$ SCFT is a sum of various contributions (see \cite{ohmori-shimizu-tachikawa-yonekura} and appendix \ref{app:aFT}), which can be summarized as follows:\footnote{$c_2(R)=\frac{1}{4}\Tr F_R^2$ denotes the second Chern class of the (background) $\SU(2)$ R-symmetry bundle, and $p_1,p_2$ the first and second Pontryagin classes of the tangent bundle of a formal eight-manifold.}
\begin{equation}
\mathcal{I} = \alpha \, c_2(R)^2 + \beta\, c_2(R) p_1(T) + \gamma \,p_1(T)^2 + \delta\, p_2(T)\ + \mathcal{I}_\text{flavor} \ .
\end{equation}
The coefficients $\alpha,\ldots,\delta$ are functions of the group theory data, the number of tensor multiplets $N_\text{T}=N-1$,\footnote{In the F-theory construction of these $(1,0)$ theories, $N_\text{T}=N-1$ coincides with the number of $-2$ curves in the base (after having blown down all possible $-1$ curves).} and the so-called Dirac pairing defined in \eqref{eq:dirac}, namely the matrix
\begin{equation}
\eta_{ij}=\mathbf{n}\, \delta_{ij} - \delta_{i\,i-1}- \delta_{i\,i+1}\ , \quad \mathbf{n}=\{n_0, \ldots, n_{N_\text{T}}\}\ .
\end{equation}
Explicitly, they are given by:
\begin{subequations}\label{eq:coeffpoly}
\begin{align}
\alpha &= \frac{1}{24}\left(N_\text{T}- N_\text{V} \right) +\frac{1}{2}(\eta^{-1})_{ij}\,h^\vee_{G_i}h^\vee_{G_j}\ ,\\
\beta &= \frac{1}{48}\left(N_\text{T}- N_\text{V} \right) +\frac{1}{12}(\eta^{-1})_{ij} K_i h^\vee_{G_j}\ ,\\
\gamma &= \frac{1}{5760}\left(23 N_\text{T}- 7 N_\text{V} +7 N_\text{H} \right)+\frac{1}{288}(\eta^{-1})_{ij} K_i K_j\ ,\\
\delta &= \frac{1}{5760}\left(-116 N_\text{T}+ 4 N_\text{V} - 4 N_\text{H} \right)\ ,
\end{align}
\end{subequations}
where $N_\text{V}$ and $N_\text{H}$ are the total numbers of vector and hypermultiplets respectively,
\begin{align}
& N_\text{V}= \sum_{i=1}^{N_\text{T}} d_{G_i}\ ,\\
& N_\text{H}=\sum_{i=1}^{N_\text{T}}\left( \epsilon_i d_i +\epsilon^\text{flv}_i  f_i d_i + \epsilon_{i\,i+1} d_i d_{i+1} \right)\ ,
\end{align}
and
\begin{equation}\label{eq:KK}
K_i := h^\vee_{G_i}-\epsilon_i \, \Ind( \rho_i)-s_{G_i}\left( \epsilon_{i\,i-1} d_{i-1}+\epsilon_{i\,i+1}d_{i+1} + \epsilon^\text{flv}_i f_i\right)\ .
\end{equation}
$h^\vee_{G_i}$ is the dual Coxeter number of the $i$-th gauge group $G_i$, $s_{G_i}$ the constant defined in table \ref{tab:groupconst}, and the coefficients ${\epsilon}_i, \epsilon_{i\,i+1}, \epsilon_i^\text{flv}=\{1,\frac{1}{2},0\}$ account for the presence of full hypermultiplets (1) as appropriate for $\SU$ quivers, half-hypermultiplets ($\tfrac{1}{2}$) as appropriate for alternating $\SO$-$\USp$ quivers, or no hypermultiplets at all (0). $\Ind(\rho_i)$ is the  index of the hypermultiplet representation $\rho_i$ of real dimension $d_i :=\dim_\rr (\rho_i)$.\footnote{\label{foot:index}By index we mean the eigenvalue of the quadratic Casimir in the representation, normalized such as to be an integer. If $\rho$ is an irreducible representation of the Lie algebra $\mathfrak{g}$ associated with $G$, then $\Tr_\rho (T^a T^b) = \Ind_\rho \, \delta^{ab}$. More intrinsically, it can be defined in terms of $d:=\dim_\rr \rho$ and $\ord \mathfrak{g} := \dim_\rr (\text{adj})$ (i.e. the number of roots of $\mathfrak{g}$) through $\Ind_\rho := \frac{d}{\ord \mathfrak{g}} (\Lambda,\Lambda+\delta)$ \cite{lieart}, where $\Lambda$ is the highest weight of $\rho$ and $\delta=(1,1,\ldots,1)$ is (half) the sum of all positive roots in the so-called Dynkin basis (the one in which the rows of the Cartan matrix of $\mathfrak{g}$ give its simple roots).} (The dimension is called $f_i$ for flavor hypermultiplets.)

Finally, the $a$ conformal anomaly is given by the following combination of anomaly polynomial coefficients \cite[Eq. (1.6)]{cordova-dumitrescu-intriligator}:
\begin{equation}
a=\frac{384}{7}(\alpha - \beta + \gamma) + \frac{144}{7}\delta\ .
\end{equation}
Plugging the expressions \eqref{eq:coeffpoly} into the above equation we obtain the very general formula (in which a sum over repeated indices is understood):
\begin{multline} \label{eq:a}
a= \frac{1}{210} (199 N_\text{T}- 251 N_\text{V}+11 N_\text{H}) \ + \\ +\frac{16}{7} \left( 12  (\eta^{-1})_{ij}h^\vee_{G_i} h^\vee_{G_j} -2 (\eta^{-1})_{ij} K_i h^\vee_{G_j}+\frac{1}{12}(\eta^{-1})_{ij} K_i K_j\right)\ .
\end{multline}
In this section we shall compute explicitly the leading term of the $a$ conformal anomaly in field theory for a few linear quivers as $h^\vee_{G_i} \sim N \to \infty$, which we claim to be
\begin{equation}\label{eq:adominant}
a \sim \frac{192}{7} (\eta^{-1})_{ij} \, h^\vee_{G_i} h^\vee_{G_j} \ .
\end{equation}
This leading behavior can be proven by showing that the last two terms in parentheses in \eqref{eq:a} are subdominant w.r.t. the first, namely \eqref{eq:adominant}. This is easily done as follows.

First of all, as explained above \eqref{eq:F4canc}, the cancellation of the gauge anomaly involving the term $\frac{1}{16} \Tr(F_i^4)$ implies the constraint
\begin{equation}\label{eq:ti-gaugeanom}
t_{G_i} = {\epsilon}_i \alpha_{\rho_i} +\left(  \epsilon_{i\, i-1} d_{i-1}+\epsilon_{i\,i+1} d_{i+1} + \epsilon^\text{flv}_i f_i\right)\ ,
\end{equation}
where the constants $t_{G_i}$ have been defined in \eqref{eq:Tr4},
\begin{equation}\label{eq:adj-fund}
\tr_\text{adj}(F_i^4) =t_{G_{i}} \tr_\text{fund} (F^4_i) + \ldots\ ,
\end{equation}
and $\alpha_{\rho_i}$ is the quartic Casimir of $G_i$ in the representation $\rho_i$, which is defined in \eqref{eq:Tr2Tr4gen}. (Notice that in the pure $\SU(k)$ case \eqref{eq:ti-gaugeanom} precisely reduces to \eqref{eq:flavors}.)

As one can see from tables \ref{tab:groupconst} and \ref{tab:ac} by direct inspection, the above constants satisfy the following relations for $h^\vee_{G_i}\sim N \to \infty$:
\begin{equation}\label{eq:ratio-const}
t_{G_i}\sim \frac{h^\vee_{G_i}}{s_{G_i}}\ , \quad \alpha_{\rho_i} \sim \frac{\Ind_{\rho_i}}{s_{G_i}}\ .
\end{equation}
Using \eqref{eq:ti-gaugeanom} inside \eqref{eq:KK}, and subsequently plugging in \eqref{eq:ratio-const}, we immediately realize that $K_i $ is independent of $N$ (i.e. it tends to a constant as $N\to \infty$), hence any term in \eqref{eq:a} with a bilinear involving $K_i $ is subdominant w.r.t. $(\eta^{-1})_{ij} \, h^\vee_{G_i} h^\vee_{G_j} $.\\

%bhardwaj's $\alpha_R$ becomes tachi's $t_G = 2n$ for the adjoint.}\\

We now turn to the computation of $a$ in some important classes of theories. By specializing the general formulae provided below, one can easily obtain the leading contribution to the $a$ anomaly of any $(1,0)$ linear quiver with massive type IIA dual (including the so-called ``formal'' quivers of \cite{mekareeya-rudelius-tomasiello}).

\subsection{SU quivers on the tensor branch}
\label{sub:regD6poles}
The possible brane configurations realizing linear quivers with only $\SU(r_i)$ groups are depicted in figures \ref{fig:NS5D6} (without D8-branes) and \ref{fig:NS5D6D8} (with D8-branes -- the latter is a specific example, easily generalizable to others). In the first, we have semi-infinite flavor D6's extending beyond the left- and rightmost NS5-branes; in the second, stacks of magnetized D8-branes. As explained in section \ref{subsub:pureSUk}, in the supergravity AdS$_7$ solution we then have D6 ($r_0, r_N \neq 0$) or regular ($r_0=r_N=0$) poles respectively (see table \ref{tab:configurations}); $\alpha_0=\alpha_N=0$ in both cases.

The regular poles case has already been treated in \cite{cremonesi-tomasiello}. Those results carry through to the case with D6 poles, i.e. the computation of $a$ is totally equivalent in both cases. The ultimate reason is that $r_0, r_N$ only appear in the gravity computation as coefficients of subleading terms (w.r.t. the dominant $\mathcal{O}(N^5)$ order), and are also washed away in the field theory computation as $r_i\sim k, N \to \infty$. Thus we may completely neglect them.

The quivers, depicted in figures \ref{fig:quiverNS5D6} and \ref{fig:quiverNS5D6D8}, are given by a collection of $\SU(r_i)$ gauge groups (flavor groups for $i=0,N$); therefore $h^\vee_{G_i} = r_i$ for $i=1,\ldots,N-1=N_\text{T}$. The (inverse) Dirac pairing \eqref{eq:eta-1D6} is the (inverse) Cartan matrix of $A_{N-1}$. Therefore:
\begin{equation}\label{eq:aD8-D6FTholo}
a \sim \frac{192}{7} (\eta^{-1}_\text{D6})_{ij}\, r_i r_j\ .
\end{equation}
Dividing the sum over $i,j$ into left massive region, massless plateau, and right massive region, and keeping only the leading terms in $N$, we find:
\begin{subequations}\label{eq:aFTholD6}
\begin{align}
a \sim&\ \frac{192}{7}\left(\sum_{i=1}^L + \sum_{i=L+1}^{N-R-1} + \sum_{i=N-R}^{N-1} \right)\left(\sum_{j=1}^L + \sum_{j=L+1}^{N-R-1} + \sum_{j=N-R}^{N-1} \right) (\eta^{-1}_\text{D6})_{ij} \, r_i r_j \\
\frac{7}{192} a \sim &\ \frac{k^2}{N} \frac{1}{12}(N-L-R)^2(N^2+2(L+R)N-3(L-R)^2)\ + \nonumber \\
&\ + \frac{k}{N}(N-L-R)\left( (N-L+R) \sum_{i=1}^L i r_i + (N+L-R) \sum_{i=1}^R i r_{N-i}\right) + \nonumber
\end{align}
\begin{align}
&\ + \frac{1}{N}\left(2 \sum_{i=1}^L\sum_{j=1}^{R} i j r_i r_{N-j} +  \sum_{i=1}^L i(N-i) r_i^2 + 2  \sum_{j=1}^L\sum_{i=1}^{j-1} i(N-j) r_i r_j \right. + \nonumber\\
& \qquad \quad\  + \left. \sum_{i=1}^R i(N-i) r_{N-i}^2 + 2  \sum_{j=1}^R\sum_{i=1}^{j-1} i(N-j) r_{N-i} r_{N-j} \right)\ ,
\end{align}
\end{subequations}
which is exactly \cite[Eq. (3.15)]{cremonesi-tomasiello}. The only nontrivial identity that one needs is the following:
\begin{equation}
\sum_{i,j=1}^L  i(N-j)\, r_i r_j + \sum_{i=1}^{N}\sum_{j=1}^{i-1} N(j-i)r_i r_j = \sum_{i=1}^L i(N-i)r_i^2 + 2\sum_{i=1}^L \sum_{i<j}i(N-j)r_i r_j\ ,
\end{equation}
and likewise for $L \leftrightarrow R$ (in the summation extrema) and $i,j \leftrightarrow N-i, N-j$ (inside the sums).\\

To get a more explicit result one can specify a linear quiver corresponding to a chosen brane configuration. E.g. selecting the theory in \cite[Fig. 6]{cremonesi-tomasiello}, we have to impose $N=L=k$, $R=0$, $\{ r_i \}_{i=0}^{k} = i+1$ (that is, we only have a left massive region occupying the whole interval $I=[0,N]$). Notice that, for $i=0,N$, the groups $\SU(r_i)$ are flavor symmetries; we trade $r_0=1$ D6 for a D8 on the left via a Hanany--Witten move, whereas we keep $r_N=k+1$ semi-infinite D6's on the right. Plugging this into \eqref{eq:aFTholD6} gives 
\begin{equation}
a \sim \frac{16}{7}\frac{4}{15}k^5 \quad \text{as}\quad k \to \infty\ ,
\end{equation}
which is the holographic $a$ conformal anomaly of the ``simple massive solution'' of \cite[Sec. 5.2]{afrt} and \cite[Sec. 5.5]{afpt}, sourced by a single D8 and defined (in the sense of section \ref{sub:gensol}) by the function
\begin{equation}\label{eq:simplemassive}
\alpha(z) = (3\pi)^3 F_0 (N z -z^3)\ \Rightarrow\ y_0 = -\frac{3}{2} \pi^2 F_0 N^2, \ s_1 = 2\pi F_0 = 2\pi = r_1\ ,
\end{equation}
supported on $[0,N]$. Such a vacuum is characterized by a regular pole at $z=0$ (where $r_0=\alpha_0=0,\ \sigma_0 \propto N^4$) and a D6 pole (with $k+1$ branes) at $z=N$ (where $r_N \propto N,\ \alpha_N=0,\  \sigma_N \propto N^4$). Its integration constants and boundary data fall into the class of section \ref{appsubsub:genasym} without massless plateau and with $\alpha_0=\alpha_N=0$.

The theory in \cite[Fig. 7]{cremonesi-tomasiello} requires instead taking $L=R=k$, $\{ r_i \}_{i=1}^k = \{ r_{N-i} \}_{i=1}^k = i$, and plugging this information into \eqref{eq:aFTholD6} gives \cite[Eq. (3.18)]{cremonesi-tomasiello} (with $\mu = k$ -- see also \cite[Eq. (5.71)]{afpt} for an earlier computation):
\begin{equation}
a \sim \frac{16}{7}k^2 \left( N^3 - 4 k N + \frac{16}{5} k^3\right) \quad \text{as}\quad k,N \to \infty\ ,
\end{equation}
which is the holographic $a$ conformal anomaly of a typical massive theory with two equal tails and a massless plateau of height $k$. (The supergravity solution is defined by a function $\alpha(z)$ whose integration constants and boundary data fall into the class of section \ref{appsubsub:gensym}.)

Notice that, for both theories, $a$ is of order $\mathcal{O}(N^5)$ as anticipated in section \ref{sec:intro}, and all terms come from supergravity (not stringy corrections).

\subsection{Alternating SO-USp quivers on the tensor branch}
\label{sub:O6poles}

The possible brane configurations realizing linear quivers with alternating $\SO$-$\USp$ gauge groups are depicted in figures \ref{fig:NS5D6O6} (without D8-branes) and \ref{fig:NS5D6O6D8} (with D8-branes). For the unHiggsed quiver in \ref{fig:quiverNS5D6O6}, since $k \geq 4$, the left- and rightmost flavor symmetries correspond to D6-O6$^-$ stacks of total positive D6 charge. Therefore the dual massless vacuum has D6 poles at $z=0,N$:  $r_0 = r_N=k$ (whereas $\tilde{r}_0=\tilde{r}_N= 0$) and the fibers are $\rr\pp^2$'s. For the generic Higgsed quiver of figure \ref{fig:quiverNS5D6O6D8}, the poles of the supergravity dual are of O6 type whenever the flavor symmetry has rank low enough (i.e. $\SO(0,\ldots,3)$) as explained in section \ref{subsub:alternatingSOUSp}; hence $\alpha_0, \alpha_N \neq 0 \Rightarrow \tilde{r}_0,\tilde{r}_N \neq 0$ (but $r_0= r_N= 0$).

For D6 poles the large $N$ field theory computation falls into the pure $\SU$ case. Therefore here we will only discuss the O6 poles one. The inverse Dirac pairing is given by \eqref{eq:eta-1O6D6}. Therefore:
\begin{equation}\label{eq:aO6FTholo}
a \sim \frac{192}{7} (\eta^{-1}_\text{O6})_{ij}\, h^\vee_{G_i} h^\vee_{G_j} = \frac{192}{7\cdot 2} (\eta^{-1}_\text{D6})_{ij}\, v_i v_j h^\vee_{G_i} h^\vee_{G_j} \ .
\end{equation}
Its entries depend on the components $v_i$ of an auxiliary vector $\mathbf{v}$ of dimension $N_\text{T}$, which are all equal to either 1 or 2: We have 1 for an $\SO$ group, and 2 for a $\USp$ one. Using table \ref{tab:groupconst}, we see that $v_i h^\vee_{\SO(2r_i)} = 2r_i-2 \sim 2r_i$ as $r_i \sim N \to \infty$, and similarly $v_i h^\vee_{\USp(2r_i)} = 2r_i+2 \sim 2r_i$, where $r_i$ can be viewed as the number of brane pairs on top of the O6-planes as in figures \ref{fig:quiverNS5D6O6} and \ref{fig:quiverNS5D6O6D8}. For this reason and because of \eqref{eq:aO6FTholo}, the large $N$ computation of the $a$ conformal anomaly from field theory is analogous to the case without O6-planes.

%This simple observation shows that, in the $N \to \infty$ limit, \eqref{eq:aO6FTholo} has the same behavior as \eqref{eq:aD8-D6FTholo}:
%\begin{equation}\label{eq:aO6FTholo2}
%\frac{192}{7} (\eta^{-1}_\text{D6})_{ij}\, v_i v_j h^\vee_{G_i} h^\vee_{G_j} \overset{N \to \infty}{\sim}\ \frac{192}{7\times 2} (\eta^{-1}_\text{D6})_{ij}\, r_i r_j \ ,
%\end{equation}
%with $r_i=2k$ the ranks of $N_\text{T}$ ``effective'' $\SU(r_i)$ groups. In fact, the only differences in the computation are subleading contributions (such as those coming from terms proportional to $\alpha_0,\alpha_N$), which can be neglected at large $N$. Therefore the result \eqref{eq:aFTholD6} holds in the O6 pole case, too.

\subsection{Quivers from brane configurations with an O8-plane}
\label{sub:O8pole}

The brane configurations realizing $\USp(r_0)$-$\SU(r_i)$ (resp. $\SO(r_0)$-$\SU(r_i)$) linear quivers with an $\SO(2n_0)$ (resp. $\USp(2n_0)$) flavor symmetry, engineered by a D8-O8 stack, are depicted in figure \ref{fig:NS5D6D8O8unstuck-stuck}. (We will defer the computation of the $a$ anomaly in the case with O6-planes to section \ref{sub:O8-O6gravdual} for clarity of exposition.) The gravity duals will be characterized by an O8 pole with D6 charge at $z=0$, i.e. $r_0,\alpha_0 \neq 0$ but $y_0=0$, and by a regular (or D6 pole) at $z=N$, i.e. $\alpha_N=0$ and $r_N=0$ ($r_N\neq0$).\footnote{For the case without D6 charge at $z=0$, i.e. $r_0=0$, see the discussion in section \ref{subsub:O8pureSUk}.}

When the O8 sits at $x^6=0$ (corresponding to $z=0$ in the near-horizon limit) between an NS5 and its image, the quiver are of the type of the ones given by figure \ref{fig:quiverNS5D6D8O8-unstuck}, where the introduction of more D8 stacks is allowed. The Dirac pairing $\eta_\text{O8}$ is given by \eqref{eq:DPfAP}, and its inverse is given below in \eqref{eq:etaO8+-inv}. We will use the latter to compute the leading term of $a$. The dual Coxeter number $h^\vee_{G_0}$ of the first gauge group $G_0$ is given by $\frac{r_0}{2}+1$ (for $\USp(r_0)$) or $r_0-2$ (for $\SO(r_0)$). For $i>0$ the gauge groups are all $\SU(r_i)$, and $h^\vee_{G_i}=r_i $. When $r_i\sim k \sim N \to \infty$, all numbers $h^\vee_{G_i}$ scale like $N$. The holographic $a$ anomaly can be computed as follows:
\begin{subequations}\label{eq:aFTholO8}
\begin{align}
a \sim &\ \frac{192}{7}\left(\sum_{i=1}^L + \sum_{i=L+1}^{R'-1} + \sum_{i=R'}^{N-1} \right)\left(\sum_{j=1}^L + \sum_{j=L+1}^{R'-1} + \sum_{j=R'}^{N-1} \right) (\eta^{-1}_\text{O8})_{ij}\, h^\vee_{G_i} h^\vee_{G_j} \label{eq:aFTholO8-sum}\\
\frac{7}{192} a \sim &\ \frac{k^2}{3}(N-L-R)^2(N-L+2R) + k(N-L-R)\left(k(N-L+R)\sum_{i=1}^{L} r_i \right. + \nonumber \\
& + \left.  2\sum_{i=1}^{R} i r_{N-i} \right) + \sum_{i=1}^{L} (N-i) r_i^2 + 2  \sum_{i=1}^{L}\sum_{j=1}^{i-1} (N-i) r_i r_j  + \sum_{i=1}^{R} i r_{N-i}^2\ + \nonumber \\ 
& + 2  \sum_{i=1}^{R}\sum_{j=1}^{i-1} i r_{N-i} r_{N-j} + 2 \sum_{j=1}^{R}j r_{N-j} \sum_{i=1}^L r_i\ .
\end{align}
\end{subequations}
The only nontrivial identities needed to obtain the above result are the following:\footnote{We are not being careful about the summation extrema due to $L-1 \sim L$, $R-1 \sim R$ in the holographic limit.}
\begin{equation}
\begin{split}
&\sum_{i=1}^{L}r_i \sum_{j=1}^{L}(N-j)r_j - \sum_{i=1}^{L} \sum_{j=1}^{i-1}(i-j)r_j r_i = \sum_{i=1}^{L} (N-i)r_i^2 + 2 \sum_{i=1}^L \sum_{j<i} (N-i)r_i r_j  \ ,\\
&\sum_{i=1}^{R}r_{N-i} \sum_{j=1}^{R}j r_{N-j} - \sum_{i=1}^{R} \sum_{j=1}^{i-1}(i-j)r_{N-j} r_{N-i} = \sum_{i=1}^{R} i r_{N-i}^2 + 2 \sum_{i=1}^L \sum_{j<i} i r_{N-i} r_{N-j}\ .
\end{split}
\end{equation}

When the O8$^\pm$ is stuck on an NS5 at $x^6=0$ ($z=0$ in the near-horizon) the quivers are the ones in figure \ref{fig:quiverNS5D6D8O8-stuck}, and the corresponding brane configurations are given in figures \ref{fig:NS5D6D8O8+stuck}, \ref{fig:NS5D6D8O8-stuck} respectively. This case is slightly more subtle, being entirely characterized by $\SU(r_i)$ groups: $h^\vee_{G_i}=r_i \sim k$ as $k \sim N \to \infty$ for all $i$. However the Dirac pairing (which can again be found by applying \eqref{eq:DPfAP}) turns out to be equivalent to $\eta_\text{O8}^-$ in \eqref{eq:etaO8+-noF} for both O8$^\pm$. Hence the computation \eqref{eq:aFTholO8} holds in this case, too. Moreover (as will be explained in greater detail at the end of the next subsection) the leading order of the $a$ conformal anomaly cannot distinguish between the two configurations.

\subsubsection{\texorpdfstring{Computation of $a$ in O8$^+$ and D8-O8$^-$ theories}{Computation of a in O8+ and D8-O8- theories}}
\label{sub:O8ramps}

We will now make the result \eqref{eq:aFTholO8} much more explicit. Consider the quivers in figure \ref{fig:quiverNS5D6D8O8-unstuck}. They are engineered by having an O8-plane sit between the first NS5 and its image.

If we overlay $n_0=16$ D8 pairs onto the O8$^-$, this stack has the same D8 charge as a single O8$^+$ without overlaid D8-branes (hence no flavor symmetry). The theory on the right in figure \ref{fig:quiverNS5D6D8O8-unstuck} has two flavor symmetries: $\SO(32)$ with $2n_0=32$ half-hypermultiplets in the fundamental of the first gauge group $\USp(2k)$, and $\SU(2k+N(8-16))$ with $2k-8N$ hypermultiplets in its antifundamental representation and in the fundamental of the last gauge group $\SU(2k-8(N-1))$. The theory on the left only has an $\SU(2k-8N)$ flavor symmetry leading to $2k-8N$ hypermultiplets in its antifundamental and in the fundamental of the gauge group $\SU(2k-8(N-1))$. The two product gauge groups are:
\begin{subequations}
\begin{align}
& \text{O8}^- + 32\ \text{D8's}: &\USp(2k) \times \prod_{i=1}^{N-1} \SU(2k-8i) \ ,\\
& \text{O8}^+:&\SO(2k) \times \prod_{i=1}^{N-1} \SU(2k-8i)\ .
\end{align}
\end{subequations}
The Dirac pairings for these two theories are the following $(N-1) \times (N-1)$ matrices:
\begin{equation}\label{eq:etaO8+-noF}
 \eta_\text{O8$^-$}= \begin{bmatrix}
1 & -1& 0 & \cdots & 0\\
-1 & 2 & -1& \cdots &0 \\
0& -1 & 2& \cdots &0\\
  \vdots  & \vdots & \vdots  & \ddots & \vdots  \\
0 & 0 & -1& \cdots & 2 
 \end{bmatrix}\ , \quad
  \eta_\text{O8$^+$}= \begin{bmatrix}
4 & -2& 0 & \cdots & 0\\
-2 & 2 & -1& \cdots &0 \\
0& -1 & 2& \cdots &0\\
  \vdots  & \vdots & \vdots  & \ddots & \vdots  \\
0 & 0 & -1& \cdots & 2 
 \end{bmatrix}\ ,
\end{equation}
Notice that for $\eta_\text{O8$^+$}$ there is a discrepancy between what we would obtain from formula \eqref{eq:dirac} and the gauge-anomaly freedom requirement \eqref{eq:DPfAP} which we derived from the six-dimensional anomaly polynomial. The former gives the adjacency matrix of base curves (of negative self-intersection) $422\ldots2$ in an F-theory engineering of the same SCFT. The disagreement, which has to do with a subtle effect due to the presence of a frozen $\widehat{I}_4^*$ Kodaira fiber over an O7$^+$, disappears once the F-theory formula is modified so as to include the O7$^+$ case \cite{bhardwaj-morrison-tachikawa-tomasiello}.\footnote{We thank T.~Rudelius and A.~Tomasiello for discussion on this point.}

In order to compute $a$ we need to evaluate the inverses of \eqref{eq:etaO8+-noF}. We find:
\begin{small}
\begin{equation}\label{eq:etaO8+-inv}
 \eta_\text{O8$^-$}^{-1}= \begin{bmatrix}
N-1 & N-2 & N-3 & \cdots & 1\\
N-2 & N-2 & N-3& \cdots &1 \\
N-3& N-3 & N-3& \cdots &1 \\
  \vdots  & \vdots & \vdots  & \ddots & \vdots  \\
1 & 1 & 1& \cdots & 1 
 \end{bmatrix}\ , \quad  \eta^{-1}_\text{O8$^+$}= \begin{bmatrix}
\frac{N-1}{4} & \frac{N-2}{2} & \frac{N-3}{2} & \cdots & \frac{1}{2}\\
\frac{N-2}{2} & N-2 & N-3& \cdots &1 \\
\frac{N-3}{2}& N-3 & N-3& \cdots &1 \\
  \vdots  & \vdots & \vdots  & \ddots & \vdots  \\
\frac{1}{2} & 1 & 1& \cdots & 1 
 \end{bmatrix}\ .
\end{equation}
\end{small}%
Thus:
\begin{subequations}
\begin{align}
(\eta_\text{O8$^-$}^{-1})_{ij}\,h^\vee_{G_i}h^\vee_{G_j}  = & \   h^\vee_{G_0}\left((N-1)h^\vee_{G_0}+\sum_{j=2}^{N-1}(N-j) (2k-8(j-1))\right) +   \nonumber \\
 & +\sum_{i=2}^{N-1} (2k-8(i-1)) \left((N-i)h^\vee_{G_0}+(N-i)\sum_{j=2}^{i-1}  (2k-8(j-1))\right. \nonumber \\
 & +\left. \sum_{j=i}^{N-1}(N-j) (2k-8(j-1)) \right)\ ;\\ 
%\end{align}
%\begin{align}
(\eta_\text{O8$^+$}^{-1})_{ij}\,h^\vee_{G_i}h^\vee_{G_j} =& \  h^\vee_{G_0}\left(\frac{(N-1)}{4}h^\vee_{G_0}+\sum_{j=2}^{N-1}\frac{(N-j)}{2}(2k-8(j-1))\right)+ \nonumber\\
 & +\sum_{i=2}^{N-1}(2k-8(i-1))\left(\frac{(N-i)}{2}h^\vee_{G_0}+(N-i)\sum_{j=2}^{i-1}(2k-8(j-1)) \right. + \nonumber \\
& + \left. \sum_{j=i}^{N-1}(N-j)(2k-8(j-1))\right) \ ,
\end{align}
\end{subequations}
where $h^\vee_{G_0}= k+1$ for $\USp(2k)$ and $h^\vee_{G_i} = r_i=2k + i(8-16)$ for $\SU(r_i)$ (in case of $n_0=16$ pairs of D8's on O8$^-$), or $h^\vee_{G_0}=2k-2$ for $\SO(2k)$ and $h^\vee_{G_i} = r_i=2k -  8i$ for $\SU(r_i)$ (in case of a single O8$^+$), with $i=1,\ldots,N-1$ in both cases. Therefore, all ranks scale like $2k$ at large $k\sim N\to \infty$. All in all we obtain:
\begin{subequations}\label{eq:aFTO8+-}
\begin{small}
\begin{align}
a_\text{O8$^-$} &= \frac{192}{7} \left[ \left(\frac{16}{5}N^5 -4 k N^4 + \frac{4}{3}k^2 N^3 \right) +\left(- 16 N^4 +16 k N^3 -4 k^2 N^2\right) + \mathcal{O}(N^3) + \ldots \right] \ , \label{eq:aFTO8-}\\
 a_\text{O8$^+$} &= \frac{192}{7} \left[ \left(\frac{16}{5}N^5 -4 k N^4 + \frac{4}{3} k^2 N^3\right) +\left(- 16 N^4+16 k N^3- 4 k^2 N^2 \right)+ \mathcal{O}(N^3) + \ldots \right]\ . \label{eq:aFTO8+}
\end{align}
\end{small}%
\end{subequations}%
We observe that $a_\text{O8$^-$}$ and $a_\text{O8$^+$}$ are equal up to order $\mathcal{O}(N^4)$. Therefore the dual AdS$_7$ vacua, which only capture the leading $\mathcal{O}(N^5)$ contributions, cannot distinguish between the two field theories and will be defined by the same solution $\alpha(z)_\text{O8}$ to \eqref{eq:third-der-alpha}. However we can compute $a_\text{O8$^-$}$ and $a_\text{O8$^+$}$ to all orders in field theory, by evaluating the exact formula \eqref{eq:a}. Calling $a_5$ and $a_4$ the order $\mathcal{O}(N^5)$ and $\mathcal{O}(N^4)$ terms in \eqref{eq:aFTO8+-} respectively, we obtain:
\begin{subequations}\label{eq:aFTO8+-tot}
\begin{small}
\begin{align}
a_\text{O8$^-$} =\ & \frac{192}{7} \left[ a_5 + a_4 + \left( \frac{202}{9} N^3 -\frac{101}{6} kN^2 + \frac{7}{2} k^2 N \right) - \left( \frac{829}{180} N^2 -\frac{829}{360} kN + \frac{5051}{5760}k^2 \right)\right. + \nonumber \\
&\left. - \left( \frac{9787}{2880} N -\frac{5921}{2304} k \right) - \frac{105}{64} \right] \ , \label{eq:aFTO8-tot}\\
a_\text{O8$^+$} =\ & \frac{192}{7} \left[ a_5 + a_4 + \left( \frac{262}{9} N^3 -\frac{131}{6} kN^2 + \frac{7}{2} k^2 N \right) - \left( \frac{4429}{180} N^2 -\frac{4429}{360} kN + \frac{5051}{5760}k^2 \right)\right. + \nonumber \\
&\left. +\left( \frac{28613}{2880} N -\frac{3133 k}{1280}k \right) - \frac{105}{64} \right]\ . \label{eq:aFTO8+tot}
\end{align}
\end{small}%
\end{subequations}%
We expect that the exact subleading $\mathcal{O}(N^4)$ and lower contributions be reproduced by stringy and curvature corrections to the supergravity solution $\alpha(z)_\text{O8}$. We leave this for future investigation.

\subsection{Quivers from brane configurations with O6-planes and an O8-plane}
\label{sub:O8O6pole} 

We shall now add O6-planes to the configuration considered in the previous subsection. The allowed brane setups and resulting quivers are depicted in figure \ref{fig:NS5D6O6D8O8unstuck}.

As can  be seen there, we have an $\SU$ gauge group followed by a chain of alternating $\SO$-$\USp$ groups. As we explained towards the end of section \ref{subsub:O8pureSUk},  the former is engineered through a combined O6$^\pm$-O8$^\mp$ projection: $\SU(r_0) \to \USp(2r_0) \to \SU(r_0)$ (resp. $\SU(r_0) \to \SO(2r_0) \to \SU(r_0)$). The dual supergravity solutions are characterized by an O8 pole at $z=0$ with effective D6 charge provided by the one of the D6-O6 stack (that is, $y_0=0$ but $\alpha_0\neq0$) which can be either positive or negative, and a D6 or O6 pole at $z=N$ where $r_N\neq 0$, $y_N\neq0$ but $\alpha_N=\tilde{r}_N=0$, respectively $r_N=0$ but $y_N\neq 0, \alpha_N \neq 0 \Rightarrow\tilde{r}_N \neq 0$, depending on the total effective D6 charge. In either case, this information will be washed away in the holographic limit, and the two leading field theory results are equivalent.

In this case, the inverse Dirac pairing is given by \eqref{eq:eta-1O8O6}. Therefore:
\begin{equation}\label{eq:aO8O6FTholo}
a \sim \frac{192}{7} (\eta^{-1}_\text{O6O8})_{ij}\, h^\vee_{G_i} h^\vee_{G_j} = \frac{192}{7 \cdot 2} (\eta^{-1}_\text{O8$^-$})_{ij}\, v_i v_j \, h^\vee_{G_i} h^\vee_{G_j} \ .
\end{equation}
Its entries again depend on the components $v_i$ of an auxiliary vector $\mathbf{v}$ of dimension $N_\text{T}$, which are all equal to either 1 or 2: We have 1 for an $\SO$ group and the first $\SU$ group, and 2 for a $\USp$ one, i.e. $\mathbf{v}=\{1,2,1,2,1,\ldots\}$ or $\mathbf{v}=\{1,1,2,1,2,\ldots\}$. From table \ref{tab:groupconst} we also see that  $v_0 h^\vee_{\SU(r_0)} = r_0$, $v_i h^\vee_{\SO(2r_i)} = 2r_i-2 \sim 2r_i$ as $r_i \sim N \to \infty$, and similarly $v_i h^\vee_{\USp(2r_i)} = 2r_i+2 \sim 2r_i$, where $r_i$ can be viewed as the effective number of D6-branes in a D6-O6 stack as in figures \ref{fig:NS5D6O6D8O8+unstuck} and \ref{fig:NS5D6O6D8O8-unstuck}. For this reason and because of \eqref{eq:aO8O6FTholo}, the computation of the $a$ conformal anomaly from field theory is analogous to the O8 case without O6-planes.

% fold sec (aexamples)

%%%%%%%%%%%%%%%%%%%%%%%%%%%%%%%%%%%
\section{Holographic match} % sec (match)
\label{sec:match}
%%%%%%%%%%%%%%%%%%%%%%%%%%%%%%%%%%%

In this section we will perform the $a$ conformal anomaly holographic match for the brane configurations and quivers of section \ref{sub:dictionary}. Namely, we will match the holographic limit of the exact field theory results we have computed in section \ref{sec:aexamples} to the supergravity results (called $a_\text{hol}$ in the following) at large $r_i,k,N$ that we derive below.\\

The leading order of $a$ can be computed in supergravity as an integral over the internal space $M_3$ of the AdS$_7$ vacuum \cite[Eq. (5.67)]{afpt}:\footnote{See appendix \ref{app:agrav} for an expanded discussion.}
\begin{equation}\label{eq:aholo-sec}
a_\text{hol} =\frac{128}{189\,\pi^2} \int_0^N \alpha(z) q(z)\, dz \ ,
\end{equation}
where $\alpha(z)$ is the cubic polynomial in \eqref{eq:newalphanew} by which the vacuum is defined.

\subsection{Solutions with regular or D6 poles}
\label{sub:regD6polesgrav}

In this subsection we aim to match the supergravity computation of the $a$ conformal anomaly with the holographic limit of the field theory result obtained in section \ref{sub:regD6poles}.\\

In this case $\alpha_0=\alpha_N=0$ and we must use the integration constants $y_0, y_N$ in \eqref{eq:y0limit}, \eqref{eq:yNlimit}. The parameters $r_0,r_N$ may or may not be zero, according to the type of poles (regular or D6, respectively) we want that the internal space $M_3$ have. However, as it turns out, the holographic match is completely insensitive to $r_0,r_N$, which enter in subleading terms w.r.t. the leading $\mathcal{O}(N^5)$ order (i.e. they are subleading as $k\sim N \to \infty$).

All is left to do is to straightforwardly match \eqref{eq:aFTholD6} to \eqref{eq:aGRhol} -- which we reproduce below for reference -- with $y_0, y_N$ given by \eqref{eq:y0limit}, \eqref{eq:yNlimit} respectively. (The comparison is easier order-by-order in the parameter $k$, i.e. the height of the central plateau.)  
\begin{subequations}\label{eq:aGRhol-mainbody}
\begin{align}
a_\text{hol} \sim
&-\frac{192}{7}\left(\sum_{i=1}^{L}\sum_{k=1}^{i-1}(i -k)r_i r_k + \frac{2}{9\pi}y_0 \sum_{i=1}^{L} i r_i  \right)  \nonumber \\
&-\frac{192}{7}\left(\sum_{i=1}^{R}\sum_{k=1}^{i-1}(i - k)r_{N-i} r_{N-k} - \frac{2}{9\pi}y_N \sum_{i=1}^{R} i r_{N-i}  \right)  \nonumber \\
&-\frac{192}{7}\frac{1}{3(9\pi)}\,(N-R-L) \left[ 3k\left( Ly_0-Ry_N +\frac{9\pi}{2} \sum_{i=1}^{L} (L-i) r_i \right. \right. + \nonumber \\
&+ \left. \left.  \frac{9\pi}{2}\sum_{i=1}^{R} (R-i) r_{N-i}\right) - \frac{9\pi}{4}k^2(N-R-L)^2 \right]\ . 
\end{align}
\end{subequations}
The only nontrivial identity needed to carry out the comparison is the following:
\begin{equation}
\sum_{i,j=1}^L j(N-i)r_i r_j  + \sum_{i=1}^L \sum_{j<i} N (j-i)r_i r_j= \sum_{i=1}^L i(N-i)r_i^2 + \sum_{i=1}^L \sum_{j<i}i(N-j)r_i r_j\ ,
\end{equation}
and likewise for $L \leftrightarrow R$ (in the summation extrema) and $i,j \leftrightarrow N-i, N-j$ (inside the sums).

\subsection{Solutions with O6-planes}
\label{sub:O6polesgrav}

In this subsection we aim to match the supergravity computation of the $a$ conformal anomaly with the holographic limit of the field theory result obtained in section \ref{sub:O6poles}.\\

Here we have three cases to distinguish depending on the boundary data (i.e. type of poles) compatible with the presence of the O6-planes:
\begin{itemize}
\item $r_0=r_N=0$ but $\alpha_0,\alpha_N \neq 0$, the latter being defined in terms of the (negative) D6 charges $\tilde{r}_0, \tilde{r}_N$ of a D6-O6$^-$ source localized at the poles $z=0,N$ of $M_3$;
\item $r_0=r_N\neq 0$ but $\alpha_0,\alpha_N = 0$, that is the total charge of the D6-O6 system at each of the two poles is positive;
\item One pole has a D6-O6$^-$ source with negative effective charge and the other has positive effective D6 charge.
\end{itemize}

The holographic $a$ conformal anomaly from the gravity dual is obtained by plugging \eqref{eq:y0efflimit} and \eqref{eq:yNefflimit} into \eqref{eq:aGRhol}. As is the case with only D6-branes, the numbers $\tilde r_0, \tilde r_N$ or $r_0,r_N$ do not play any role at leading $\mathcal{O}(N^5)$ order: They appear only in subleading terms and hence are washed away in the holographic limit. The match then works equally for the three cases mentioned above, and it is exactly equivalent to the one in the previous section, with the following subtlety though: Due to the orientifold projections, all ranks are effectively multiplied by two, $r_i \to 2r_i$, but the volume of the $\mathbb{RP}^2$ fibers in the solution with O6-planes is half of that of $S^2$ in the solution with only D6's.\footnote{\label{foot:k2k}Another subtlety is the following. In appendix \ref{app:int-const} we have derived the boundary data and integration constants of a generic supergravity vacuum assuming the latter describes the near-horizon limit of an NS5-D6-D8(-O8) brane configuration. In such a case, the height of the plateau (i.e. the maximum rank) is $k$. Notice however that upon introducing O6-planes, the height of the plateau becomes $2k$ as explained towards the end of section \ref{sub:boundary} (see also figure \ref{fig:quiverNS5D6O6}). Therefore we should also send $k \to 2k$ in those formulae for the alternating $\SO$-$\USp$ case. The same applies to boundary data and integration constants of sections \ref{sub:O8O6polegrav}, \ref{sub:formal}, and \ref{sub:O8-O6gravdual}.}

\subsection{\texorpdfstring{Solutions with an O8 at $z=0$, regular or D6 pole at $z=N$}{Solutions with an O8 at z=0, regular or D6 pole at z=N}}
\label{sub:O8polegrav}

In this subsection we aim to match the supergravity computation of the $a$ conformal anomaly with the holographic limit of the field theory result obtained in section \ref{sub:O8pole}.\\

In this case $r_0,\alpha_0 \neq 0$ but $y_0=0$, whereas at the other pole of $M_3$ ($z=N$) we have $\alpha_N=0$ but $y_N\neq 0$, and $r_N$ may or may not be zero (in case of a regular, resp. D6, pole). We have already determined the appropriate integration constants in appendix \ref{sub:O8intconst} (we simply need to plug $\alpha_N=0$ in there). Taking their holographic limit yields:
\begin{subequations}\label{eq:intconstO8}
\begin{align}
& y_j \sim \frac{9\pi}{2} \sum_{i=1}^{j-1}r_i\ ,\quad j\in\left[1,L\right] \ , \\
& y_{N-j} \sim y_N- \frac{9\pi}{2} \sum_{i=1}^{j-1}r_{N-i}\ ,\quad j\in\left[1,R \right] \ , \\
%&y_L \sim \frac{9\pi}{2}\sum_{i=1}^{L-1} r_i\ , \\
%&y_{N-R} \sim y_N - \frac{9\pi}{2}\sum_{i=1}^{R-1} r_{N-i}\ , \\
&y_N \sim \frac{(9\pi)}{2} \left(k(N-L-R)+  \sum_{i=1}^{L-1} r_i + \sum_{i=1}^{R-1}  r_{N-i}\right)\ ; \\
& \alpha_j \sim \alpha_0 - (9\pi)^2\sum_{i=1}^{j-1}(j-i) r_i \ , \quad j \in \left[1,L \right]\ , \\
&\alpha_{N-j} \sim (9\pi) (2 j) y_N -(9\pi)^2\sum_{i=1}^{j-1} (j-i) r_{N-i}\ ,\\
&\alpha_0 \sim \frac{(9\pi)^2}{6} \left(3k(N-L-R)(N+L-R) +  6\left( \sum_{i=1}^{L-1} (N-i) r_i + \sum_{i=1}^{R-1}i r_{N-i} \right)\right)\ .
\end{align}
\end{subequations}
We can now sum \eqref{eq:left-integral}, \eqref{eq:right-integral}, and \eqref{eq:int-massless-fin1}, plug in the integration constants given by \eqref{eq:intconstO8}, and keep only the leading terms. Doing so yields:
\begin{subequations}\label{eq:aO8hollim}
\begin{align}
a_\text{hol} \sim & -\frac{192}{7}\left( -\frac{1}{(9\pi)^2} \alpha_0 \sum_{i=1}^{L-2}r_i -\frac{2}{(9\pi)} y_N \sum_{i=1}^{R-1}ir_{N-i} \right. \ + \nonumber \\
& +\left. \sum_{i=1}^{L-1}\sum_{k=1}^{i-2}(i -k)r_i r_k + \sum_{i=1}^{R-1}\sum_{k=1}^{i-2}(i -k)r_{N-i} r_{N-k} \right) +  \nonumber \\
& + \frac{2^7}{3^6\,7\, \pi ^3 k} \,(y_{N-R} - y_L) \left( \frac{3}{2}k(\alpha_L + \alpha_{N-R}) + (y_{N-R} -y_L)^2 \right) \ , \\
\frac{7}{192}a_\text{hol} \sim &\left( \sum_{i=1}^{L-1}(N-i) r_i^2 + 2 \sum_{i=1}^{L-1}\sum_{j<i} (N-i) r_i r_j + 2\sum_{i=1}^{L-1}r_i \sum_{j=1}^{R-1} j r_{N-j} + \sum_{i=1}^{R-1}i r_{N-i}^2 
\right. +  \nonumber \\
& + 2 \sum_{i=1}^{R-1}\sum_{j<i} i r_{N-i} r_{N-j} + k(N-L-R)\left(\sum_{i=1}^{R-1}i r_{N-i} +k(N-L+R)\sum_{i=1}^{L-1}r_i  \right) + \nonumber \\
&+ \big. \frac{k^2}{3} (N-L-R)^2 (N-L+2R)\Bigg) \ .
\end{align}
\end{subequations}
It is a simple exercise to match this expression to the field theory one \eqref{eq:aFTholO8} term by term.

\subsection{\texorpdfstring{Solutions with an O8 at $z=0$, and O6-planes}{Solutions with an O8 at z=0, and O6-planes}}
\label{sub:O8O6polegrav}

In this subsection we aim to match the supergravity computation of the $a$ conformal anomaly with the holographic limit of the field theory result obtained in section \ref{sub:O8O6pole}.\\

Here $y_0=0$, but $r_0 \neq 0$ and $\alpha_0\neq0$, while at $z=N$ we can have two possibilities:
\begin{itemize}
\item $r_N= 0$ but $\alpha_N \neq 0 \Rightarrow\tilde{r}_N \neq 0$, when the total charge of the D6-O6$^-$ system at $z=N$ is negative. The number $\tilde{r}_{N}$ is interpreted as the effective D6 charge of a D6-O6$^-$ source localized at the pole $z=N$ of $M_3$;
\item $r_N \neq 0$ but $\alpha_N = 0$, when the total charge of the D6-O6 system at $z=N$ is positive.
\end{itemize}
We can define the rank $2n_2$ of a flavor $\SO$ group as explained in section \ref{subsub:alternatingSOUSp}. 

The holographic $a$ conformal anomaly is obtained by plugging the expression for $y_N$ and the expression \eqref{eq:alpha0O8} for $\alpha_0$ into \eqref{eq:aO8hollim} with $r_i \to 2r_i$ for $i>1$. As in the case without O6-planes, the numbers $\tilde r_N$ or $r_N$ do  not appear directly at leading $\mathcal{O}(N^5)$ order, and hence do not affect the holographic computation. The match of the gravity calculation of $a$ with the field theory result \ref{sub:O8O6pole} works just as in the previous section, keeping in mind the caveat concerning the factor of $1/2$ from the integrated volume of $\mathbb{RP}^2$ with respect to $S^2$, and that $r_i\rightarrow  2r_i$ as well as $k \rightarrow 2k$ due to the orientifold action from the O6-planes.

% fold sec (match)

%%%%%%%%%%%%%%%%%%%
\section{New examples} % sec (newex)
\label{sec:newex}
%%%%%%%%%%%%%%%%%%%

In this section we shall compute the holographic $a$ anomaly from supergravity for a few novel examples of $(1,0)$ theories. Given that the latter fall into the classes treated in section \ref{sec:aexamples}, the holographic result is guaranteed to match the one obtained in field theory. Along the way, we will also construct their AdS$_7$ supergravity duals for the first time.\\

The first theory we focus on is an example of $(1,0)$ linear quiver engineered by a so-called ``formal'' massive type IIA configuration \cite{mekareeya-rudelius-tomasiello}, which we have already defined towards the end of section \ref{subsub:alternatingSOUSp}. The second example is the theory engineered by the NS5-D6-D8-O8$^-$ brane configuration depicted in figure \ref{fig:NS5D6D8O8-unstuck} (in particular, we will match the gravity result to \eqref{eq:aFTO8-}). Finally, we will tackle the case characterized by a combined O6$^+$-O8$^-$ orientifold action: The quiver of which we compute the holographic $a$ anomaly is the right one in figure \ref{fig:quiverNS5D6O6D8O8-unstuck}, and is engineered by the brane configuration in figure \ref{fig:NS5D6O6D8O8-unstuck}.

\subsection{A formal massive IIA quiver and its dual vacuum}
\label{sub:formal}

In presence of O6-planes, the perturbative type IIA description of an alternating $\SO$-$\USp$ quiver may sometimes break down due to the appearance of hypermultiplet spinor representations in the F-theory engineering of the same SCFT \cite{heckman-rudelius-tomasiello}. One must then turn to the latter description to reliably compute field theory observables. However \cite{mekareeya-rudelius-tomasiello} made a remarkable observation: One may still use a type IIA description to compute the $a$ conformal anomaly of the quiver theory, at the price of using a so-called ``formal'' brane configuration, where some gauge groups have a non-positive rank. As anticipated in section \ref{subsub:alternatingSOUSp}, this always happens for a $(1,0)$ theory one of whose tails is labeled by the principal orbit $\mathcal{O}_{[2k-1,1]}$ of $\mathfrak{so}(2k)$ (i.e. $2k=(2k-1)+1$ with $2k-1$ odd, since $2k$ is even). Using the notation of the right quiver in figure \ref{fig:quiverNS5D6O6D8}, we see that the $(1,0)$ theory is given by
\begin{subequations}\label{eq:formalF}
\begin{small}
\begin{align}
\Node{\ver{}{1}}{-6}-\node{}{3}-\Node{}{-4}-&\node{}{5}-\Node{}{-2}-\node{}{7}-\Node{}{0} - \node{}{9} - \Node{}{2} -  \cdots -  \node{}{2k-1} - \Node{\ver{}{1}}{2k-8}  -  \node{}{2k}-  \Node{}{2k-8} -  \cdots -   \Node{\ver{}{2k}}{2k-8}   \label{fig:formalquiv55}  \\
& \nonumber \\
&  2 \quad \, \, \overset{\mathfrak{su_2}}2 \quad  \overset{\mathfrak{g_{2}}}3  \quad   1 \quad  \overset{\mathfrak{so_{9}}}4  \,\,\,   \overset{\mathfrak{usp_2}}1  \,\,\,\,    \cdots \,\,\,  \overset{\mathfrak{so}_{2k-1}}4  \,\,\,  \underset{\left[N_f=\frac{1}{2} \right]}{\overset{\mathfrak{usp}_{2k-8}}1} \,\,\,\,   \overset{\mathfrak{so}_{2k}}4 \,\,\,  \overset{\mathfrak{usp}_{2k-8}}1\,\,\,\, \cdots \,\,\,\, \overset{\mathfrak{usp}_{2k-8}}1 \,\,\,\, [\SO(2k)]   \label{fig:Ftheoryquiv55}
\end{align}
\end{small}%
\end{subequations}%
Notice that \eqref{fig:formalquiv55} corresponds to a Higgsing of the quiver in figure \ref{fig:quiverNS5D6O6}, with a left massive tail, then a massless plateau. We have also overlaid the formal type IIA construction onto the F-theory engineering \eqref{fig:Ftheoryquiv55} of the same quiver. We see that, superficially, the two are very different. For instance, the latter features the exceptional gauge algebra $\mathfrak{g}_2$, a half-hypermultiplet of the leftmost $\mathfrak{usp}(2k-8)$ algebra (at the beginning of the massless plateau), and an $\SO(2k)$ flavor symmetry (denoted by square brackets) corresponding to the gray square of rank $2k$ in \eqref{fig:formalquiv55}. Nonetheless in \cite[Eqs. (4.42)-(4.48)]{mekareeya-rudelius-tomasiello} it was checked that the coefficients $\alpha,\ldots,\delta$ \eqref{eq:coeffpoly} through which $a$ is defined agree on both sides at finite $k,N$. 

Here we will extract the large $k,N$ behavior of $a$, and also construct the AdS$_7$ vacuum dual to \eqref{fig:formalquiv55}. We propose that the latter provide the gravity dual to the nonperturbative F-theory configuration \eqref{fig:Ftheoryquiv55}. This partially reduces the scarcity of AdS$_d$ ``solutions'' of F-theory, extending the class of those claimed to exist in \cite{garciaetxebarria-regalado,garciaetxebarria-regalado-proc,aharony-fayyazuddin-maldacena,aharony-tachikawa} (for $d=5$), \cite{couzens-lawrie-martelli-schafernameki-wong} (for $d=3$) and \cite{inverso-samtleben-trigiante} (for $d=4$) to the case $d=7$.\footnote{\label{foot:AdS-F}Notice however a crucial difference. In the cited examples, the AdS vacua are type IIB supergravity solutions with varying axiodilaton (with or without seven-brane monodromies). Here we have a type IIA vacuum producing the same $a$ conformal anomaly as an F-theory configuration, in the holographic limit.}\\

The supergravity solution is constructed as usual by reading off the combinatorial data from the quiver in \eqref{fig:formalquiv55}. We have, for $2k-1$ odd:
\begin{equation}\label{eq:ranks-formal}
L=2k-1\ , \ R= 0\ ; \quad \varrho_i:=\begin{cases} -8+(2i-1)+1 & \text{$i$ odd} \\ 2i+1 & \text{$i$ even} \end{cases}\quad \text{for}\ i=1,\ldots,k-1\ .
\end{equation}
Notice that in the holographic limit, i.e. when $\varrho_i \sim k \sim N \to \infty$, we can simply put $\varrho_i:=2r_i+1=2i+1$ for all $i$, as constant shifts are unimportant. We have $\rho_1:=r_0=0$, but $\tilde{r}_0=-3$, given the leftmost flavor $\SO(1)$ which is engineered by an $\widetilde{\text{O6}}^-$ (see the discussion in section \ref{subsub:alternatingSOUSp}). This effective charge will define an $\alpha_0 \propto \tilde{r}_0 y_0 \neq 0$ via \eqref{eq:effectivealphas}. On the other hand, the rightmost $\SO(2k)$ group corresponds to a D6 pole (i.e. a source with positive D6 charge) given that $k\geq 4$ (remember that this is a Higgsing of the theory of $N$ M5's probing $\cc^2/D_k$, $k\geq 4$); therefore we will take $y_N \neq 0, \alpha_N =0$. Finally $L=2k-1$ but $R=0$, since we only have a left massive tail before the plateau.

We have already computed the integration constants and boundary data in the generic (i.e. $r_{0,N}$, $\alpha_{0,N}$ not necessarily vanishing) asymmetric (i.e. $L \neq R$, $L,R <N$) case in appendix \ref{appsubsub:genasym}. However since $R=0$ this falls into the limiting case treated in section \eqref{appsub:limcasO8}, with $\varrho_i\rightarrow 2i+1$ from $[0,L]$ (the left massive tail) and $\varrho_i\rightarrow 2k$ from $[L+1,N]$ (the massless plateau). With the above choices, \eqref{eq:y0-asym-generic-eff} and \eqref{eq:effectiver0} become:\footnote{Notice that \eqref{eq:y0formal-mar} is indeed negative (as required by the argument in footnote \ref{foot:alpha-signs}), once we fix the dependence of $k$ on $L\sim N$ via $k:=\sum_{i=1}^L s_i =r_L-r_0$, which is of order $\mathcal{O}(N^1)$. In particular $k=\kappa N$ with $0<\kappa <\sqrt{3}/2$. This region may seem ``small'' and nongeneric; however notice that it becomes large if we invert the dependence as $N=1/\kappa\, k$, which is just another admissible way of achieving \eqref{eq:hololimit}.}
\begin{subequations}
\begin{align}
y_0 &= \frac{9 \pi^3  (k (4 (k-3) k-3 (N-1) N+10)-3)}{4+6\pi^2 N}\ , \label{eq:y0formal-mar}\\
y_N &=\frac{3 \pi^3 k \left(4 k^2-6 k+3 N (N+3)\right)+ 18 \pi (k N+k-1)}{4+4\pi^2 N}\ , \\
\alpha_0 &= -\frac{81\pi^2 (k (4 (k-3) k-3 (N-1) N+10)-3)}{2+3\pi^2 N}\ . \label{eq:a0formal-mar}
\end{align}
\end{subequations}
The integration constants, in the subintervals $z\in[i,i+1]$ for $i=1,\ldots,L-1$, are given by:
\begin{small}
\begin{subequations}
\begin{align}
y_i =&\ \frac{9 \pi^3 (3 (i (i+2)-1) N+2 k (4 (k-3) k-3 (N-1) N+10)-6)+18\pi (i (i+2)-1)}{8+ 12 \pi^2  N}\ , \\
\alpha_i =&\ -\frac{54 (i (i (i+3)-3)+3 k (4 (k-3) k-3 (N-1) N+10)-8)}{4+6\pi^2 N}\ + \nonumber \\ &\ + \frac{81 \pi ^2 (i ((i (i+3)-3) N-2 k (4 (k-3) k-3 (N-1) N+10)+6)+N)}{4+6\pi^2 N}\ .
\end{align}
\end{subequations}
\end{small}%
In the massless region $z\in [i,i+1]$ with $i\in[L,N-2]$ we have instead:
\begin{subequations}
\begin{small}
\begin{align}
y_i =&\ \frac{1}{8+12\pi^2 N}\left[ 9 \pi ^3 \left(2 k \left(6 (i+1) N+4 k^2-6 k (N+2)-3 N^2+10\right)+ 3 (i-1) N-6\right) \right. + \nonumber \\ &+ \left. 18\pi\left(4 i k+i-4 k^2+2 k-1\right) \right]\ , \\
\alpha_i =&\ -\frac{1}{4+6\pi^2 N}\left[ 81 \pi ^2 \left(6 i^2 k N+i \left(2 k \left(-4 k^2-6 k (N-2)+3 N (N+1)-10\right)-5 N+6\right) + \right.\right. \nonumber\\
&\ + \left.\left. 4 (k-1) k (2 k-1) N+N\right) + 54 \left(6 i^2 k+i (-12 (k-1) k-5)+k (4 k (5 k-12)\ + \right.\right. \nonumber \\
&\ - \left. \left. 9 (N-1) N+34)-8\right) \right]\ .
\end{align}%
\end{small}%
\end{subequations}%
Plugging the above integration constants and the ranks $\varrho_i=2r_i+1$ into \eqref{eq:newalphanew} defines the corresponding supergravity solution $\alpha(z)_\text{formal}$. Performing the internal space integral \eqref{eq:aGRhol} (and setting to zero by hand the contribution from the right massive region), the leading order of the $a$ conformal anomaly is found to be
\begin{equation}
\boxed{a_\text{formal} \sim \frac{32}{7}k^2 \left(N^3 +8 k^2 N - \frac{64}{5}k^3-\frac{16}{3} \frac{k^4}{N} \right)}\ , \quad \text{as $k,N \to \infty$}\ .
\end{equation}
One can check that this indeed agrees with \eqref{eq:aO6FTholo}, specialized to the present case.

\subsection{\texorpdfstring{The gravity dual of the O8$^-$}{The gravity dual of the O8-}}
\label{sub:O8-gravdual}

\begin{figure}[htp]
\centering
\includegraphics[scale=.25]{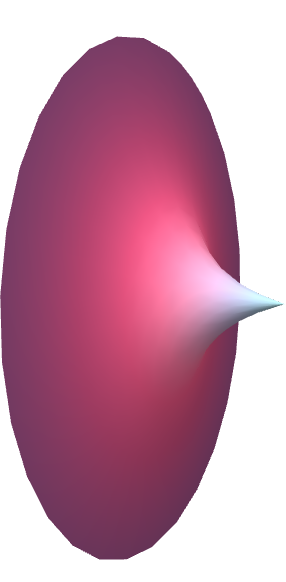}
\label{fig:singleO8}
\caption{An artist's impression of the internal space $M_3$ of the AdS$_7$ vacuum dual to the quiver in \eqref{eq:singleO8-productgauge} with $N=80$ and $k=10$. The topology is that of a half-$S^3$. (This is somewhat reminiscent of the half-$S^4$ internal space of the vacuum in \cite{brandhuber-oz}, dual to the D4-D8-O8$^-$ configuration of \cite{seiberg-5d}.). The O8$^-$ source wraps its equator (i.e. is localized at the $z=0$ pole). The $z=N$ pole is of D6 type, i.e. is a singular point for the metric $ds^2_{M_3}$. Notice that, in contrast to figure \ref{fig:solNS5D6D8}, there are no black creases representing D8's wrapping $S^2$ fibers, since here $n_0=0$.}
\end{figure}
In this section we shall construct the AdS$_7$ dual to the right quiver in figure \ref{fig:quiverNS5D6D8O8-unstuck} without flavors, and extract the holographic $a$ conformal anomaly. We decide to focus on the theory engineered by a single O8$^-$ (that is, $n_0=0$ in the notation of figure \ref{fig:NS5D6D8O8-unstuck} but $n_0=8$ in the notation of figure \ref{fig:massiveE}), instead of a D8-O8$^-$ stack.

The O8 source sits at the $z=0$ pole of the internal space $M_3$, and enforces the conditions $r_0=2k$, $\alpha_0 \neq 0$ (but $y_0=0$); at the other pole ($z=N$) we have a stack of $2k+8N$ semi-infinite D6-branes, therefore $r_N=2k+8N$, $\alpha_N=0$, $y_N \neq 0$ (here $k$ should not be confused with the maximum rank -- i.e. maximum number of D6-branes in the configuration, which is $2k+8N$ in this case). The product gauge group is
\begin{equation}\label{eq:singleO8-productgauge}
\USp(r_0=2k) \times \prod_{i=1}^{N-1} \SU(r_i=2k+8i)\ .
\end{equation}
There is only a left massive region, occupying the whole base interval $I=[0,N]$. Therefore $L=N$ and $R=0$.

The boundary data fall into the class of section \ref{appsub:limcasO8}, since we are in a limiting case (i.e. $L=N$ and $R=0$). In other words the solution is parameterized by a single increasing ramp of $r_i$. The boundary data are found to be
\begin{subequations}
\begin{align}
y_N &=9 \pi  \left( k N+2 N^2\right)\ ,\\
\alpha_0 &=27 \pi ^2 N^2 (3 k+4 N)\ .
\end{align}
\end{subequations}
In each subinterval $z\in [i,i+1]$ with $i\in[1,N-1]$ the integration constants \eqref{eq:yialphai} become
\begin{subequations}
\begin{align}
y_i &=9 \pi  i (2 i+k)\ , \\
\alpha_i &= -27 \pi ^2 \left(4 i^3+3 i^2 k-N^2 (3 k+4 N)\right)\ .
\end{align}
\end{subequations}
We can now use the above boundary data and integration constants to define a function $\alpha(z)_\text{O8$^-$}$ characterizing the vacuum dual to the right quiver in figure \ref{fig:quiverNS5D6D8O8-unstuck} (with $f_i=n_0=0$). Performing the internal space integral \eqref{eq:aO8hollim} produces, at leading order, the following $a$ conformal anomaly:
\begin{equation}\label{eq:aUSpO8-}
\boxed{a_\text{single O8$^-$} \sim \frac{192}{7}\left( \frac{16}{5} N^5 +4 kN^4 +\frac{4}{3}k^2 N^3 \right)}\ , \quad \text{as $k,N \to \infty$}\ .
\end{equation}
Notice that this is exactly the $\mathcal{O}(N^5)$ order of the field theory expression \eqref{eq:aFTO8+-} albeit with a sign difference in the second summand, which is simply due to the different D8 charge of a single O8$^-$ w.r.t. what considered there. More generally, for a D8-O8$^\mp$ stack (with $n_0$ pairs of image branes) the first two coefficients in parenthesis read $\frac{1}{5}(4-n_0)^2$ and $\pm(4-n_0)$ respectively. Thus when $n_0 \neq 0$ but $k=0 \Rightarrow r_0=0$, \eqref{eq:aUSpO8-} precisely matches \cite[Eq. (5.13)]{bpt} under the identification $n_8^\text{there} := 8-n_0^\text{there} \equiv 2 n_0^\text{here}$.\footnote{Notice that \cite{bpt} only counts physical branes in the reduced space, whereby the D$p$ charge of an O$p^\pm$ is $\pm2^{p-5}$. So for $n_0 \neq 0$ and $k=0$ \eqref{eq:aUSpO8-} gives the conformal anomaly $a\sim \frac{16}{7}\frac{9}{15} (8-2n_0^\text{here})^2 N^5 = \frac{16}{7}\frac{9}{15} (n_0^\text{there})^2 N^5$ of the massive \mbox{$E_{1+n_0^\text{here}/2}$-string} theory of $n_0^\text{here}/2$ pairs of D8's overlaid onto an O8$^-$.} Our expression generalizes that formula to cases with nonzero D6 brane charge $r_0=2k$. A single O8$^-$ engineers a so-called massive $E_1$-string theory.

\subsection{\texorpdfstring{The gravity dual of the O8$^-$ with O6-planes}{The gravity dual of the O8- with O6-planes}}
\label{sub:O8-O6gravdual}

In this section we shall compute the holographic $a$ conformal anomaly of the right quiver in figure \ref{fig:quiverNS5D6O6D8O8-unstuck}, engineered by the brane configuration of figure \ref{fig:NS5D6O6D8O8-unstuck} featuring a combined O6$^+$-O8$^-$ orientifold projection on the first gauge group $\SU(k) \to \SO(2k) \to \SU(r_0=k)$. As done in the previous subsections, by computing the appropriate integration constants and boundary conditions we are able to construct the dual AdS$_7$ solution $\alpha(z)_\text{O6-O8}$.

Given the presence of the D8-O8$^-$ stack, the quiver has an $\SO(32)$ flavor symmetry of the 0-th gauge group (and the physical D8's contribute $n_0=16$ full hypermultiplets). If we do not insert any other D8 in the brane configuration, the rightmost D6-O6$^-$ stack escaping off to infinity engineers an $\SO(g_{N})$ flavor symmetry. Using $r_0=k$, $g_0=2n_0=32$ and applying condition \eqref{eq:ti-gaugeanom} repeatedly at each node, the product gauge group is found to be
\begin{equation}
\SU(k) \times \prod_{i=1}^{N-1} \SO(p_i=2k-24i +8) \times \USp(q_i=2k-24 i)\ .
\end{equation}
In the holographic computation ($k \sim N \to \infty$) we will use $2r_i = 2k-24 i+8$ for the ranks. Moreover since the latter are decreasing as $i$ increases, the $z=N$ pole will be of O6 type if the effective D6 charge becomes negative. This means $q_N=0$ but $g_N:= p_N= 2r_N \neq 0$. To have a meaningful rightmost flavor symmetry we must impose
\begin{equation}
2r_N=2k -24 N+8 \geq 0\ ,
\end{equation}
which is saturated by an empty flavor group $\SO(2r_N=0=2 n_2)$, i.e. $k = 12 N-4$. In this case $\tilde{r}_N=-4+2n_2=-4$, since there are $2r_N=0$ D6-branes on top of the semi-infinite O6$^-$. Thus, $\alpha_N=\frac{3}{4\pi}y_N$ by \eqref{eq:effectiverN}. Moreover notice that $2r_i = 2k-24i+8=24(N-i)$, so we might as well label the gauge groups starting from the right, $2r_{N-i}=24 i$.  We only have a right massive tail filling up the whole base interval $I=[0,N]$. Hence $L=0$ and $R=N$. Also, in the supergravity solution we have to use $2r_0=2k$ at $z=0$, which corresponds to the total number of D6-branes on top of the first O6-plane.

At $z=0$, the O8 enforces the conditions $y_0=0, \alpha_0 \neq 0$. Therefore, we simply need to determine $\alpha_0$ and $y_N$. This has already been done in appendix \ref{appsub:limcasO8}, but again we are in a limiting case, i.e. the massless region is absent and we just have a ramp of decreasing ranks $r_i$. We find:
\begin{subequations}
\begin{align}
\alpha_0 &= \frac{27}{2} \left(3 N^2+8 \pi ^2 (3 N (N (2 N-3)+2)-1)-1\right)\ ,\\
y_N &=18 \pi  \left(3 N^2-1\right) \ , \\
\alpha_N &= \frac{3}{4\pi} y_N =\frac{27}{2} \left(3 N^2-1\right)\ .
\end{align}
\end{subequations}
The integration constants in the subintervals $[N-i,N-(i-1)]$ for $i=1,\ldots,N-1$ are given by:
\begin{subequations}
\begin{align}
y_{N-i} &= -18 \pi  \left(3 i^2-3 N^2+1\right) \ , \\
\alpha_{N-i} &=-\frac{27}{2} \left(24 \pi ^2 i^3-3 N^2+1\right)\  .
\end{align}
\end{subequations}
We can now use the above boundary data and integration constants to define a function $\alpha_\text{O6-O8}$. Performing the internal space integral \eqref{eq:aO8hollim} finally yields:
\begin{equation}\label{eq:aO6O8holo}
\boxed{a_\text{O6-O8} \sim \frac{16^2}{7} \frac{12^2}{5}N^5}\ , \quad \text{as $N \to \infty$}\ .
\end{equation}
One can check that this agrees with the leading order of \eqref{eq:aO8O6FTholo}, which reads
\begin{equation}\label{eq:aO6O8FT}
12^2\, \frac{16}{7} \left( \frac{16}{5}N^5 -2 N^4 +\mathcal{O}(N^3) \right) \ .
\end{equation}

% fold sec (newex)

%%%%%%%%%%%%%%%%%%%%%%%%%%%%%%%%%%%%%%%%%%%%%%%%
\section{\texorpdfstring{On the holographic $a$-theorem}{On the holographic a-theorem}} % sec (athm)
\label{sec:athm}
%%%%%%%%%%%%%%%%%%%%%%%%%%%%%%%%%%%%%%%%%%%%%%%%

In this section we would like to provide evidence for the existence of a holographic $a$-theorem for Higgs branch RG flows.\\

As explained in sections \ref{subsub:pureSUk} and \ref{subsub:alternatingSOUSp}, any quiver belonging to the pure $\SU$ class, respectively alternating $\SO$-$\USp$ one, is obtained by Higgsing the theory of $N$ M5's probing the $\cc^2/A_{k-1}$ singularity, respectively $\cc^2/D_k$. In either class, the supergravity dual of the unHiggsed theory is the $\text{AdS}_7 \times S^4/\Gamma$ Freund--Rubin solution of eleven-dimensional supergravity, that can be reduced to a massless (i.e. $F_0=0$) type IIA $\text{AdS}_7 \times M_3$ vacuum. The gravity dual of a Higgsed quiver is instead a massive vacuum, the Romans mass being sourced by flavor D8-branes. The quiver is then labeled by (two) nilpotent orbit(s) of $\mathfrak{su}(k)$, respectively $\mathfrak{so}(2k)$, specifying the way color D6-branes end on the D8's at its two tails. 

A six-dimensional $a$-theorem for tensor branch flows has been proven in \cite{cordova-dumitrescu-intriligator}. For Higgs branch flows, \cite{mekareeya-rudelius-tomasiello} computed $a$ exactly at finite $k,N$ for these two classes of quivers, and established an $a$-theorem:\footnote{Other evidence for its existence was previously given in \cite{heckman-rudelius}, where three monotonically-decreasing functions along the flow are identified.} The ``massless'' quiver is characterized by an anomaly $a_\text{UV}$, whereas any ``massive'', that is Higgsed, quiver by $a_\text{IR}$ such that $a_\text{UV}>a_\text{IR}$. Moreover, $\Delta a >0$ for any two massive quivers, one lower than the other on the so-called (nilpotent orbits) Hasse diagram.

It is then natural to ask whether this statement has a holographic counterpart. We believe the answer is positive. In the pure $\SU$ case, our AdS$_7$ massive type IIA solutions can be consistently truncated to minimal gauged supergravity vacua \cite{prt}, and therefore a holographic $a$-theorem can be established following the arguments of \cite{freedman-gubser-pilch-warner,gppz1}. As usual, the seven-dimensional solutions which interpolate between two critical points (of the scalar potential) along the holographic flow will be obtained by giving appropriate vev's to scalar fields. In the alternating $\SO$-$\USp$ case, in principle one has to worry about the presence of orientifolds in a Romans mass background, which are sources of repulsive ``attraction'' due to their negative tension (at least the O6$^-$'s). However for any physically sensible effective theory in seven dimensions (the gauged supergravity), the kinetic term of the scalar fields should be positive definite, as the O-planes cannot contribute any ghosts (being nondynamical objects). Therefore the positive energy conditions and considerations of \cite{freedman-gubser-pilch-warner} are unscathed, and the holographic $a$-theorem holds true as without O6's. 

It remains to be understood what is the function that decreases monotonically along the holographic flow. A natural candidate is obviously provided by 
\begin{equation}
a_\text{hol} \sim \int_I dz \, \alpha(z) q(z) \propto \int_{M_3} e^{5A-2\phi} \text{vol}_3\ ,
\end{equation}
given that this integral captures the leading order of the holographic $a$ conformal anomaly (see \eqref{eq:aholo-sec}).
%A natural guess would be the double derivative $\ddot{\alpha}$ of the cubic polynomial $\alpha(z)$ in \eqref{eq:newalphanew} defining the AdS$_7$ vacuum. It is easier to show this in the pure $\SU$ case.
We observe however that there is an even simpler function that satisfies the required monotonicity property, namely $q$. Moreover $q \geq 0$ implies that $\int_I dz\, q(z) \geq 0$, which can be understood as a ``volume function'' that decreases along the flow. Indeed in any theory (massless or massive) we have \cite[Eq. (4.41)]{afrt}:
\begin{equation}
e^{-\phi(z)} R_{S^2}(z) = q(z) \geq 0,
\end{equation}
where $R_{S^2}$ is the radius of the $S^2$ fiber of $M_3$ over $z\in I=[0,N]$, which was given in \eqref{eq:radius}. ($S^2$ is replaced by $\rr\pp^2$ in presence of O6-planes, i.e. in the $\mathfrak{so}(2k)$ case.) Focusing on the simpler $\mathfrak{su}(k)$ case, in a Higgsed quiver $q$ is defined in terms of a (transposed) partition in each of the massive tails $[0,L]$ and $[N-R,N]$, and is constant across the massless plateau $[L,N-R]$ (where it equals $\frac{k}{2}$). The partitions can be naturally inverse-ordered starting from $\rho^\text{t}=[1^k]$, i.e. $\rho=[k]$ (which corresponds to the regular orbit $\mathcal{O}_{[k]}$ of maximum dimension), and moving one box at the end of a row to a lower row until we reach the trivial partition $\rho^\text{t}=[k] \leftrightarrow \rho=[1^k]$ (corresponding to the trivial orbit $\{ 0 \}$). E.g. for $k=4$:\footnote{This observation has been heavily exploited in \cite{mekareeya-rudelius-tomasiello}, where the dimension of the Higgs branch of the SCFT has been related to the dimension of the orbits $\mathcal{O}_\text{L,R}$.}
%\begin{subequations}
\begin{equation}\label{eq:young-order}
%\mathcal{O}_{\rho^\text{t}=[4]} = \{ 0 \}\leftrightarrow \rho= &\ {\tiny \yng(1,1,1,1)} \leftarrow {\tiny \yng(2,1,1)} \leftarrow {\tiny \yng(2,2)} \leftarrow {\tiny \yng(3,1)} \leftarrow {\tiny \yng(4)} \leftrightarrow \mathcal{O}_{\rho^\text{t}=[4]}\ , \\
\mathcal{O}_{\rho=[4]} \leftrightarrow \rho^\text{t}= {\tiny \yng(1,1,1,1)} \rightarrow {\tiny \yng(2,1,1)} \rightarrow {\tiny \yng(2,2)} \rightarrow {\tiny \yng(3,1)} \rightarrow {\tiny \yng(4)} \leftrightarrow \mathcal{O}_{\rho=[1^4]}=\{ 0 \}\ .
\end{equation}
%\end{subequations}
Notice that the ``graphical'' ordering prescription $\rightarrow$ on the transposed partitions precisely corresponds to the (partial) order on the nilpotent orbit Hasse diagram.\footnote{For classical Lie algebras $\mathfrak{g}$ the ordering is only partial, i.e. the orbits (and associated partitions) form a poset whereby some may have equal dimension. However for $\mathfrak{g}=\mathfrak{su}(k)$ (or rather its complexification $\mathfrak{sl}(k)$) \cite[Thm. 6.3.2]{collingwood-mcgovern} proves that the transposition of partitions is indeed an order-reversing involution on the Hasse diagram. For $\mathfrak{g}=\mathfrak{so}(2k)$ one needs to be more careful, and must apply the so-called Spaltenstein map \cite[Thm. 6.3.5]{collingwood-mcgovern}.} 

The massless theory has $\rho^\text{t}_\text{L}=\rho^\text{t}_\text{R}=[1^k]$ corresponding to biggest orbits $\mathcal{O}^\text{L,R}_{[k]}$, whereas the ``most massive'' quiver will have $\rho^\text{t}_\text{L}=\rho^\text{t}_\text{L}=[k]$, i.e. is characterized by two tails labeled by $\mathcal{O}^\text{L,R}_{[1^k]}$ at the bottom of the Hasse diagram. In particular it is easy to convince oneself that, for $k,N \to \infty$, the graph of the piecewise linear function $2q(z)= r_i + s_{i+1}(z-i)$ with $z\in[0,N]$ (see e.g. \cite[Fig. 2(b)]{cremonesi-tomasiello}) corresponding to two partitions lower on the $\mathfrak{su}(k)$ Hasse diagram is always dominated from the above by that of one higher on the diagram, and in particular by the massless theory. Consider e.g. $\mathcal{O}_{[s_1,s_2,\ldots,s_m]}$ and $\mathcal{O}'_{[s'_1,s'_2,\ldots,s'_n]}$: If $\mathcal{O} \geq \mathcal{O}'$ then by definition $\sum_{i=1}^j s_i \geq \sum_{i=1}^j s'_i \Leftrightarrow r_j -r_0 \geq r'_j - r_0$ for $1\leq j \leq m,n$, that implies $r_j \geq r'_j$. Then for at least one $j$ we will have $r_j > r'_j$, implying that the graph of $q_{\mathcal{O}'}$ is dominated by that of $q_{\mathcal{O}}$.\footnote{A similar observation using the nilpotent orbit hierarchy, albeit from the field theory perspective, was made in \cite{heckman-rudelius-tomasiello}. (See Eqs. (3.10) and (3.19) in that paper.)} (This is also true for $\mathfrak{so}(2k)$ orbits.) Given the positivity of $q$ throughout the base interval $I$, we also have 
\begin{equation}\label{eq:monot-int}
\int_I dz \,q(z)_\text{massless} > \int_I dz \,q(z)_{\mathcal{O}} \geq \int_I dz \,q(z)_{\mathcal{O}'} > \int_I dz \,q(z)_{\{ 0\} }\ ,
\end{equation}
which proves the monotonicity of the integral along the holographic flow. 

% fold sec (athm)

%%%%%%%%%%%%%%%%%%%%%%%%%%%%
\section{Conclusions} % sec (conc)
\label{sec:conc}
%%%%%%%%%%%%%%%%%%%%%%%%%%%%

In this paper we have computed exactly in field theory the $a$ conformal anomaly of a vast class of six-dimensional $(1,0)$ SCFT's admitting a holographic dual in massive type IIA supergravity. We have done so by leveraging the six-dimensional anomaly polynomial. On the tensor branch each such field theory is described by a linear quiver of $\SU$, $\SO$, and $\USp$ gauge and flavor groups (and matter in various representations). The last two possibilities are engineered through orientifolds inserted in the brane configurations.

We have extracted the leading behavior of $a$ as the number $N$ of gauge groups as well as the maximum rank $k$ become large, and compared this result to the one obtained in supergravity. The latter can be computed as an internal space integral of a cubic polynomial called $\alpha(z)$ by which the dual AdS$_7$ vacuum is defined.

We have provided general formulae for all classes of theories engineered by the brane configurations of figures \ref{fig:SUk}, \ref{fig:SOk}, \ref{fig:NS5D6D8O8unstuck-stuck}, and \ref{fig:NS5D6O6D8O8unstuck}. We have then specialized them to a few important examples, such as the formal massive type IIA quiver \eqref{eq:formalF}, and to the theory engineered by inserting a single O8$^-$-plane, or a combined O6$^+$-O8$^-$ orientifold projection, in a suspended NS5-D6 brane setup. By exploiting the general formalism laid down in the paper we were also able to explicitly construct their AdS$_7$ dual vacua for the first time.

Finally we have given evidence for the existence of a holographic $a$-theorem for Higgs branch RG flows among the quiver theories, and we have also identified a function that decreases monotonically along the holographic flow.\\

\noindent In the following we wish to propose a few possible avenues of future investigation. 
\begin{itemize}
\item It would be interesting to compute stringy corrections or use the brane on-shell action (along the lines of \cite{aharony-tachikawa}) to distinguish between the two configurations considered in section \ref{sub:O8ramps}, namely a single O8$^+$ and an O8$^-$ overlaid onto sixteen pairs of image D8's. This should reproduce the subleading terms in the exact $a$ conformal anomaly \eqref{eq:aFTO8+-tot} that do not come from supergravity, and can therefore be trusted. Computing the subleading terms directly from string theory could also shed light on the nature of the O8$^+$, which is related by T-duality to the O7$^+$ with a frozen singularity in F-theory \cite{bhardwaj-morrison-tachikawa-tomasiello,tachikawa-frozen} (i.e. an $\widehat{I}_4^*$ Kodaira fiber -- engineering a $\mathfrak{usp}(0)$ algebra -- that cannot be resolved, probably because of a discrete flux \cite{witten-toroidal}).

\item It would be possible to adapt our general formalism to compute the holographic $a$ anomaly of six-dimensional conformal matter of type $(E_8, \Gamma)$ \cite[Sec. 6]{delzotto-heckman-tomasiello-vafa}, that is the theory of $N$ M5's probing the intersection between an $E_8$ Ho\v{r}ava--Witten wall and a $\cc^2/\Gamma$ singularity, with $\Gamma=A_{k-1},D_k$. The (rank-$N$) massive $E_8$-string theory is an example thereof, and we have explicitly shown in section \ref{sub:O8-gravdual} how to extract the large $N$ behavior of $a$ for all $E_{1+(8-n_0)}$-string theories, $1 \leq n_0 \leq 8$ (with nonzero D6-brane charge $r_0$). By modifying the massive tail, we could easily accommodate an alternating sequence of $\SO$-$\USp$ groups, ending on the D8-O8$^-$ wall.

\item Finally, it would now be a simple exercise to extend the computation in section \ref{sub:formal} to all formal type IIA quivers derived in \cite{mekareeya-rudelius-tomasiello}, in order to enlarge the class of massive AdS$_7$ vacua producing at large $k,N$ the same $a$ conformal anomaly as nonperturbative F-theory quivers.

\end{itemize}

% fold sec (conc)

%%%%%%%%%%%%%%%%%%%%%%%%%%%%%%
\section*{Acknowledgments} % sec (acknow)
%%%%%%%%%%%%%%%%%%%%%%%%%%%%%%

We wish to thank A.~Tomasiello for guidance and valuable insight, N.~Mekareeya for bringing to our attention the formal constructions of \cite{mekareeya-rudelius-tomasiello}, S.~Cremonesi and G.~Dibitetto for useful correspondence, and O.~Bergman for helpful comments on issues with orientifolds. We have benefited from discussions with A.~Amariti, M.~Del Zotto, J.~J.~Heckman, S.~S.~Razamat, T.~Rudelius, L.~Tizzano, F.~Yagi, and A.~Zaffaroni. F.A.~is supported by the NSF CAREER grant PHY-1756996 and by the NSF grant PHY-1620311. M.F.~is supported in part by the Israel Science Foundation under grant No. 1696/15 and 504/13, and by the I-CORE Program of the Planning and Budgeting Committee. We gratefully acknowledge support from the Simons Center for Geometry and Physics, Stony Brook University at which some of the work for this paper was performed during the X Simons Summer Workshop.

% fold sec (acknow)

\appendix 

%%%%%%%%%%%%%%%%%%%%%%%%%%%%%%%%%%%%%%%
\section{\texorpdfstring{Change of variables: From $y$ to $z$}{Change of variables: From y to z}} % sec (var)
\label{app:var}
%%%%%%%%%%%%%%%%%%%%%%%%%%%%%%%%%%%%%%%

The AdS$_7$ solutions that we consider in this paper are direct generalizations of those first constructed numerically in \cite{afrt}, and then given an analytic description in \cite{afpt}. (The analytic form was obtained by leveraging the existence of a one-to-one correspondence between AdS$_5$ and AdS$_7$ vacua of massive IIA, as summarized in \cite{ads7prl}.) In the latter paper, the analytic vacua depend on a single variable $y$ parameterizing the base interval $I$ of $M_3 \cong S^3$.

All physical fields (metric, dilaton $\phi$, warping $A$, fluxes) can then be written in terms of a single function $\beta(y)$ (a prime denotes differentiation w.r.t. $y$):
\begin{subequations}
\begin{equation}
e^{2A(y)}= \frac49 \left(-\frac {\beta'}y\right)^{1/2}\ ,\quad e^{\phi(y)}=\frac{(-\beta'/y)^{5/4}}{12\sqrt{4 \beta - y \beta'}}\ , \quad q(y)= -4y \frac{\sqrt{\beta}}{\beta'} \ ,
\end{equation}
\begin{equation}
 ds_{10}^2 = e^{2A} \left( ds^2_{\text{AdS}_7} -\frac{1}{16} \frac{\beta'}{y \beta} dy^2 + \frac{\beta/4}{4\beta -y\beta'} ds_{S^2}^2\right)\ ,
\end{equation}
\end{subequations}
and more complicated expressions hold for RR and NSNS fluxes (see e.g. \cite[Eq. (2.9)]{cremonesi-tomasiello} and \cite[Eq. (5.11)]{afpt}). The function $\beta(y)$ satisfies a nonlinear ODE: Any solution to the latter (with appropriate boundary conditions) produces an AdS$_7$ vacuum. Moreover said ODE can be conveniently recast in a much simpler form if one introduces an auxiliary function $q(y)$, that turns out to govern the position of D8-brane sources. It simply reads
\begin{equation}\label{eq:PDE}
(q(y)^2)' = \frac{2}{9} F_0\ ,
\end{equation}
where $F_0$ is the Romans mass of the solution.

In this paper we have characterized the vacua in terms of quiver gauge theory data, e.g. in the pure $\SU$ case the ranks $r_i$ and their differences $s_i := r_i -r_{i-1}$ (i.e. the depths of the $\rho_\text{L,R}$ Young tableaux's columns). As explained \cite[Sec. 2.2.2]{cremonesi-tomasiello} and reviewed in section \ref{subsub:pureSUk}, one can further relate these data to the supergravity ones, such as the value of the Romans mass $F_0 = \frac{n_0}{2\pi}$ (with $n_0 \in \mathbb{Z}$) in a certain region of the internal space $M_3$, the number of D8-branes (sourcing  $F_0$) in each stack, and their D6 charge. The relation is quite simple, and reads
\begin{equation}\label{eq:relquiver-sugra}
s_{i+1} = n_{0,i+1}\ , \quad q_i = \frac{1}{2}r_i\ ,
\end{equation}
with $n_{0,i+1}$ the value of the Romans mass between the $i$-th and $(i+1)$-th D8 stack, and $q_i$ the value of $q(y)$ at the location $y_i$ of the $i$-th D8 stack. The most general solution to \eqref{eq:PDE} then reads
\begin{equation}
q^2(y) = \frac{1}{9\pi} s_{i+1}(y-y_i)+\frac{1}{4}r_i^2\quad \text{with} \quad y \in \left[y_i,y_{i+1}\right]\ , \label{eq:q2}
\end{equation}
and $y_i$ further satisfying $y_{i+1}-y_i = \frac{9\pi}{4} (r_{i+1}+r_i)$. In \cite{cremonesi-tomasiello} a clever change of variables was found that simplifies quite a bit the solutions to \eqref{eq:PDE}. It reads:
\begin{equation}\label{eq:z}
\boxed{dz :=\frac{1}{9\pi} \frac{dy}{q(y)}}\ \Leftrightarrow \  q(z) = \frac{1}{9\pi}  \dot{y}(z)\ , \quad y(z) = -\frac{1}{2(9 \pi)} \dot{\sqrt{\beta(z)}}\ ,
\end{equation}
where a dot means differentiation w.r.t. $z$. (Notice that \eqref{eq:z} then implies $\dot{\sqrt{\beta(z)}} <0$.) In effect, using the above definition and calling $\alpha(z):= \sqrt{\beta(z)}$, we have 
\begin{equation}
q(z) = \frac{1}{2}r_i + \frac{1}{2}s_{i+1}(z-i) \ ,\quad z\in \left[i,i+1\right]\ ,
\end{equation}
with 
\begin{equation}
q(z)=-\frac{1}{2(9\pi)^2}\ddot{\alpha}(z)\ .
\end{equation}
Notice that $q(z)$ has become piecewise linear, a fact that is interpreted as a (supergravity) continuum version of the discrete (quantum) field theory group data (the $r_i$'s and their differences). As $r_i \sim N \to \infty$, the piecewise function will be characterized by a smooth graph (see e.g. \cite[Fig. 2b]{cremonesi-tomasiello}). Once a solution to \eqref{eq:PDE} (which becomes \eqref{eq:third-der-alpha} in the $z$ coordinate) is found, the function $\alpha(z)$ can be obtained by double integration, which produces \eqref{eq:alphaz}.

% fold sec (var)

%%%%%%%%%%%%%%%%%%%%%%%%%%%
\section{Integration constants and boundary data} % sec (int-const)
\label{app:int-const}
%%%%%%%%%%%%%%%%%%%%%%%%%%%

In this appendix we will determine the integration constants $y_i, \alpha_i$ appearing in \eqref{eq:newalphanew} for $i=1,\ldots,N-1$, as well as the ``boundary data'' $y_0,y_N,\alpha_0,\alpha_N$.\footnote{Notice that in \cite{cremonesi-tomasiello} $y_0,y_N$ are called integration constants rather than boundary data (whereas $\alpha_0,\alpha_N$ are assumed to be identically zero).} We first determine the former by direct computation; we then attack the cases $i=0,N$ by exploiting some extra physical input.\newline

We start by evaluating \eqref{eq:alphaz} and \eqref{eq:yz} at $z=i+1$ (call $y_{i+1}:= y(i+1)$ and $\alpha_{i+1}:=\alpha(i+1)$), $i \in \left[0,N-1\right]$:
\begin{subequations}\label{eq:boundyalpha}
\begin{align}
\alpha_{i+1} - \alpha_i &= -2(9\pi) y_i -\frac{(9\pi)^2}{2} r_i -\frac{(9\pi)^2}{6} s_{i+1}\ , \label{eq:bound-alpha} \\
y_{i+1}-y_i &= \frac{9\pi}{4} (r_{i+1}+r_i)\ . \label{eq:bound-y}
\end{align}
\end{subequations}
Summing \eqref{eq:bound-y} over $i$ from $0$ to $j-1$ and solving for $y_j$ gives \eqref{eq:yjLbis} here below; summing instead $y_{N-i} - y_{N-(i+1)}$ over $i$ from $0$ to $j-1$ and solving for $y_{N-j}$ gives \eqref{eq:yjRbis}:
\begin{subequations}\label{eq:yjbis}
\begin{align}
&\boxed{\frac{2}{9\pi} y_j = \frac{2}{9\pi} y_0 + \frac{1}{2}(r_0 +r_j) +\sum_{i=1}^{j-1}r_i\ ,\quad j\in\left[1,L\right]} \ ; \label{eq:yjLbis}\\
&\boxed{\frac{2}{9\pi} y_{N-j} = \frac{2}{9\pi} y_N - \frac{1}{2}(r_N+r_{N-j}) -\sum_{i=1}^{j-1}r_{N-i}\ ,\quad j\in\left[1,R \right]} \ . \label{eq:yjRbis}
\end{align}
\end{subequations}
Evaluating \eqref{eq:yjbis} at $z=L$ and $z=N-R$ gives:
\begin{subequations}\label{eq:yLRbis}
\begin{align}
&\boxed{y_L = y_0 + \frac{9\pi}{4}(r_0 + r_L) + \frac{9\pi}{2}\sum_{i=1}^{L-1} r_i}\ , \label{eq:yLbis} \\
&\boxed{y_{N-R} = y_N - \frac{9\pi}{4}(r_N + r_{N-R}) - \frac{9\pi}{2}\sum_{i=1}^{R-1} r_{N-i}}\ . \label{eq:yRbis}
\end{align}
\end{subequations}
In the solutions of \cite{cremonesi-tomasiello} the left and right Young tableaux must have the same number of boxes. If we generalize this to the case where $r_0, r_N \neq 0$, we have:
\begin{equation}\label{eq:kbis}
k= \sum_{i=1}^L s_i = r_L - r_0\ , \quad k = -\sum_{i=1}^{R} s_{N-(i-1)}=r_{N-R}-r_N \ . %= - \sum_{i=0}^{R-1} s_{N-i} = \ .
\end{equation}
Given that $r_L = r_{N-R}$ by construction -- this number in fact defines the height of the central plateau, which must of course be constant across it -- we also seem to have the constraint
\begin{equation}\label{eq:r0rNconstraint}
r_0 = r_N\ .
\end{equation}
However the above condition need not hold in the generic setup (this is the case when the numbers of branes in the left and right stacks of semi-infinite D6's differ). Therefore we conclude that the total number of boxes in the left and right tableaux are unrelated in generic (massive) solutions. We define
\begin{subequations}\label{eq:ks}
\begin{align}
k_\text{L} &:= \sum_{i=1}^L s_i = r_L - r_0\ , \label{eq:kLbis} \\ 
k_\text{R} &:= -\sum_{i=0}^{R-1} s_{N-i} = r_{N-R}-r_N \  \label{eq:kRbis} 
\end{align}
\end{subequations}
with $k_\text{L} \neq k_\text{R}$ generically. If we now call $k:=r_L=r_{N-R}$, we can re-express $k_\text{L}$ and $k_\text{R}$ in terms of the former, which will play the role of the (constant) height of the plateau as in \cite{cremonesi-tomasiello}:
\begin{equation}\label{eq:single-k}
\boxed{k:=r_L=r_{N-R}} \quad \Rightarrow \quad \boxed{k_\text{L} = k -r_0\ , \quad k_\text{R} = k - r_N}\ .
\end{equation}
In passing we note that in the holographic setup $k$, rather than $k_\text{L,R}$, corresponds to the order of the orbifold in the eleven-dimensional supergravity solution $\text{AdS}_7 \times S^4/\mathbb{Z}_k$ (which we reduce on an $S^1 \subset S^4$ to obtain the massless ten-dimensional one \cite{afrt}). $k_\text{L,R}$ are only defined for ten-dimensional massive $\text{AdS}_7 \times M_3$ vacua (corresponding to Higgsed quivers of pure $\SU$ type, as explained in section \ref{subsub:pureSUk}).\newline

Let us now sum the $y_i$ in \eqref{eq:yjLbis} over $i$ from $0$ to $j-1$ (with $j=1,\ldots,L$), and let us do the same with the $y_{N-i}$ in \eqref{eq:yjRbis} from $0$ to $j-1$ (with $j=1,\ldots,R$):
\begin{subequations}
\begin{equation}
\sum_{i=0}^{j-1} y_i = y_0 + \sum_{i=1}^{j-1} y_i \overset{\eqref{eq:yjLbis}}{=} jy_0 + \frac{9\pi}{4}\left[ (j-1)\, r_0 +\sum_{i=1}^{j-1} r_i +2\sum_{i=1}^{j-1} \sum_{k=1}^{i-1} r_k\right] \ , \label{eq:sumparLbis}
\end{equation}
\begin{equation}
\sum_{i=0}^{j-1} y_{N-i} = y_N + \sum_{i=1}^{j-1} y_{N-i} \overset{\eqref{eq:yjRbis}}{=} jy_N - \frac{9\pi}{4}\left[(j-1) r_N + \sum_{i=1}^{j-1} r_{N-i} +2\sum_{i=1}^{j-1} \sum_{k=1}^{i-1} r_{N-k}\right] \ . \label{eq:sumparRbis} 
\end{equation}
\end{subequations}
These expressions will be needed to determine the $\alpha_i$, to which we now turn. Summing \eqref{eq:bound-alpha} over $i$ form $0$ to $j-1$ (and trading $\sum_{i=0}^{j-1} y_i$ for \eqref{eq:sumparLbis}) and solving for $\alpha_j$ yields:
\begin{equation}\label{eq:alphajLbis}
\boxed{\alpha_j = \alpha_0 - (9\pi) (2 j) y_0 -\frac{(9\pi)^2}{6} ((3j-1)r_0+r_j)- (9\pi)^2\sum_{i=1}^{j-1}(j-i) r_i \ , \quad j \in \left[1,L \right]}\ .
\end{equation}
More explicitly, \eqref{eq:alphajLbis} has been determined as follows:
\begin{enumerate}
\item evaluate \eqref{eq:alphaz} at $z=i+1$ to obtain \eqref{eq:bound-alpha} (remember that in \eqref{eq:alphaz} $z \in \left[i, i+1\right]$, so we are evaluating the expression at the interval upper endpoint);
\item sum \eqref{eq:bound-alpha} from $0$ to $j-1$ over $i$, with $j=1\ldots,L$;
\item the left-hand side of this sum is $\alpha_j - \alpha_0$, whereas the right-hand side entails summing the quantities $y_i,r_i,s_{i+1}$ over $i$ with appropriate coefficients;
\item we trade the sum of the $y_i$ over $i$ for the right-hand side of \eqref{eq:sumparLbis};
\item summing all contributions, the ensuing expression is the part of the right-hand side of \eqref{eq:alphajLbis} that does not depend on $\alpha_0$.
\end{enumerate}
To determine $\alpha_{N-j}$ with $j=1,\ldots,R$ we can proceed in two different ways: We can either repeat the procedure outlined above starting from the right endpoint at $z=N$ and summing towards the interior until we hit $z=N-R=:R'$, or we sum from $z=R'$ to $z=N-1$. In fact
\begin{equation}\label{eq:sums-alpha}
- \sum_{i=0}^{R-1} (\alpha_{N-(i+1)} - \alpha_{N-i}) = -\alpha_{N-R} + \alpha_N = -\alpha_{R'} + \alpha_N = \sum_{i=R'}^{N-1} \alpha_{i+1}-\alpha_i\ .
\end{equation}
(Notice the crucial sign in front of the summand in the first sum.) The left (right) sum in \eqref{eq:sums-alpha} corresponds to the first (second) way we just explained. We will show how the first works. Given \eqref{eq:bound-alpha}, we have to compute the sum
\begin{align}\label{eq:sumRsum}
&-(9\pi)^2 \sum_{i=0}^{j-1} \left(\frac{2}{9\pi} y_{N-(i+1)} + \frac{1}{2}r_{N-(i+1)} + \frac{1}{6}s_{N-i}\right) = \nonumber
\end{align}
\begin{align}
&= -(9\pi)^2 \left[\frac{2}{9\pi}\sum_{i=1}^{j}  y_{N-i} + \sum_{i=0}^{j-1}  \frac{1}{2}r_{N-(i+1)} + \frac{1}{6}s_{N-i}\right] \nonumber \\
&= -(9\pi)^2 \left[\frac{2}{9\pi}\left(\sum_{i=0}^{j-1}  y_{N-i}-y_N+y_{N-j} \right) + \sum_{i=0}^{j-1}  \frac{1}{2}r_{N-(i+1)} + \frac{1}{6}s_{N-i}\right] \ , 
\end{align}
for $j=1,\ldots,R$. We must now trade the sum $\sum_{i=0}^{j-1}  y_{N-i}$ for \eqref{eq:sumparRbis} and $y_{N-j}$ for \eqref{eq:yjRbis}. Doing so yields:
\begin{equation}\label{eq:alphaRjbis-true}
\boxed{\alpha_{N-j} = \alpha_N +(9\pi) (2 j) y_N - \frac{(9\pi)^2}{6} ((3j-1)r_N+r_{N-j})-(9\pi)^2\sum_{i=1}^{j-1} (j-i) r_{N-i}}\ .
\end{equation}
Using the above results for the integration constants, we will now show how to determine the boundary data $y_{0,N}$ and $\alpha_{0,N}$.

\subsection{\texorpdfstring{Recovering \cite[App. A]{cremonesi-tomasiello}: Only regular poles}{Recovering [1]: Only regular poles}}
\label{appsub:int-const-alecrem}

To recover the boundary data of \cite[App. A]{cremonesi-tomasiello} we simply set $r_0=r_N=\alpha_0=\alpha_N=0$, given that only the regular pole case is treated in that paper. Plugging all this into \eqref{eq:yjbis}, \eqref{eq:alphajLbis}, and \eqref{eq:alphaRjbis-true}, and evaluating at $z=L$, $z=N-R$, we find:
\begin{subequations}\label{eq:expr-alecrem}
\begin{align}
&k  = k_\text{L} = r_L = r_{N-R} = k_\text{R} \ ; \label{eq:k-alecrem} \\
&y_L = y_0 + \frac{9\pi}{4} r_L + \frac{9\pi}{2}\sum_{i=1}^{L-1} r_i\ , \label{eq:yL-alecrem} \\
&y_{N-R} = y_N - \frac{9\pi}{4}r_{N-R} - \frac{9\pi}{2}\sum_{i=1}^{R-1} r_{N-i}\ ; \label{eq:yR-alecrem} \\ 
%y_{R'} &= y_N - \frac{9\pi}{4}r_{R'} - \frac{9\pi}{2}\sum_{i=R'+1}^{N-1} r_i\quad (R'+R=N)\ ; \label{eq:yLright-alecrem} \\ 
&\alpha_L =  - (9\pi) (2 L) y_0 -(9\pi)^2\frac{1}{6}r_L - (9\pi)^2\sum_{i=1}^{L-1}(L-i) r_i \ , \label{eq:alphaL-alecrem}\\
&\alpha_{N-R} = (9\pi) (2 R) y_N -(9\pi)^2\frac{1}{6}r_{N-R}-(9\pi)^2\sum_{i=1}^{R-1} (R-i) r_{N-i}\ . \label{eq:alphaR-alecrem}%\\
%\alpha_{R'} &= (9\pi) (2(N- R')) y_N - (9\pi)^2\frac{1}{6}r_{R'}-(9\pi)^2\sum_{i=R'+1}^{N-1} (i-R') r_i\ . \label{eq:alphaLright-alecrem}
\end{align}
\end{subequations}

We need now to determine the boundary data $\{y_0,y_N,\alpha_0,\alpha_N\}$, in terms of the $N,L,R,r_i$, and in the following sections we will assume that $L,R\neq N$. When this happens, we need to take care of these cases separately since the constraint that fix the boundary data will be slightly modified. We will call these limiting case, i.e. no massless region and just an increasing or decreasing ramp of $r_i$.

\subsubsection{Symmetric case}
\label{appsubsub:regsym}

In the symmetric case we have $\rho_\text{L} = - \rho_\text{R}$ and $L=R$, and the equations we need to solve in order to determine $y_0$ and $y_N$ are \cite{cremonesi-tomasiello}:\footnote{\label{foot:typo1}Notice that there is a typo in \cite[Eq. (2.20)]{cremonesi-tomasiello}. However \cite[Eqs. (A.2), (A.5)]{cremonesi-tomasiello} are correct.}
\begin{subequations}\label{eq:sym-alecrem}
\begin{align}
& y_{N-R} - y_L = \frac{9\pi}{2}k(N-R-L) \ , \\
& y_{N-R} + y_L =0 \ .
\end{align}
\end{subequations}
Plugging \eqref{eq:expr-alecrem} into \eqref{eq:sym-alecrem} and solving for $y_0$ and $y_N$ we obtain:
\begin{equation}\label{eq:y0Npre-sym-alecrem}
y_0 = \frac{9\pi}{4} \left[ (2L-N-1)k-2 \sum_{i=1}^{L-1} r_i \right]\ , \quad y_N =- \frac{9\pi}{4} \left[ (2L-N-1) k-2 \sum_{i=1}^{L-1} r_{N-i} \right]\ .
\end{equation}
We shall now make use of the fact $\rho_\text{L} = - \rho_\text{R}$: This simply means that the collections of ranks $r_i$ that define the two tableaux are identical term-by-term, that is $r_i = r_{N-i}$, for $i=1,\ldots,L$.\footnote{The minus in $\rho_\text{L} = - \rho_\text{R}$ accounts for the fact that in the right Young tableau the columns have ``negative depth'', given that $r_{i+1} < r_i$ (for $i=N-R,\ldots,N-1$) implies $s_{i+1} < 0$ for every $i$. However the ranks themselves are obviously positive, hence the meaningful identifications $r_{N-i}=r_i$ for $i=1,\ldots,L=R$.} Therefore:
\begin{equation}\label{eq:y0N-sym-alecrem}
\boxed{y_0 = -y_N = \frac{9\pi}{4} \left[ (2L-N-1) k-2 \sum_{i=1}^{L-1} r_i \right]}\ ,
\end{equation}
which is precisely \cite[Eq. (A.2)]{cremonesi-tomasiello}.

\subsubsection{No-massless symmetric case}
\label{appsubsub:regnosym}

A subcase of the above is when there is no massless region: $N-L-R=0$ i.e. $L=R=\frac{N}{2}$, and $y_L = y_{N-R}$ (given that $r_L=r_{N-R}=k$). Therefore:
\begin{equation}\label{eq:sym-nomass-alecrem}
y_{N-R} - y_L = 0\ , \quad y_{N-R} + y_L = 0
\end{equation}
obviously yielding $y_L = y_{N-R}=0$. Solving the latter equations for $y_0,y_N$ gives
\begin{equation}\label{eq:y0N-sym-nomass-alecrem}
\boxed{ y_0 = -y_N = -\frac{9\pi}{4} \left[ k + 2 \sum_{i=1}^{L-1} r_i \right]}\ ,
\end{equation}
which is the $N=2L$ limit of \eqref{eq:y0N-sym-alecrem}.

\subsubsection{Asymmetric case} 
\label{appsubsub:regasym}

We have $\rho_\text{L} \neq \rho_\text{R}$ and $L \neq R$, and the relevant equations we need to solve are \cite{cremonesi-tomasiello}:
\begin{equation}\label{eq:asym-alecrem}
\begin{split}
y_{N-R} - y_L &= \frac{9\pi}{2} k(N-R-L) \ ,\\
\alpha_{N-R} - \alpha_L &=\frac{2}{k}(y_L^2-y_{N-R}^2)\ ;
\end{split}\quad  \Leftrightarrow \quad
\begin{split}
y_{N-R} - y_L &= \frac{9\pi}{2}  k(N-R-L) \ ,\\
\alpha_{N-R} - \alpha_L &=-9\pi (N-R-L)(y_L+y_{N-R})\ .
\end{split}
\end{equation}
Plugging \eqref{eq:expr-alecrem} into the latter and solving for $y_0,y_N$ we obtain the expressions \cite[Eq. (A.5)]{cremonesi-tomasiello}:\footnote{\label{foot:typo7}Notice that there is a typo in the upper extremum of the sum $\sum_{i=1}^{R-1} r_{N-i}$, in the second line of \cite[Eq. (A.5)]{cremonesi-tomasiello}: $L-1$ should read $R-1$.}
\begin{subequations}\label{eq:y0N-asym-alecrem}
\begin{align}
&\boxed{\frac{4}{9\pi} y_0  = \ \frac{k}{N} (L-N-R) (N+1-L-R) -2 \sum_{i=1}^{L-1} r_i +\frac{2}{N}\left( \sum_{i=1}^{L-1} i r_i - \sum_{i=1}^{R-1} i r_{N-i} \right)} \ , \\
&\boxed{\frac{4}{9\pi} y_N  = \ \frac{k}{N} (L+N-R) (N+1-L-R) +2 \sum_{i=1}^{R-1} r_{N-i} +\frac{2}{N}\left( \sum_{i=1}^{L-1} i r_i - \sum_{i=1}^{R-1} i r_{N-i} \right)} \ .
\end{align}
\end{subequations}

\subsection{\texorpdfstring{Generic poles: None among $r_0,r_N,\alpha_0,\alpha_N$ is zero}{Generic poles: None among r0, rN, alpha0, alphaN is zero}}
\label{appsub:int-const-generic}

We will now determine the boundary data both in the symmetric and asymmetric case for the generic setup, i.e. when $r_0,r_N,\alpha_0,\alpha_N$ are not necessarily zero. The specific AdS$_7$ solution might require setting some (or none) of them to zero. We have:
\begin{subequations}\label{eq:expr-generic}
\begin{align}
&k_\text{L} = k-r_0 = r_L -r_0 \ , \quad k_\text{R} = k-r_N = r_{N-R} - r_N \ ; \label{eq:k-generic} \\
&y_L = y_0 + \frac{9\pi}{4}(r_0 + k) + \frac{9\pi}{2}\sum_{i=1}^{L-1} r_i\ ,  \label{eq:yL-generic}\\
&y_{N-R}= y_N - \frac{9\pi}{4}(r_N + k) - \frac{9\pi}{2}\sum_{i=1}^{R-1} r_{N-i}\ ; \label{eq:yR-generic}\\
%&y_{R'} = y_N - \frac{9\pi}{4}(r_N + k) -\frac{9\pi}{2} \sum_{i=R'+1}^{N-1} r_i \quad (R'+R=N)\ ;
\end{align}
\begin{align}
&\alpha_L = \alpha_0 - (9\pi) (2 L) y_0 -(9\pi)^2 \left(\frac{1}{6}(3L-1)r_0+\frac{1}{6}k \right) - (9\pi)^2\sum_{i=1}^{L-1} (L-i) r_i \ , \label{eq:alphaL-generic}\\
&\alpha_{N-R} = \alpha_N +(9\pi) (2 R) y_N - (9\pi)^2 \left(\frac{1}{6}(3R-1) r_N +\frac{1}{6}k\right)-(9\pi)^2\sum_{i=1}^{R-1} (R-i) r_{N-i}\ . \label{eq:alphaR-generic}%\\
%&\alpha_{R'} = \alpha_N + (9\pi) (2(N- R')) y_N - (9\pi)^2 \left(\frac{1}{6}(3(N-R')-1)r_N +\frac{1}{6}k\right) -(9\pi)^2\sum_{i=R'+1}^{N-1} (i-R') r_i\ . \label{eq:alphaLright-generic}
\end{align}
\end{subequations}

\subsubsection{Symmetric case} 
\label{appsubsub:gensym}

We have $\rho_\text{L} = - \rho_\text{R}$ (i.e. $r_i = r_{N-i}$ for $i=1,\ldots,L=R$) and $L=R$. Upon solving \eqref{eq:sym-alecrem} for $y_0,y_N$ we find:
\begin{subequations}\label{eq:y0N-sym-generic}
\begin{align}
& \boxed{y_0 = \frac{9\pi}{4} \left[ \left(2L-N-1 \right)k-r_0 - 2 \sum_{i=1}^{L-1} r_i \right]}\ , \label{eq:y0-sym-gen}\\
& \boxed{y_N = - \frac{9\pi}{4} \left[ \left(2L-N-1 \right) k-r_N - 2 \sum_{i=1}^{L-1} r_i \right]}\ . \label{eq:yN-sym-gen}
\end{align}
\end{subequations}
Clearly, in the subcase $r_0 = r_N$ (which implies $k_\text{L} = k_\text{R} = k - r_0$) we have $y_0 = - y_N$, mimicking \eqref{eq:y0N-sym-alecrem}. Notice that $y_0,y_N$ do not depend on $\alpha_0,\alpha_N$. This situation corresponds to having $r_0$ D6-branes at one pole of the solution, and $r_N$  at the other, if $\alpha_0=\alpha_N=0$, and to a D6-O6 stack if $\alpha_0,\alpha_N\neq 0$ -- see table \ref{tab:configurations}.

\subsubsection{No-massless symmetric case}
\label{appsubsub:gennosym}

In the no-massless-region subcase (i.e. $N-L-R=0$, $N=2L$) we have:
\begin{equation}\label{eq:y0N-sym-nomass-gen}
\boxed{y_0 = -\frac{9\pi}{4} \left[ k+r_0 + 2 \sum_{i=1}^{L-1} r_i \right]}\ , \quad 
\boxed{y_N = \frac{9\pi}{4} \left[ k+r_N + 2 \sum_{i=1}^{L-1} r_i \right]}\ ,
\end{equation}
%\begin{subequations}\label{eq:y0N-sym-nomass-generic}
%\begin{align}
%y_0 &= -\frac{9\pi}{4} \left[ k+r_0 + 2 \sum_{i=1}^{L-1} r_i \right]\ , \label{eq:y0-sym-nomass-gen}\\
%y_N &= \frac{9\pi}{4} \left[ k+r_N + 2 \sum_{i=1}^{L-1} r_i \right]\ , \label{eq:yN-sym-nomass-gen}
%\end{align}
%\end{subequations}
which is just the $N=2L$ limit of \eqref{eq:y0N-sym-generic}.

\subsubsection{Asymmetric case}
\label{appsubsub:genasym}

We have $\rho_\text{L} \neq \rho_\text{R}$ and $L \neq R$. Solving \eqref{eq:asym-alecrem} for $y_0,y_N$ yields:
\begin{subequations}\label{eq:y0N-asym-gen}
\begin{align}
\frac{4}{9\pi} y_0 =\ &  \frac{k}{N}(L-N-R)(N+1-L-R) -2 \sum_{i=1}^{L-1} r_i + \frac{2}{N} \left( \sum_{i=1}^{L-1} i r_i - \sum_{i=1}^{R-1} i r_{N-i} \right) + \nonumber \\ & + 
\frac{4}{9\pi} \frac{\alpha_0 - \alpha_N}{2N(9\pi)} - \frac{1}{3N}\left((3N-1)r_0 +r_N \right) \ , \label{eq:y0-asym-gen} \\
\frac{4}{9\pi} y_N = \ &  \frac{k}{N}(L+N-R)(N+1-L-R) +2 \sum_{i=1}^{R-1} r_{N-i} + \frac{2}{N} \left( \sum_{i=1}^{L-1} i r_i - \sum_{i=1}^{R-1} i r_{N-i} \right) + \nonumber \\ & + 
\frac{4}{9\pi} \frac{\alpha_0 - \alpha_N}{2N(9\pi)} + \frac{1}{3N}\left((3N-1)r_N +r_0 \right) \ . \label{eq:yN-asym-gen}
\end{align}
\end{subequations}
These expressions correctly reduce to \eqref{eq:y0N-asym-alecrem} once we plug in $r_{0}=r_{N}=\alpha_{0}=\alpha_{N}=0$.

\subsection{\texorpdfstring{Using $F_2$ to determine $\alpha_0,\alpha_N$: Physical interpretation}{Using F2 to determine alpha0, alphaN: Physical interpretation}}
\label{sub:F2alpha0N}

We would like to express the boundary data $\alpha_0,\alpha_N$ in terms of the physical ranks which define both the brane configuration and the supergravity solution.

Consider the expression \eqref{eq:fluxesBF} for $F_2$:
\begin{equation}\label{eq:F2z}
F_2(z) = \left( \frac{\ddot{\alpha}(z)}{162\pi^2}+\pi F_0 \, \frac{\alpha(z) \dot{\alpha}(z)}{\sigma(z)}\right) \text{vol}_{S^2}\ .
\end{equation}
Evaluating \eqref{eq:alphaz} in the first interval $z \in \left[0, 1\right]$ gives 
\begin{equation}\label{eq:alphaz01}
\alpha(z) = \alpha_0 -(9\pi)(2 y_0) \,z -\frac{(9\pi)^2}{2}r_0 \,z^2 -\frac{(9\pi)^2}{6} s_{1} \, z^3\ .
\end{equation}
We also know that the Romans mass is given in that interval by $F_0 = 2\pi n_0 = 2\pi s_1 = 2\pi(r_1 - r_0)$, so we can replace the coefficient of the cubic term in \eqref{eq:alphaz01} by $F_0$. Plugging \eqref{eq:alphaz01} into \eqref{eq:F2z} then gives us the expression for $F_2$ in the first interval $z \in \left[0,1\right]$:
\begin{footnotesize}
\begin{align}
F_2(z) =&\ \frac{1}{2} \! \left( \frac{4 \pi  (r_0-r_1) (9 \pi  z (r_0 (z-2)-r_1 z)-4 y_0) (2 \alpha_0+9 \pi  z (3 \pi  z (r_0 (z-3)-r_1 z)-4 y_0))}{9 \left(72 \pi  (r_1-r_0) y_0 z^2-8 \left(\alpha_0 (r_0 (-z)+r_1 z+r_0)+2 y_0^2\right)+27 \pi ^2 (r_0-r_1) z^3 (r_0 (z-4)-r_1 z)\right)} \right.  \nonumber\\
& \bigg. \, +(r_0-r_1) z-r_0 \bigg) \text{vol}_{S^2} \ .
\end{align}
\end{footnotesize}
Taylor expanding in $z$ around $z=0$ gives:
\begin{equation}\label{eq:F20}
F_2(z) \sim \left[\frac{1}{2} \left(\frac{4 \pi  \alpha_0 (r_0-r_1) y_0}{9 \left(\alpha_0 r_0+2 y_0^2\right)}-r_0\right) + \mathcal{O}(z)\right] \text{vol}_{S^2}\ .
\end{equation}
For an O6 pole we have $r_0 = 0$ whereas $\alpha_0 \neq 0$ (see the third row of table \ref{tab:configurations}), thus we might hope to determine the latter parameter by using the flux $F_2$. Plugging $r_0 = 0$ into \eqref{eq:F20} yields
\begin{equation}\label{eq:valueF2pole0}
-\frac{\pi  \alpha_0 r_1}{9 y_0}\ ,
\end{equation} 
whereas plugging in $\alpha_0= 0$ (which is appropriate for a regular or D6 pole) would give $-\frac{1}{2}r_0$. We interpret this fact as saying that the $F_2$ flux close to the pole gives the D6-brane charge; along the same lines, we interpret the value \eqref{eq:valueF2pole0} as the ``effective charge'' $-\frac{1}{2}\tilde{r}_0$ in presence of an O6$^\mp$-plane (which has itself $\mp 2^{6-4}=\mp 4$ D6 charge, and is overlaid onto image pairs of D6-branes). This suggests the following definition:
\begin{equation}\label{eq:effectiver0}
\boxed{\alpha_0 := \frac{9}{2\pi} \frac{\tilde{r}_0 y_0}{r_1}}\ .
\end{equation}
$\tilde{r}_0$ will be determined case by case, i.e. it is specified by the brane configuration with a leftmost D6-O6$^-$ stack of negative charge. Similarly, we can  define
\begin{equation}\label{eq:effectiverN}
\boxed{\alpha_N := -\frac{9}{2\pi} \frac{\tilde{r}_N y_N}{r_{N-1}}}\ ,
\end{equation}
which is obtained from the expression of $F_2$ valid in the subinterval $z \in \left[N-1, N\right]$. 

Equations \eqref{eq:effectiver0} and \eqref{eq:effectiverN} are only valid for solutions with $\alpha_0,\alpha_N \neq 0$ due to O6 poles. Moreover, since the $\alpha_0,\alpha_N$ we just defined depend on the $y_0,y_N$, the expressions \eqref{eq:y0N-asym-gen} are not valid anymore (as they assumed that the former be independent of the latter), and we should solve \eqref{eq:asym-alecrem} keeping this fact in mind. Doing so yields:
\begin{subequations}\label{eq:y0N-asym-generic-eff}
%\begin{empheq}[box=\mathbox]{align}
\begin{align}
 y_0^\text{eff} = & \left(4\tilde{r}_N r_1 + 4(\tilde{r}_0 -4N\pi^2 r_1)r_{N-1} \right)^{-1} \Biggl( -9\pi r_1\tilde{r}_N \Biggl[r_0+r_N  +  2k(N+1-L-R) \ + \nonumber \\ & + 2  \left( \sum_{i=1}^{L-1} r_i + \sum_{i=1}^{R-1}  r_{N-i} \right) \Biggr] + 12\pi^3 r_1 r_{N-1} \Biggl[ -3k(L-N-R)(N+1-L-R) \ + \nonumber \\
&+(3N-1)r_0 + r_N  + 6\sum_{i=1}^{L-1} (N-i) r_i  +6\sum_{i=1}^{R-1} i r_{N-i} \Biggr) \Biggr]  \Biggr)\ , \label{eq:y0-asym-generic-eff}
\end{align}
\begin{align}
y_N^\text{eff} = & \left(4\tilde{r}_N r_1 + 4(\tilde{r}_0 -4N\pi^2 r_1)r_{N-1} \right)^{-1} \Biggl( 9\pi r_{N-1}\tilde{r}_0 \Biggl[r_0+r_N  +  2k(N+1-L-R) \ + \nonumber \\ & + 2  \left( \sum_{i=1}^{L-1} r_i + \sum_{i=1}^{R-1}  r_{N-i} \right) \Biggr] - 12\pi^3 r_1 r_{N-1} \Biggl[ 3k(L+N-R)(N+1-L-R) \ + \nonumber \\
&+(3N-1)r_N + r_0  + 6\sum_{i=1}^{L-1} i r_i  +6\sum_{i=1}^{R-1} (N-i) r_{N-i} \Biggr) \Biggr]  \Biggr)\ . \label{eq:yN-asym-generic-eff}
\end{align}
%\end{empheq}
\end{subequations}
Whenever $\alpha_0,\alpha_N \neq 0$ the expressions \eqref{eq:y0N-asym-gen} should not be used, and the above should be used instead. Therefore \eqref{eq:y0N-asym-gen} makes sense only when $\alpha_0=\alpha_N=0$.

\subsection{Limiting cases}
\label{appsub:limcas}
Suppose e.g. that $R=0$; then we have the following constraints from \eqref{eq:asym-alecrem}:
\begin{equation} \label{eq:Lrampconst}
y_L+\frac{9\pi}{2}(N-L)=y_N \ , \quad \alpha_L-9\pi(N-L)(y_L+y_N)=\alpha_N \ .
\end{equation}
Plugging \eqref{eq:yL-generic} and \eqref{eq:alphaL-generic}, with $k=r_L=r_N$, into \eqref{eq:Lrampconst}, we get:
\begin{subequations}\label{eq:y0N-asym-Lramp-eff}
%\begin{empheq}[box=\mathbox]{align}
\begin{align}
 y_0^\text{eff} = & \left(4\tilde{r}_N r_1 + 4(\tilde{r}_0 -4N\pi^2 r_1)r_{N-1} \right)^{-1} \Biggl( -9\pi r_1\tilde{r}_N \Biggl[r_0+k +2k(N-L)  + 2  \left( \sum_{i=1}^{L-1} r_i  \right) \Biggr] +  \nonumber \\
&12\pi^3 r_1 r_{N-1}\Biggl[(3L-1)r_0 + k + 3k(N-L)(N-L+1)+ 6\sum_{i=1}^{L-1} (N-i) r_i   \Biggr]  \Biggr)\ , \label{eq:y0-asym-Lramp-eff} \\
y_N^\text{eff} = & \left(4\tilde{r}_N r_1 + 4(\tilde{r}_0 -4N\pi^2 r_1)r_{N-1} \right)^{-1} \Biggl( 9\pi r_{N-1}\tilde{r}_0 \Biggl[r_0+k +2k(N-L) + 2  \left( \sum_{i=1}^{L-1} r_i \right) \Biggr] - \nonumber \\
&12\pi^3 r_1 r_{N-1} \Biggl[ (3L-1)k + r_0 +3k(N-L+1)(N+L)  + 6\sum_{i=1}^{L-1} i r_i   \Biggr]  \Biggr)\ . \label{eq:yN-asym-Lramp-eff}
\end{align}
\end{subequations}
where we also used the definitions \eqref{eq:effectivealphas} of $\alpha_{0,N}$ in terms of $\tilde r_0, \tilde r_N$. Similarly when $L=0$, we have from \eqref{eq:asym-alecrem}:
\begin{equation} \label{eq:Rrampconst}
y_{N-R}=y_0+\frac{9\pi}{2}(N-R) \ , \quad \alpha_{N-R}=\alpha_0-9\pi (N-R)(y_{N-R}+y_0)\ .
\end{equation}
Plugging \eqref{eq:yR-generic} and \eqref{eq:alphaR-generic}, with $k=r_{N-R}=r_0$, into \eqref{eq:Rrampconst}, we get:
\begin{subequations}\label{eq:y0N-asym-Rramp-eff}
%\begin{empheq}[box=\mathbox]{align}
\begin{align}
 y_0^\text{eff} = & \left(4\tilde{r}_N r_1 + 4(\tilde{r}_0 -4N\pi^2 r_1)r_{N-1} \right)^{-1} \Biggl( -9\pi r_1\tilde{r}_N \Biggl[k+r_N  +2k(N-R)+ 2  \left( \sum_{i=1}^{R-1} r_i  \right) \Biggr] +  \nonumber \\
&12\pi^3 r_1 r_{N-1}\Biggl[(3N-1)k + r_N + 3k(N+R)(N-R+1) + 6\sum_{i=1}^{R-1} i r_{N-i}  \Biggr]  \Biggr)\ , \label{eq:y0-asym-Rramp-eff} \\
y_N^\text{eff} = & \left(4\tilde{r}_N r_1 + 4(\tilde{r}_0 -4N\pi^2 r_1)r_{N-1} \right)^{-1} \Biggl( 9\pi r_{N-1}\tilde{r}_0 \Biggl[r_N+k+2k(N-R)  + 2  \left( \sum_{i=1}^{R-1} r_i \right) \Biggr] - \nonumber \\
&12\pi^3 r_1 r_{N-1} \Biggl[ (3N-1)k+3k(N-R)(N-R+1)+ r_N  + 6\sum_{i=1}^{R-1} (N-i) r_{N-i}   \Biggr]  \Biggr)\ . \label{eq:yN-asym-Rramp-eff}
\end{align}
\end{subequations}

Notice that in the large $k,N$ limit the expressions \eqref{eq:y0N-asym-Lramp-eff} and \eqref{eq:y0N-asym-Rramp-eff} approximate nicely \eqref{eq:y0N-asym-generic-eff}, since they only differ from the latter by subleading $\mathcal{O}(N^1)$ terms.

\subsection{\texorpdfstring{Special case: O8 at $z=0$}{Special case: O8 at z=0}}
\label{sub:O8intconst}

As summarized in table \ref{tab:configurations}, an O8 pole with D6 charge at $z=0$ requires $y_0=0$, but $r_0,\alpha_0 \neq 0$. 

%Given the definition \eqref{eq:2qz}, the first condition is equivalent to requiring that $q(z)$ (which is piece-wise linear throughout $\left[0,N\right]$) have zero intercept in $\left[0,1\right]$. In addition, the second condition forces $\alpha(z)$ (the double integral of $q$) to be a perfect cubic polynomial.

To determine the boundary data $y_N, \alpha_0, \alpha_N$ we start by plugging $y_0 = 0$ into the expressions \eqref{eq:yjbis}, \eqref{eq:alphajLbis}, and \eqref{eq:alphaRjbis-true}. The latter have to satisfy the conditions \eqref{eq:asym-alecrem} for $i=L,N-R$, and we can use them to relate $y_N$ and  $\alpha_0, \alpha_N$. In fact
\begin{subequations}
\begin{align} 
y_L &= \frac{9\pi}{4}(r_0+k) + \frac{9\pi}{2}\sum_{i=1}^{L-1} r_i\ , \label{eq:yLO8gen}\\ 
y_{N-R} &= y_N - \frac{9\pi}{4}(r_N + k) - \frac{9\pi}{2}\sum_{i=1}^{R-1} r_{N-i}\ , \label{eq:yRO8gen}
\end{align}
\end{subequations}
and solving $y_{N-R}-y_L = \frac{9\pi}{2}k(N-L-R)$ for $y_N$ yields
\begin{equation}\label{eq:yNO8}
\boxed{y_N = \frac{(9\pi)}{4} \left(2k(N+1-L-R) + r_0 + r_N +  2\left( \sum_{i=1}^{L-1} r_i + \sum_{i=1}^{R-1} r_{N-i} \right)\right)}\ .
\end{equation}
On the other hand
\begin{subequations}
\begin{align}
\alpha_L &= \alpha_0  - (9\pi)^2 \left(\frac{1}{6}(3L-1) r_N +\frac{1}{6}k\right) - (9\pi)^2\sum_{i=1}^{L-1} (L-i) r_i \ ,\label{eq:alphaLO8gen} \\
\alpha_{N-R} &= \alpha_N +(9\pi) (2 R) y_N - (9\pi)^2 \left(\frac{1}{6}(3R-1) r_N +\frac{1}{6}k\right)-(9\pi)^2\sum_{i=1}^{R-1} (R-i) r_{N-i}\ , \label{eq:alphaRO8gen}
\end{align}
\end{subequations}
and solving $\alpha_{N-R}-\alpha_L = -(9\pi)(N-L-R)(y_{N-R}+y_L)$ for $\alpha_0$ yields
\begin{align}\label{eq:alpha0O8}
\alpha_0 = &\ \alpha_N + \frac{(9\pi)^2}{6} \Bigg( 3k(N+1-L-R)(N+L-R) + (3N-1)r_0+ r_N +\ \Bigg. \nonumber \\ 
&+ \Bigg.  6\left( \sum_{i=1}^{L-1} (N-i) r_i + \sum_{i=1}^{R-1}i r_{N-i} \right) \Bigg)\ .
\end{align}%%
%\begin{align}\label{eq:alpha0O8}
%\alpha_0 = &\ \alpha_N + \frac{(9\pi)^2}{6} \Bigg( 3k(N+1-L-R)(N+L-R) + (3N-1)r_0+ r_N +\ \Bigg. \nonumber \\ 
%&+ \Bigg.  6\left( \sum_{i=1}^{L-1} (N-i) r_i - \sum_{i=1}^{R-1}i r_{N-i} \right) \Bigg)\ .
%\end{align}%
A remark is in order here. In section \ref{sub:F2alpha0N} we used extra physical input to determine $\alpha_0,\alpha_N$ in terms of the defining data of a generic solution (i.e. the ranks $r_i$ and the effective charges $\tilde{r}_{0,N}$). Here we are simply relating the two via the conditions \eqref{eq:asym-alecrem}, in case $\alpha_N\neq0$. The point is that the latter are two linear equations in $y_0,y_N$, but given that $y_0 = 0$ we only need one to determine $y_N$. The other can instead be used to relate $\alpha_0$ to $\alpha_N$, which is \eqref{eq:alpha0O8}. In case of a regular or D6 pole at $z=N$, $\alpha_N=0$; in case of an O6 pole $\alpha_N$ can be defined via \eqref{eq:effectiverN}. (Moreover notice that in the holographic limit $\alpha_N$ drops out of \eqref{eq:alpha0O8} as $N \to \infty$, since it is subleading w.r.t. the $\mathcal{O}(N^3)$ contributions.)

\subsection{Limiting cases with an O8}
\label{appsub:limcasO8}

Suppose now we have $R=0$, i.e. the quiver is characterized by a single increasing ramp of $r_i$, and a massless region. Once again we must impose \eqref{eq:Lrampconst}. \eqref{eq:yLO8gen} and \eqref{eq:alphaLO8gen} with $k=r_L=r_N$ yield the following boundary data:
\begin{subequations}
\begin{align}
y_N &= \frac{9\pi}{4}(2k(N-L) +r_0+k) + \frac{9\pi}{2}\sum_{i=1}^{L-1} r_i\ , \\
\alpha_0 &= \alpha_N + \frac{(9\pi)^2}{6} \Bigg( (3N-1)r_0+ k +  3k(N+1-L)(N+L)+ 6\left( \sum_{i=1}^{N-1} (N-i) r_i\right) \Bigg)\ .
\end{align}
\end{subequations}
Similarly when $L=0$ we must impose \eqref{eq:Rrampconst}. By using \eqref{eq:yRO8gen} and \eqref{eq:alphaRO8gen} with $k=r_{N-R}=r_0$ we find:
\begin{subequations}
\begin{align}
y_0 &=- \frac{9\pi}{4}(k+r_N+2k(N-R)) - \frac{9\pi}{2}\sum_{i=1}^{N-1} r_{N-i}\ , \\
\alpha_0 &=  \alpha_N + \frac{(9\pi)^2}{6} \Bigg( (3N-1)k+ r_N + 3k(N+1-L)(N+L) + 6\left( \sum_{i=1}^{N-1} i r_{N-i}\right) \Bigg)\ .
\end{align}
\end{subequations}

\begin{comment}
\subsection{Mixed cases: $\alpha_0 \neq 0$ but $\alpha_N= 0$ (one O8 pole, one D6 pole)}
\label{sub:mixed}

In the mixed case where one pole (say at $z=N$) is either regular or appropriate for D6-branes, while the other (at $z=0$) corresponds to a D8-O8 stack, we have $\alpha_0 \neq 0$, but $\alpha_N =0$ (see table \ref{tab:configurations}). $y_N$ will be nonvanishing (whereas $y_0 = 0$), and we can simply substitute $\tilde{r}_0=\tilde{r}_N=0$ in \eqref{eq:y0N-asym-generic-eff} to obtain the former. $r_0$ will be the D6 charge of the D8-O8 stack (the rank of the 0-th gauge group), while $r_N \neq 0$ that of the semi-infinite flavor D6-branes (if present, $r_N=0$ for a regular pole). $\alpha_0$ is given by \eqref{eq:alpha0O8} with $\alpha_N=0$.
\end{comment}

% fold sec (int-const)

%%%%%%%%%%%%%%%%%%%%%%%%%%%%%%%%%%%%
\section{\texorpdfstring{Gravity side: The $a$ conformal anomaly}{Gravity side: The a conformal anomaly}} % sec (agrav)
\label{app:agrav}
%%%%%%%%%%%%%%%%%%%%%%%%%%%%%%%%%%%%

The leading order of the $a$ conformal anomaly can be computed in supergravity as an integral over the internal space $M_3$ of the AdS$_7$ vacuum (see \cite[Eq. (5.67)]{afpt}, \cite[Eq. (4.6)]{cremonesi-tomasiello} or \cite[Eq. (D.9)]{bpt}). This integral is then to be compared with the holographic limit (i.e. $N,L,R,k \to \infty$ with $\frac{L}{N},\frac{R}{N},\frac{k}{N}$ finite) of the field theory result, which can be extracted from the six-dimensional anomaly polynomial.\\

The relevant integral is the following:
\begin{align}\label{eq:aholo}
a_\text{hol} &= \frac{3}{56\, \pi^4} \int_{M_3} e^{5A(z)-2\phi(z)} \,\text{vol}_3 = - \frac{192}{7} \int_0^N 2q(z) \left[\frac{1}{\partial_z^2} 2q(z) \right] dz \nonumber \\ &= \frac{128}{189\,\pi^2} \int_0^N \alpha(z) q(z)\, dz \ ,
\end{align}
where by $\partial_z^{-2}$ we mean the second primitive. The strategy to tackle this computation is as follows:
\begin{enumerate}
\item Divide the integral over $I$ parameterized by $z\in \left[0,N\right]$ into subintegrals, one for each subinterval $z \in \left[l-1,l\right]$, $l=1,\ldots,N$. In each of the latter $q(z)=\frac{1}{2}r_{l-1} +\frac{1}{2}s_{l}(z-(l-1))$ and $\alpha(z)$ is given by the expression \eqref{eq:alphaz} if the subinterval corresponds to a ``massive'' region of the supergravity solution (i.e. $F_0 \neq 0$ there). If the interval corresponds instead to a massless region ($F_0=0$) we put $s_l = 0$: $\alpha(z)$ becomes quadratic in $z$ whereas $q(z)$ is constant, $q(z)=\frac{1}{2}r_{l-1}$.
\item Letting the subintervals start from the left, i.e. $l=1,\ldots,L$, $\alpha(z)$ (supported in $\left[l-1,l\right]$) depends on all ranks $\left\lbrace r_i \right\rbrace_{i=0}^l$ (through $y_{l-1}$ -- see \eqref{eq:yjLbis} -- and $s_l$) and on the boundary data $y_0$, $\alpha_0$, $\alpha_{l-1}$. If $r_0\neq 0$, the first is given by \eqref{eq:y0-sym-gen} in the symmetric case ($L=R$), by \eqref{eq:y0N-sym-nomass-gen} in the symmetric no-massless case ($N-L-R=0$), and by \eqref{eq:y0-asym-gen} in the asymmetric case ($L\neq R$). The second can be either zero, when there is no O6-plane at the pole $z=0$, or given by the effective rank $\tilde{r}_0$ according to \eqref{eq:effectiver0}, when an O6-plane is present. The third is given by \eqref{eq:alphajLbis} for $l=2,\ldots,L$; for $l=1$ (i.e. in $\left[0,1\right]$) it is given by $\alpha_0$. Moreover, if $\tilde{r}_0 \neq 0$, $y_0$ is given by  $y_0^\text{eff}$ in \eqref{eq:y0-asym-generic-eff} in the asymmetric case (in the symmetric case the equations determining $y_{0,N}$ do not depend on $\alpha_{0,N}$ at all).
\item If we perform the integration starting from the right, the subintervals can be written as $z \in \left[N-l,N-(l-1)\right]$ with $l=1,\ldots,R$. Then $\alpha(z)$ depends on the ranks $\left\lbrace r_{N-i} \right\rbrace_{i=0}^l$ and on $y_N, \alpha_N, \alpha_{N-(l-1)}$. If $r_N \neq 0$, the first is given by \eqref{eq:yN-sym-gen} in the symmetric case ($L=R$), by \eqref{eq:y0N-sym-nomass-gen} in the symmetric no-massless case ($N-L-R=0$), and by \eqref{eq:yN-asym-gen} in the asymmetric case ($L\neq R$). The second is given in terms of the effective rank $\tilde{r}_N$ according to \eqref{eq:effectiverN}. The third is given by \eqref{eq:alphaRjbis-true} for $l=2,\ldots,R$; for $l=1$ by $\alpha_N$. If $\tilde{r}_N \neq 0$ then $y_N$ is given by $y_N^\text{eff}$ in \eqref{eq:yN-asym-generic-eff}, in the asymmetric case.
\item We perform the sums 
\begin{equation}\label{eq:intsum}
\sum_{l=1}^L \int_{l-1}^l \alpha(z) q(z)\, dz\ , \quad \sum_{l=1}^R \int_{N-l}^{N-(l-1)} \alpha(z) q(z)\, dz\ ;
\end{equation}
the integrands are given by the expressions of $\alpha(z)$ and $q(z)$ appropriate for the different subintervals, as explained at points 1 through 3.
\item Finally we have to compute the contribution from the massless region $z\in \left[L, N-R\right]$. This splits into subintervals $z\in \left[i,i+1\right]$, $i=L,\ldots,N-R-1$. In this (sum of) subinterval(s) we have $s_{i+1}=0$ (essentially because the latter is given by the value of $F_0$ in the interval), and therefore
\begin{equation}\label{eq:alphamassless}
\alpha(z)_{F_0=0}=\alpha_i-(9\pi)(2y_i)(z-i)-\frac{(9\pi)^2}{2}r_i (z-i)^2\ , \quad q(z)_{F_0=0}=\frac{1}{2}r_i\ .
\end{equation}
Given that the ranks do not change (the massless region coincides with the constant plateau), $\alpha(z)$ in \eqref{eq:alphamassless} is supported on the whole $\left[ L, N-R\right]$ interval, and it suffices to compute one ``full'' integral
\begin{equation}\label{eq:intless}
\int_L^{N-R} q\,\alpha(z)_{F_0=0} \, dz\ .
\end{equation}
The integration constants $\alpha_L, \alpha_{N-R}$ are respectively given by \eqref{eq:alphaL-generic} and \eqref{eq:alphaR-generic} if $r_0, r_N, \alpha_0,\alpha_N \neq 0$ (or by \eqref{eq:alphaL-alecrem} and \eqref{eq:alphaR-alecrem} if these numbers are all vanishing). The former also depend on $y_{0,N}$ which in the generic symmetric case are given by \eqref{eq:y0N-sym-generic}, and in the asymmetric one by \eqref{eq:y0N-asym-generic-eff}.

Actually, we will find it more convenient to perform the change of variables \eqref{eq:z} backwards, and compute \eqref{eq:intless} over the integration variable $y$, namely:
\begin{equation}\label{eq:intless-y}
\int_{y_L}^{y_{N-R}} q(y) \sqrt{\beta(y)}\, \frac{1}{9\pi q(y)} dy = \frac{1}{9\pi} \int_{y_L}^{y_{N-R}} \sqrt{\beta(y)}\, dy \ ,
\end{equation}
with $q(y)$ and $\sqrt{\beta(y)}$ in the massless region as in appendix in \ref{app:var}.

\end{enumerate}

\subsection{The contribution from the left massive tail}

The left massive region is defined by the left Young tableau $\rho_\text{L}$ in terms of its length $L$ and column depths $s_{i+1}$ (giving the differences between the ranks $\left\lbrace r_i \right\rbrace_{i=0}^L$).

Let us first compute the integral
\begin{equation}
\text{int}^\text{L}_l:=\int_{l-1}^l \alpha(z) q(z)\, dz\ , \quad l\in\left[1,L\right]\ .
\end{equation}
The result is the following:\footnote{\label{foot:typo3}Notice that there is a typo in \cite[Eq. (4.11)]{cremonesi-tomasiello}: There should be a $y_0$ in front of $\frac{4}{9\pi}$, which is instead missing. This propagates to \cite[Eq. (4.12)]{cremonesi-tomasiello} too, so that a direct comparison between the latter and our \eqref{eq:left-integral} must be done with care.}
\begin{align}\label{eq:intL-l}
-\frac{7}{16}\, \text{int}^\text{L}_l = &\sum_{k=1}^{l-2} r_k \left[ 2r_{l-1}+4r_l +6(l-k-1)(r_l+r_{l-1})\right] +\frac{1}{5}(12 r^2_{l-1} +21 r_l r_{l-1} + 2r_l^2)\ + \nonumber \\
& +\frac{4}{9\pi} y_0 \left[ 3(l-1) (r_l+r_{l-1}) +2r_l +r_{l-1}\right] +r_0 \left[ 3(l-1)(r_l+r_{l-1}) +r_l \right] \ + \nonumber \\
& -\frac{6}{(9\pi)^2}\alpha_0 (r_l+r_{l-1})\ .
\end{align}
Now we have to sum these contributions from $l=1$ to $l=L$:
\begin{equation}\label{eq:intL-l-sum}
\text{Int}_\text{L} := \int_0^L \alpha(z) q(z)\, dz = \sum_{l=1}^L \int_{l-1}^l \alpha(z) q(z)\, dz = \sum_{l=1}^L \text{int}^\text{L}_l\ .
\end{equation}
In the first identity we used the fact that there is one $\alpha(z)$ supported in each interval $\left[l-1,l\right]$ defined by \eqref{eq:alphaz}. 

The first summand in \eqref{eq:intL-l} yields the following contribution to the sum on the right-hand side of \eqref{eq:intL-l-sum} (remember that $k:=r_L = r_{N-R}$ as per \eqref{eq:single-k}):
\begin{equation}\label{eq:contrib1}
2k\sum_{l=1}^{L-2}r_l \left[3(L-l)-1\right] +12\sum_{l=1}^{L-1}\sum_{k=1}^{l-2}lr_l r_k -12\sum_{l=1}^{L-1}\sum_{k=1}^{l-2}k r_l r_k + 8 \sum_{l=1}^{L-1}r_l r_{l-1}\ .
\end{equation}
Notice that the first summand (i.e. for $l=1$) in the last sum in \eqref{eq:contrib1} is nonzero due to $r_0 \neq 0$ generically (contrarily to what happens in \cite{cremonesi-tomasiello}). The second summand gives:
\begin{equation}\label{eq:contrib2}
\frac{1}{5}\left(2k^2+12r_0^2+21k r_{L-1} + 14\sum_{l=1}^{L-1}r_l^2+21\sum_{l=1}^{L-1}r_l r_{l-1}\right)\ ;
\end{equation}
the third gives:
\begin{equation}\label{eq:contrib3}
\frac{4}{9\pi}y_0 \left(k(3L-1)+r_0+6\sum_{l=1}^{L-1} l r_l\right)\ ,
\end{equation}
due to various canceling contributions. \eqref{eq:contrib1}, \eqref{eq:contrib2} and \eqref{eq:contrib3} match exactly\footnote{\label{foot:metatypo}Modulo the typo reported in footnote \ref{foot:typo3}.} with the respective terms in \cite[Eq. (4.12)]{cremonesi-tomasiello} once we impose $r_0=0$. We now take care of the last two terms in \eqref{eq:intL-l}, which were not present in \cite{cremonesi-tomasiello}. The fourth summand (i.e. the term proportional to $r_0$ in \eqref{eq:intL-l}) gives
\begin{equation}\label{eq:contrib4}
r_0 \left(k(3L-2)+6\sum_{l=1}^{L-1} l r_l - 2\sum_{l=1}^{L-1} r_l\right)\ ,
\end{equation}
whereas the fifth
\begin{equation}\label{ee:contrib5}
-\frac{6}{(9\pi)^2}\,\alpha_0 \left(k+r_0+2\sum_{l=1}^{L-1}  r_l\right)\ .
\end{equation}
All in all we get (renaming the dummy index $l \to i$ for ease of comparison with the field theory result):
\begin{equation}\label{eq:left-integral}
\begin{split}
-\frac{7}{16}\text{Int}_\text{L} = & \ 2\sum_{i=1}^{L-2}r_i \left[ 3k(L-i)-k-r_0-\frac{6}{(9\pi)^2} \alpha_0 \right] +12\sum_{i=1}^{L-1}\sum_{k=1}^{i-2}(i -k)r_i r_k \ + \\
&+\frac{61}{5} \sum_{i=1}^{L-1}r_i r_{i-1} +\frac{14}{5} \sum_{i=1}^{L-1}r_i^2 + 6 \sum_{i=1}^{L-1} i r_i \left[r_0+\frac{4}{9\pi}y_0 \right] +  \\
& +k \left[\frac{2}{5}k+\frac{21}{5}r_{L-1}+\frac{4}{9\pi}y_0(3L-1)-\frac{6}{(9\pi)^2}\alpha_0 \right] + kr_0(3L-2) +\\
& + r_0 \left[ \frac{12}{5} r_0  - 2 r_{L-1} + \frac{4}{9\pi} y_0 - \frac{6}{(9\pi)^2}\alpha_0 \right] - \frac{12}{(9\pi)^2} \alpha_0 r_{L-1}\ .
\end{split}
\end{equation}
The above generalizes \cite[Eq. (4.12)]{cremonesi-tomasiello}. 

\subsection{The contribution from the right massive tail}

We have to compute the integral 
\begin{equation}
\text{int}^\text{R}_l:=\int_{N-l}^{N-(l-1)} \alpha(z) q(z)\, dz\ , \quad l\in\left[1,R\right]\ .
\end{equation}
In each of the intervals $z \in \left[N-l,N-(l-1)\right]$, which can be equivalently written as $ \left[N-(i+1),N-i\right]$ with $i:=l-1=0,\ldots,R-1$, the function $\alpha(z)$ is given by the expression
\begin{multline}\label{eq:alphazNN-1}
\alpha(z) = \alpha_{N-i} -(9\pi)(2 y_{N-i}) \left[ z-(N-i)\right] \ + \\ -\frac{(9\pi)^2}{2}r_{N-i}\left[ z-(N-i)\right]^2 -\frac{(9\pi)^2}{6} s_{N-i+1} \left[ z-(N-i)\right]^3\ ,
\end{multline}
and not by \eqref{eq:alphaz}, as in the left region. ($y_{N-i}$ and $\alpha_{N-i}$ can be found, respectively, in \eqref{eq:yjRbis} and \eqref{eq:alphaRjbis-true}.) The function $q(z)$ simply reads
\begin{equation}\label{eq:qzNN-1}
q(z) = \frac{1}{2} r_{N-i}+ \frac{1}{2} s_{N-i+1} \left[ z-(N-i)\right]\ .
\end{equation}
We now have to sum the integrals $\text{int}^\text{R}_l$ from $l=1$ to $l=R$:
\begin{equation}\label{eq:intR-l-sum}
\text{Int}_\text{R} := \int_{N-R}^N \alpha(z) q(z)\, dz = \sum_{l=1}^R \int_{N-l}^{N-(l-1)} \alpha(z) q(z)\, dz = \sum_{l=1}^R \text{int}^\text{R}_l\ .
\end{equation}

We will not show the various steps of this computation, as the result can simply be obtained by applying the following substitutions to \eqref{eq:left-integral}:\footnote{\label{foot:typo5}Notice that in \cite[Sec. 4.3]{cremonesi-tomasiello} it is said that the contribution from the right region can be found by sending $y_0 \to y_N$. This is a typo, and the correct substitution should be $y_0 \to -y_N$, as in \eqref{eq:LRgeneric}.}
\begin{subequations}\label{eq:LRgeneric}
\begin{gather}
L \to R\ ,\quad r_0 \to r_N\ , \quad y_0 \to -y_N\ , \quad \alpha_0 \to \alpha_N\ , \\
r_{i-1} \to r_{N-(i-1)}\ , \quad r_i \to r_{N-(i-1)+1}\ ,\quad  r_k \to r_{N-k}\ .
\end{gather}
\end{subequations}
All in all
\begin{equation}\label{eq:right-integral}
\begin{split}
-\frac{7}{16}\text{Int}_\text{R} = & \ 2\sum_{i=1}^{R-2}r_{N-(i-1)+1} \left[ 3k(R-i)-k-r_N-\frac{6}{(9\pi)^2} \alpha_N \right] \ + \\
&+ 12\sum_{i=1}^{R-1}\sum_{k=1}^{i-2}(i-k) r_{N-(i-1)+1} r_{N-k} \ +  \\
&+\frac{61}{5} \sum_{i=1}^{R-1}r_{N-(i-1)+1} r_{N-(i-1)} +\frac{14}{5} \sum_{i=1}^{R-1}r_{N-(i-1)+1}^2 \ +  \\
& + 6 \sum_{i=1}^{R-1} i r_{N-(i-1)+1} \left[r_N-\frac{4}{9\pi}y_N \right] \ +\\
& +k \left[\frac{2}{5}k+\frac{21}{5}r_{N-R+1}-\frac{4}{9\pi}y_N(3R-1)-\frac{6}{(9\pi)^2}\alpha_N \right] \ + \\
& +kr_N(3R-2) + r_N \left[ \frac{12}{5} r_N  - 2 r_{N-R+1} - \frac{4}{9\pi} y_N - \frac{6}{(9\pi)^2}\alpha_N \right] \ + \\
& - \frac{12}{(9\pi)^2} \alpha_N r_{N-R+1}\ .
\end{split}
\end{equation}
Notice that $r_{N-R+1}$ is the image of $r_{L-1}$ under \eqref{eq:LRgeneric} with $i=L$.

\subsection{The contribution from the central massless plateau}

To evaluate the contribution from the massless plateau we must use a function $\alpha(z)$ supported in $z \in  \left[L, N-R\right]$. To do that we cannot simply impose $s_{i+1}=0$ on \eqref{eq:alphaz}, given that the latter expression depends on $y_i$ and $\alpha_i$, which are only defined for $i=1,\ldots,L$ and not for $i=L,\ldots,N-R$. Therefore we revert to using the coordinate $y$. In that coordinate, in the massless region we have (see \cite[Eqs. (2.10), (2.18)]{cremonesi-tomasiello} and \eqref{eq:z}):
\begin{equation}\label{eq:alpha-massless}
dz = \frac{1}{9\pi} \frac{dy}{q(y)}\ , \quad \sqrt{\beta(y)} := \frac{2}{k} (\tilde{R}_0^2-y^2)\ , \quad q(y) = \frac{1}{2} r_i = \frac{k}{2} = -4y \frac{\sqrt{\beta(y)}}{\partial_y \beta(y)} \ ,
\end{equation}
given that in the massless region $s_{i+1}=0$ and $r_i = k = r_L = r_{N-R}$ for all $i=L,\ldots,N-R$. $\tilde{R}_0 := R_0^3$ is a constant parameter the massless solution depends on, and $R_0$ may be interpreted as the radius of $S^4$ in the eleven-dimensional supergravity solution $\mathrm{AdS}_7 \times S^4/\mathbb{Z}_k$.\footnote{\label{foot:typo6}{The $\tilde{R}_0^2$ constant in \eqref{eq:alpha-massless} (which is taken from \cite[Eq. (2.10)]{cremonesi-tomasiello}) should be converted to $R_0^6$ -- see \cite[Eq. (4.13)]{cremonesi-tomasiello} and the older \cite[Eq. (C.17)]{afpt} -- hence the definition $\tilde{R}_0 := R_0^3$.}} In the ten-dimensional $\mathrm{AdS}_7 \times M_3$ massless solution of \cite{afrt} it is determined via $e^{3A(y)-\phi(y)} = -2 n_2 \frac{e^{2A(y)}}{\sqrt{1-x_7(y)^2}}=R_0^3$ (see \cite[Eq. (5.4)]{afrt} and \cite[below Eq. (5.41)]{afpt}),\footnote{These fields read $x_7(y) = \left( \frac{-y \partial_y\beta(y)}{4\beta(y)-y \partial_y \beta(y)} \right)^2$, $e^{A(y)} = \frac{2}{3} \left(- \frac{\partial_y \beta(y)}{y}\right)^{1/4}$, and $e^{\phi(y)} = \frac{1}{12} \left(- \frac{\partial_y \beta(y)}{y}\right)^{5/4}(4\beta(y)-y \partial_y \beta(y))^{-1/2}$. See \cite[Eq. (5.20)]{afpt}.} and can also be related to $k$ via $R_0^3 = 4\pi k N$ (see \cite[below Eq. (5.4)]{gaiotto-t-6d}).

We have:
\begin{equation}\label{eq:massless-contrib}
\begin{split}
\text{Int}_\text{plateau} &= \frac{128}{189 \pi ^2} \int_{y_L}^{y_{N-R}} \sqrt{\beta(y)} \, q(y) \, \frac{1}{9\pi}\frac{dy}{q(y)} =  \frac{2^8}{3^5\,7\, \pi ^3}  \sum_{i=L}^{N-R-1} \int_{y_i}^{y_{i+1}} \frac{1}{k} (\tilde{R}_0^2-y^2) \, dy  \\
&= \frac{2^8}{3^5\,7\, \pi ^3 k} \left( \tilde{R}_0^2 \sum_{i=L}^{N-R-1} (y_{i+1} -y_i) -\frac{1}{3} \sum_{i=L}^{N-R-1} (y_{i+1}^3 -y_i^3)\right)  \\
&= \frac{2^8}{3^5\,7\, \pi ^3 k} \left( \tilde{R}_0^2 (y_{N-R} -y_L) -\frac{1}{3} (y_{N-R}^3 -y_L^3)\right)  \\
&=\frac{2^8}{3^5\,7\, \pi ^3 k} (y_{N-R} -y_L) \left( \tilde{R}_0^2 -\frac{1}{3}(y_{N-R}^2+y_L^2 -2 y_{N-R}y_L)-y_{N-R}y_L \right)   \\
&=\frac{2^8}{3^5\,7\, \pi ^3 k} (y_{N-R} -y_L)  \left( \tilde{R}_0^2 -\frac{1}{3}(y_{N-R}-y_L)^2-y_{N-R}y_L \right)\ .
\end{split}
\end{equation}
We may now use the massless expression for $\sqrt{\beta(y)}$ in \eqref{eq:alpha-massless} to determine $\tilde{R}_0^2$:
\begin{equation}
\tilde{R}_0^2 = \frac{k}{4}(\alpha_L + \alpha_{N-R}) + \frac{1}{2}\left[ ( y_{N-R}-y_L)^2 + 2 y_{N-R}y_L\right]\ , \quad \alpha_{L,N-R} := \sqrt{\beta(y_{L,N-R})}\ .
\end{equation}
Trading $\tilde{R}_0^2$ for the above expression in \eqref{eq:massless-contrib} yields:\footnote{\label{foot:typo8}Notice that there is a typo in \cite[Eq. (4.13)]{cremonesi-tomasiello}: The $2^8$ factor in the numerator of that formula should read $2^7$.}
\begin{equation}\label{eq:int-massless-fin1}
\text{Int}_\text{plateau} = \frac{2^7}{3^6\,7\, \pi ^3 k} \,(y_{N-R} -y_L) \left[ \frac{3}{2}k(\alpha_L + \alpha_{N-R}) + (y_{N-R} -y_L)^2 \right]\ .
\end{equation}
We can now use \eqref{eq:asym-alecrem} in the above equation, that is $y_{N-R} - y_L = \frac{9\pi}{2}k(N-R-L)$ and $\alpha_{N-R} - \alpha_L =-9\pi (N-R-L)(y_L+y_{N-R})$. This gives:
\begin{small}
\begin{equation}\label{eq:int-massless-asym}
\boxed{\text{Int}_\text{plateau}^\text{asym} = \frac{2^6\,k}{3^3\,7\, \pi^2} \,(N-R-L) \left( \alpha_L +\frac{9\pi}{4}(N-R-L)\left(3\pi k(N-R-L)-2(y_L+y_{N-R}) \right) \right)}\ .
\end{equation}
\end{small}%
The above expression holds in the most generic situation, i.e. the asymmetric case when none among $r_0,r_N,\alpha_0,\alpha_N$ are zero. The parameters $y_L$, $y_{N-R}$, $\alpha_L$ are given by \eqref{eq:expr-generic} and depend on the boundary data $y^\text{eff}_0,y^\text{eff}_N$ in \eqref{eq:y0N-asym-generic-eff} (or on $y_0,y_N$ in \eqref{eq:y0N-asym-gen} if $\alpha_0=\alpha_N=0$).

In the symmetric case ($L=R$) we must use \eqref{eq:sym-alecrem} instead, that is $y_{N-R} - y_L = \frac{9\pi}{2} k(N-R-L) =  \frac{9\pi}{2}k(N-2L)$ and $y_{N-R} + y_L =0$. The latter also implies that $\alpha_{N-R} - \alpha_L = 0$ (see \eqref{eq:alpha-massless}). Plugging this into \eqref{eq:int-massless-fin1} yields:
\begin{equation}\label{eq:int-massless-sym}
\boxed{\text{Int}_\text{plateau}^\text{sym} = \frac{2^6\,k}{3^3\,7\, \pi^2} \,(N-2L) \left( \alpha_L +\frac{9\pi}{4}\,3\pi k (N-2L)^2 \right)}\ ,
\end{equation}
which is of course the $y_{N-R} + y_L =0$ limit of \eqref{eq:int-massless-asym}.

The parameter $\alpha_L$ can be found in \eqref{eq:expr-generic} but now depends on the boundary data $y_0,y_N$ in \eqref{eq:y0N-sym-generic}. In the subcase where the plateau shrinks to zero size ($N-2L=0$) we see that \eqref{eq:int-massless-sym} automatically vanishes, as expected.

\subsection{The full gravity result in the generic case}

We now put together the contributions \eqref{eq:left-integral}, \eqref{eq:right-integral}, and \eqref{eq:int-massless-asym} to obtain the full integral \eqref{eq:aholo} in the generic case: 
\begin{equation}
a_\text{hol} = \frac{128}{189\pi^2} \int_0^N \alpha(z)q(z)\, dz = \text{Int}_\text{L} + \text{Int}_\text{R} + \text{Int}_\text{plateau}^\text{asym} \ . 
\end{equation}
Expressing the parameters $\alpha_L, y_L, y_{N-R}$ as functions of $r_0,r_N$ and the boundary data $y_0^\text{eff},y_N^\text{eff},\alpha_0,\alpha_N$ (the latter being themselves functions of the ranks $\left\lbrace r_i \right\rbrace_{i=0}^L, \left\lbrace r_{N-i} \right\rbrace_{i=0}^R)$), we obtain the very long expression:
\begin{small}
\begin{align}\label{eq:aholtot}
a_\text{hol} = & \ \frac{12}{(9\pi)^2}\frac{16}{7}\,k(N-R-L) \left( \alpha_L +\frac{9\pi}{4}(N-R-L)\left(3\pi k(N-R-L)-2(y_L+y_{N-R}) \right) \right)\ + \nonumber \\
&+ \frac{6}{(9\pi)^2}\frac{16}{7}(k+r_0+2r_{L-1})\alpha_0-\frac{4}{9\pi}\frac{16}{7}((3L-1)k+r_0)y_0^\text{eff} \ + \nonumber \\
&-\frac{16}{7} k \left(\frac{2}{5}k+\frac{21}{5}r_{L-1} \right) -\frac{16}{7} kr_0(3L-2) -\frac{16}{7} r_0 \left( \frac{12}{5} r_0  - 2 r_{L-1} \right)\ + \nonumber \\
&+\frac{12}{(9\pi)^2}  \frac{16}{7} \alpha_0 \sum_{i=1}^{L-2}r_i -\frac{24}{9\pi} \frac{16}{7} y_0^\text{eff} \sum_{i=1}^{L-1} i r_i\  +\nonumber \\
&+ \frac{6}{(9\pi)^2}\frac{16}{7}(k+r_N+2r_{N-R+1})\alpha_N+\frac{4}{9\pi}\frac{16}{7}((3R-1)k+r_N)y_N^\text{eff} \ + \nonumber
\end{align}
\begin{align}
&-\frac{16}{7} k \left(\frac{2}{5}k+\frac{21}{5}r_{N-R+1} \right) -\frac{16}{7} kr_N(3R-2) -\frac{16}{7} r_N \left( \frac{12}{5} r_N  - 2 r_{N-R+1} \right)\ + \nonumber \\
&+\frac{12}{(9\pi)^2}  \frac{16}{7} \alpha_N \sum_{i=1}^{R-2}r_{N-i} +\frac{24}{9\pi} \frac{16}{7} y_N^\text{eff} \sum_{i=1}^{R-1} i r_{N-i}\  +\nonumber \\
&-\frac{32}{7} \sum_{i=1}^{L-2}r_i \left[ 3k(L-i)-k-r_0 \right] -\frac{192}{7}\sum_{i=1}^{L-1}\sum_{k=1}^{i-2}(i-k) r_i r_k \ + \nonumber \\
&-\frac{976}{35} \sum_{i=1}^{L-1}r_i r_{i-1} -\frac{32}{5} \sum_{i=1}^{L-1}r_i^2 -\frac{96}{7}r_0 \sum_{i=1}^{L-1} i r_i \ + \nonumber \\
&-\frac{32}{7} \sum_{i=1}^{R-2}r_{N-i} \left[ 3k(R-i)-k-r_N \right] -\frac{192}{7}\sum_{i=1}^{R-1}\sum_{k=1}^{i-2}(i-k) r_{N-i} r_k \ + \nonumber \\
&-\frac{976}{35} \sum_{i=1}^{R-1}r_{N-i} r_{N-i+1} -\frac{32}{5} \sum_{i=1}^{R-1}r_{N-i}^2 -\frac{96}{7}r_N \sum_{i=1}^{R-1} i r_{N-i}  \ .
\end{align}
\end{small}%
We remark that this expression depends only on the ranks $r_i$ describing the brane configuration, i.e. on the combinatorial data defining the quiver contained in the tableaux $\rho_\text{L},\rho_\text{R}$, and nothing else.\\

In the holographic limit $N,L,R,k,r_i \to \infty$ with $\frac{L}{N},\frac{R}{N},\frac{k}{N},\frac{r_i}{N}$ fixed (which we denote by a $\sim$), it simplifies quite a bit: $a_\text{hol} = \text{Int}_\text{L} + \text{Int}_\text{R} + \text{Int}_\text{plateau}^\text{asym}$ with
\begin{subequations}\label{eq:aGRhol}
\begin{align}
\text{Int}_\text{L} & \sim -\frac{192}{7}\left(\sum_{i=1}^{L}\sum_{k=1}^{i-2}(i -k)r_i r_k + \frac{2}{9\pi}y_0^\text{eff} \sum_{i=1}^{L} i r_i  \right) \ , \\
\text{Int}_\text{R} & \sim -\frac{192}{7}\left(\sum_{i=1}^{R}\sum_{k=1}^{i-2}(i - k)r_{N-i} r_{N-k} - \frac{2}{9\pi}y_N^\text{eff} \sum_{i=1}^{R} i r_{N-i}  \right) \ , \\
\text{Int}_\text{plateau}^\text{asym} & \sim -\frac{192}{7}\frac{1}{3(9\pi)}\,(N-R-L) \left[ 3k\left( Ly_0^\text{eff}-Ry_N^\text{eff} +\frac{9\pi}{2} \sum_{i=1}^{L} (L-i) r_i \right. \right. + \nonumber \\
&\; \quad + \left. \left.  \frac{9\pi}{2}\sum_{i=1}^{R} (R-i) r_{N-i}\right) - \frac{9\pi}{4}k^2(N-R-L)^2 \right]\ .
\end{align}
\end{subequations}
The quantity $a_\text{hol}$ is now of order $\mathcal{O}(N^5)$. To obtain its expression we only kept the highest order (i.e. $\mathcal{O}(N^2)$) terms of the boundary data:
\begin{subequations}
\begin{align}
y_0 \sim\ & \frac{9\pi}{4}\left[ -\frac{k}{N}(N+R-L)(N-L-R) -2 \sum_{i=1}^{L} r_i + \frac{2}{N} \left( \sum_{i=1}^{L} i r_i - \sum_{i=1}^{R} i r_{N-i} \right)  \right] \ ,  \label{eq:y0limit}
\end{align}
\begin{align}
 y_N \sim \ & \frac{9\pi}{4} \left[ \frac{k}{N}(N+L-R)(N-L-R) + 2 \sum_{i=1}^{R} r_{N-i} + \frac{2}{N} \left( \sum_{i=1}^{L} i r_i - \sum_{i=1}^{R} i r_{N-i} \right) \right]\ ; \label{eq:yNlimit}\\
y_0^\text{eff} \sim\ & y_0 \ \text{given by \eqref{eq:y0limit}}\ , \label{eq:y0efflimit} \\
y_N^\text{eff} \sim\ & y_N \ \text{given by \eqref{eq:yNlimit}}\ ;  \label{eq:yNefflimit}
\end{align}
%\sim\ & \frac{9\pi}{4} \left[ \frac{k}{N}(N-L+R)(N-L-R) -\frac{2}{N}\left(\sum_{i=1}^{L} (N-i) r_i  +\sum_{i=1}^{R} i r_{N-i} \right)\right]\ , \\
% \frac{2(9\pi)}{(4\pi)^2 r_{N-1}N} \Biggl( \tilde{r}_N \Biggl[k(N-L-R)+ \sum_{i=1}^{L} r_i + \sum_{i=1}^{R}  r_{N-i}\Biggr] \ + \nonumber \\ &+  2\pi^2 r_{N-1} \Biggl[ {\color{red}+}k(N+R-L)(N-L-R) -Nr_0 -2\left(\sum_{i=1}^{L} (N-i) r_i  +\sum_{i=1}^{R} i r_{N-i} \right) \Biggr]  \Biggr) \nonumber \\
%y_N^\text{eff} \sim\ & \frac{9\pi}{4} \left[ \frac{k}{N}(N+L-R)(N-L-R) +\frac{2}{N}\left(\sum_{i=1}^{L} i r_i  +\sum_{i=1}^{R} (N-i) r_{N-i} \right)\right]\ ;
% -\frac{2(9\pi)}{(4\pi)^2 r_1 N} \Biggl( \tilde{r}_0 \Biggl[k(N-L-R)+ \sum_{i=1}^{L} r_i + \sum_{i=1}^{R}  r_{N-i} \Biggr] \ + \nonumber \\ & - 2\pi^2 r_1 \Biggl[ {\color{red}+} k(N+L-R)(N-L-R) +Nr_N +2\left(\sum_{i=1}^{L} i r_i  +\sum_{i=1}^{R} (N-i) r_{N-i} \right) \Biggr]  \Biggr) \nonumber \\
\begin{align}\label{eq:alpha0limit} 
\alpha_0 \sim \ & \frac{9^2}{8} \frac{\tilde{r}_0}{r_1} \left[ \frac{k}{N}(N+R-L)(N-L-R) -\frac{2}{N}\left(\sum_{i=1}^{L} (N-i) r_i  +\sum_{i=1}^{R} i r_{N-i} \right)\right]\ , \\
\alpha_N \sim \ & -\frac{9^2}{8} \frac{\tilde{r}_N}{r_{N-1}}\left[ \frac{k}{N}(N+L-R)(N-L-R) +\frac{2}{N}\left(\sum_{i=1}^{L} i r_i  +\sum_{i=1}^{R} (N-i) r_{N-i} \right)\right]\ ; \label{eq:alphaNlimit} 
\end{align}
\begin{equation}\label{eq:yLRlimit} 
y_L \sim  y_0^\text{eff}  + \frac{9\pi}{2}\sum_{i=1}^{L} r_i\ , \quad y_{N-R}\sim  y_N^\text{eff}  - \frac{9\pi}{2}\sum_{i=1}^{R} r_{N-i} \ ,
% y_0^\text{eff} + \frac{9\pi}{4} k + \frac{9\pi}{2}\sum_{i=1}^{L} r_i\ ,  \\
%y_N^\text{eff} - \frac{9\pi}{4} k - \frac{9\pi}{2}\sum_{i=1}^{R} r_{N-i}\ ;
\end{equation}
\begin{align}
\alpha_L \sim \ & \alpha_0 - (9\pi) (2 L) y_0^\text{eff} - (9\pi)^2\sum_{i=1}^{L} (L-i) r_i \ ; \label{eq:alphaLlimit} \\
%\alpha_0 - (9\pi) (2 L) y_0^\text{eff} -\frac{(9\pi)^2}{2} \left(L r_0+\frac{1}{3}k \right) - (9\pi)^2\sum_{i=1}^{L} (L-i) r_i \ , \\
\alpha_{N-R} \sim \ & \alpha_N +(9\pi) (2 R) y_N^\text{eff} -(9\pi)^2\sum_{i=1}^{R} (R-i) r_{N-i}\ . \label{eq:alphaRlimit} 
% \alpha_N +(9\pi) (2 R) y_N^\text{eff} - \frac{(9\pi)^2}{2} \left(R r_N +\frac{1}{3}k\right)-(9\pi)^2\sum_{i=1}^{R} (R-i) r_{N-i}
\end{align}
In the four quantities \eqref{eq:yLRlimit}-\eqref{eq:alphaRlimit} $y^\text{eff}_0,y^\text{eff}_N$ are respectively given by \eqref{eq:y0efflimit}, \eqref{eq:yNefflimit} if $\alpha_0,\alpha_N \neq 0$, and should be replaced by \eqref{eq:y0limit}, \eqref{eq:yNlimit} if $\alpha_0 =\alpha_N=0$. Finally notice that $\alpha_0,\alpha_N$ are subleading terms (as they are of order $\mathcal{O}(N^2)$) in \eqref{eq:alphaLlimit}, \eqref{eq:alphaRlimit} which are of order $\mathcal{O}(N^3)$; therefore we are left with:
\begin{align}\label{eq:alphaLRlimit}
\alpha_L &\sim  - (9\pi) \left( 2 L \,y_0^\text{eff} + 9\pi\sum_{i=1}^{L} (L-i) r_i\right) \ , \\
\alpha_{N-R} &\sim  -(9\pi) \left( -2 R\, y_N^\text{eff} +9\pi \sum_{i=1}^{R} (R-i) r_{N-i} \right)\ .
\end{align}
\end{subequations}

% fold sec (agrav)

%%%%%%%%%%%%%%%%%%%%%%%%%%%%%%%%%%%%%%%%%%%%%%%%%%
\section{Field theory side: The anomaly polynomial} % sec (aFT)
\label{app:aFT}
%%%%%%%%%%%%%%%%%%%%%%%%%%%%%%%%%%%%%%%%%%%%%%%%%%

In this section we shall compute $a$ exactly in field theory by leveraging the six-dimensional anomaly polynomial. We will then extract its leading contribution in the holographic limit. 

\subsection{\texorpdfstring{Extracting $a$ from the six-dimensional anomaly polynomial}{Extracting a from the six-dimensional anomaly polynomial}}
\label{sub:aM5}
We will now extract the $a$ conformal anomaly of six-dimensional $(1,0)$ SCFT's on the tensor branch that feature $\SU$, $\SO$ or $\USp$ gauge and flavor groups, as appropriate in the presence of O6 and O8-planes. The $a$ anomaly will be given as a combination of coefficients appearing in the six-dimensional anomaly polynomial $\mathcal{I}$; the $c_i$ anomalies are all proportional to $a$ in the holographic limit \cite{beccaria-tseytlin-I,beccaria-tseytlin-II,yankielowicz-zhou} ($c_1 \sim -\frac{7}{12}a,c_2 \sim \frac{1}{4}c_1,c_3 \sim -\frac{1}{12}c_1$), so their holographic match follows from that of $a$ and will not be considered here. %{\color{red} Possibly use newer results of \cite{beccaria-tseytlin-II} to determine exact coefficient in front of $\alpha$ in $a$ and $c_i$'s.} 
We will then estimate the leading behavior of the exact formula for $a$ as $N \to \infty$.

\paragraph{Strategy} More concretely, the strategy will be the following:
\begin{enumerate}
\item Collect all contributions to the six-dimensional anomaly polynomial $\mathcal{I}$ without assuming that gauge and flavor groups are restricted to being $\SU$; use instead the general formulae provided below. $a$ can be written in terms of the coefficients of $\mathcal{I}$.
\item Estimate the behavior of $a$ as $N \to \infty$ and extract its leading term; disregard subleading terms. The former depends on the inverse of the Dirac pairing of the $(1,0)$ theory (which is given e.g. by the Cartan matrix of $A_{N-1}$ in the case without O-planes). Estimate its inverse recursively.
\item Break the leading term into contributions corresponding to central massless plateau and massive tails.
%\item Re-express everything in previous step in terms of the $f_i$, if one is interested in extracting an explicit formula more useful \& compact for the particular case at hand.
\end{enumerate}
This will greatly facilitate the comparison with the gravity result.\newline

Let us first focus on the computation of the anomaly polynomial $\mathcal{I}$ of six-dimensional $(1,0)$ theories. We will list all the supersymmetry multiplets that contribute to it.
\paragraph{Tensors} The multiplet content of $(1,0)$ six-dimensional theories comprises $N_\text{T}$ tensor multiplets
\begin{equation} \label{eq:deftens}
(\phi, B_{\mu \nu}, \psi)_i \ , \quad i=1,\ldots, N_\text{T}=N-1\ ,
\end{equation}
where $\phi$ is a real scalar, $B_{\mu \nu}$ the two-form potential, and $\psi$ the fermion superpartners. Our AdS$_7$ solutions are dual to SCFT's amenable to a weakly coupled quiver gauge theory description, which is obtained by giving the $\phi_i$'s a suitable vev. In general the configuration of gauge groups has the following form
\begin{equation}
G_1 \times G_2 \times \ldots \times G_{N_\text{T}}\ ,
\end{equation}
where on each gauge node we can have a possible flavor symmetry. (For theories on $N$ separated NS5-branes we will take $N_\text{T} = N-1$ since one tensor -- corresponding to the center-of-mass motion of the branes along direction $x^6$ -- decouples from the dynamics.) 

\paragraph{Vectors} We also have vector multiplets
\begin{equation}
(A_\mu, \lambda)_i\ , \quad i=1,\ldots, N_\text{V}\ ,
\end{equation}
where $A_{\mu}$ is the gauge potential and $\lambda_i$ the fermion superpartner.

\paragraph{Hypers} Next we consider hypermultiplets in the bifundamental of $G_i \times G_{i+1}$,
\begin{equation} \label{eq:defhyp}
(q\oplus q^\text{c}, \chi\oplus \chi^\text{c})_k \ , \quad k=1,\ldots, N_\text{H}\ ,
\end{equation}
with $q$ the complex scalar in the fundamental and $\chi$ the fermion superpartner, and $q^\text{c},\chi^\text{c}$ in the conjugate representation. We can also have hypermultiplets which are not in the bifundamental, but instead in the fundamental (antifundamental) of the gauge (flavor) group. There can also be hypermultiplets in the symmetric or antisymmetric representations.

\paragraph{One-loop polynomial} The (one-loop, gauge and mixed) anomaly polynomial eight-form of the theory is given by a sum of various contributions. We have the tensor multiplets contribution
\begin{equation}
 I_\text{tens}=\frac{N_\text{T}}{24}\left(c_2(R)^2+\frac{1}{2}c_2(R)p_1(T)+ \frac{1}{240}\left(23 p_1(T)^2 -116p_2(T)\right)  \right)\ ,
\label{eq:Itensor}
\end{equation}
and the vector multiplets contribution
\begin{align} \label{eq:apvec}
 I_\text{vec}=\sum_{i=1}^{N_\text{T}}&\biggl(-\frac{\tr_\text{adj}(F_i^4)+ 6 c_2(R) {\rm tr}(F_i^2)+ d_{G_i} c_2(R)^2 }{24}- p_1(T) \left(\frac{\tr_\text{adj}(F_i^2) + d_{G_i} c_2(R)}{48} \right)-\nonumber \\
 &d_{G_i} \left(\frac{7 p_1(T)^2-4p_2(T)}{5760} \right) \biggr)\ ,
\end{align}
where the trace $\tr_\text{adj}$ is taken over the adjoint representation of the gauge group $G_i$ (of real dimension $d_{G_i}$) and $F_i$ are the field strengths of the gauge potentials $A_i$. The hypermultiplets in the representation ${\rho}_i$ contribute:
\begin{align} \label{eq:aphyp0}
 I_\text{hyp}=&\sum_{i=1}^{N_\text{T}}\epsilon_i \biggl(d_i \frac{\left(7p_1(T)^2-4 p_2(T)\right)}{5760}  +\frac{ \text{tr}_{\rho_i}(F^2_{i} )p_1(T)}{48}  +\frac{\text{tr}_{\rho_i}(F^4_{i})}{24}\biggr)\ ,
\end{align} 
where $d_i$ is the dimension of the representation $\rho_i$. Sometimes we can have half-hypermultiplets instead of ``full'' ones.\footnote{This means that in \eqref{eq:defhyp} we do not have the conjugate representation, just $(q,\chi)$.} The associated anomaly polynomial contribution is then divided by two; for this reason (and for convenience) each hypermultiplet comes with a factor $ \epsilon_i=\left\lbrace 1,\frac{1}{2},0\right\rbrace$ in front (zero means there are no hypermultiplets). A hypermultiplet in the bifundamental $\rho_i \otimes \rho_{i+1}$ contributes
\begin{small}
\begin{align} \label{eq:aphyp}
 I_\text{hyp-bi}= & \sum_{i=1}^{N_\text{T}-1}\epsilon_{i\,i+1} \left( \frac{1}{5760} d_i d_{i+1} \left(7p_1(T)^2-4 p_2(T)\right) +\frac{1}{48} p_1(T) \left( d_i \tr_{\rho_{i+1}} (F^2_{i+1})\right. \right. \ + \nonumber\\ 
& + \left. \left. d_{i+1}\tr_{\rho_i}(F^2_{i}) \right) +  \frac{1}{24} \left( d_i\tr_{\rho_{i+1}} (F^4_{i+1}) + d_{i+1} \tr_{\rho_i}(F^4_i)+6 \tr_{\rho_{i+1}} (F^2_{i+1})\tr_{\rho_i}(F^2_{i}) \right) \right)\ ,
\end{align} 
\end{small}%
where the trace is taken over the representation $\rho_i$ of dimension $d_i$, and the number $\epsilon_{i\,i+1}=\left\lbrace 1,\frac{1}{2},0\right\rbrace$ accounts for the presence of a full, half, no hypermultiplet between two consecutive gauge groups $G_i \times G_{i+1}$. If flavor symmetries are present, we need to add the extra hypermultiplet contribution
\begin{align}\label{eq:aphypflv}
 I_\text{hyp-flv}=& \sum_{i=1}^{N_\text{T}} \frac{ \epsilon^\text{flv}_i}{24} \biggl(f_i \tr_{\rho_i}(F^4_{i})+  \frac{p_1(T)}{2} \left(f_i\tr_{\rho_i}(F^2_{i})+\frac{ f_i d_{i} }{240}\left(7p_1(T)^2-4 p_2(T)\right) \right) +   \nonumber\\
&  d_i \tr_{\rho^\text{flv}_i}(F^4_{\text{flv}\,i})+  \frac{p_1(T)}{2} d_i\tr_{\rho^\text{flv}_i}(F^2_{\text{flv}\,i}) + 6  \tr_{\rho^\text{flv}_i}(F^2_{\text{flv}\,i}) \tr_{\rho_i}(F^2_{i}) \biggr)\ ,
\end{align}
where $F_{\text{flv}\,i}$ are the flavor field strengths, $f_i$ the dimension of the hypermultiplet flavor representation, and $\epsilon^\text{flv}_i=\left\lbrace 1,\frac{1}{2},0\right\rbrace$ as for ${\epsilon}_i$ and $\epsilon_{i\,i+1}$.

Many traces over different representations appear in the various contributions. We define $\Tr F^2$ as in \cite{ohmori-shimizu-tachikawa-yonekura} to be the trace in the adjoint of $G$ divided by $h^\vee_{G}$; the former is also related to the trace in the fundamental by a constant $s_{G}$ that depends on the group:
\begin{equation}\label{eq:Tr2}
\tr_\text{adj} (F^2) =: h^\vee_{G} \Tr (F^2)\ , \quad \tr_\text{fund}(F^2) = s_{G}\, \text{Tr}(F^2)\ , \quad \tr_\text{fund}(F^4)=\Tr (F^4)\ .
\end{equation}
Moreover
\begin{equation}\label{eq:Tr4}
\tr_\text{adj} (F^4) = t_{G} \tr_\text{fund} F^4+\frac{3}{4}u_{G} \left( \Tr(F^2)\right)^2\ .
\end{equation}
We have collected the constants $h^\vee_{G}, s_{G}, t_{G}, u_{G}$ for groups $\SU, \SO, \USp$ in table \ref{tab:groupconst}. 
\begin{table}[!tb]
\centering
{\renewcommand{\arraystretch}{1.2}%
\begin{tabular}{@{}  c  c  c  c  @{}}
 &$\SU(k)$ &$ \SO(2k) $&$\USp(2k)$ \\
\toprule \toprule
rank $r_{G}$ & \multicolumn{1}{!{\vrule width 1pt}c}{$k-1$} &$ k $&$k$ \\
%\toprule
dual Coxeter number $h^\vee_{G}$ & \multicolumn{1}{!{\vrule width 1pt}c}{$k$} &$ 2k-2 $ & $k+1$ \\
%\toprule
$d_G := \dim_\rr G$ & \multicolumn{1}{!{\vrule width 1pt}c}{$k^2-1$} &$ k(2k-1)$ & $k(2k+1)$ \\
%\toprule
$d_\text{fund} := \dim_\rr (\text{fund})$ & \multicolumn{1}{!{\vrule width 1pt}c}{$k$} &$ 2k $ & $2k$ \\
%\toprule
$s_{G}$ & \multicolumn{1}{!{\vrule width 1pt}c}{$1/2$} &$1 $ & $1/2$ \\
%\toprule
$t_{G}$ & \multicolumn{1}{!{\vrule width 1pt}c}{$2k$} &$2k-8 $ & $2k+8$ \\
%\toprule
$u_{G}$ & \multicolumn{1}{!{\vrule width 1pt}c}{$2$} &$ 4$ & $1$ \\
\toprule
\end{tabular}}
\caption{The group theory constants appearing in \eqref{eq:Tr2} and \eqref{eq:Tr4}. Notice that $t_{\SO(8)} = 0$. The constants are taken from \cite{ohmori-shimizu-tachikawa-yonekura,erler}.}
\label{tab:groupconst}
\end{table}
For general $(1,0)$ theories, matter can be in other representations than just the (anti) fundamental (this is the case when O6 and O8-planes are present);\footnote{For instance, spinor representations appear in the F-theory engineering of the quiver in figure \ref{fig:quiverNS5D6O6} with $k=4$ \cite{heckman-rudelius-tomasiello}.} we need a generalization of \eqref{eq:Tr2} and \eqref{eq:Tr4} to compute the traces  $\tr_\rho (F^2)$ and $\tr_\rho (F^4)$ appearing in \eqref{eq:aphyp0}, \eqref{eq:aphyp}, and \eqref{eq:aphypflv}. This can be done at the expense of introducing two more constants:
\begin{equation}\label{eq:Tr2Tr4gen}
\tr_\rho (F^2) = \Ind_\rho \,\widetilde{\tr} (F^2)\ , \quad \tr_\rho(F^4) = \alpha_\rho \, \widetilde{\tr}F^4+c_\rho \left( \widetilde{\tr} (F^2)\right)^2\ ,
\end{equation}
where $\widetilde{\tr}=\tr_\text{fund}$ for $G=\SU(k)$ and $\widetilde{\tr}=\tr_\text{vec}$ for $G=\SO(2k),\USp(2k)$. $\Ind_\rho$ is the index of the (irreducible) representation $\rho$ (which has been defined in footnote \ref{foot:index}). When $\rho$ is the fundamental, $\text{Ind}_\text{fund}$ coincides with $s_G$. We have collected the constants $\alpha_\rho$ and $c_\rho$ in table \ref{tab:ac}.
\begin{table}[!tb]
\centering
{\renewcommand{\arraystretch}{1.2}%
\begin{tabular}{@{}  c  c  c  c  c  c @{}}
$\rho$ & type & $d:= \dim_\rr \rho$ & index & $\alpha_\rho$ & $4 c_\rho$ \\
\toprule \toprule
\multicolumn{5}{c}{$\SU(k)$, $k\geq 4$:} \\
fundamental &  \multicolumn{1}{!{\vrule width 1pt}c}{complex} & $k$ & $\frac{1}{2}$ & 1 & 0 \\
symmetric & \multicolumn{1}{!{\vrule width 1pt}c}{complex} & $\frac{k}{2}(k+1)$ & $\frac{1}{2} (k + 2)$& $k+8$ & 3 \\
antisymmetric & \multicolumn{1}{!{\vrule width 1pt}c}{complex} & $\frac{k}{2}(k-1)$ & $\frac{1}{2} (k - 2)$& $k-8$ & 3 \\
adjoint & \multicolumn{1}{!{\vrule width 1pt}c}{real}& $k^2-1$ & $k$ & $2k$ & 6 \\
\toprule
\multicolumn{5}{c}{$\SO(2k)$, $k> 4$:} \\
vector $=$ fundamental &  \multicolumn{1}{!{\vrule width 1pt}c}{real} & $2k$ & 1 & 1 & 0 \\
adjoint &  \multicolumn{1}{!{\vrule width 1pt}c}{real}& $k(2k-1)$ & $4k-4$& $2k-8$ & 3 \\
\toprule
\multicolumn{5}{c}{$\USp(2k)$, $k\geq 2$:} \\
vector $=$ fundamental &  \multicolumn{1}{!{\vrule width 1pt}c}{pseudo-real} & $2k$ & $\frac{1}{2}$ & 1 & 0 \\
antisymmetric &  \multicolumn{1}{!{\vrule width 1pt}c}{real} & $(k-1)(2k+1)$ & $k-1$ & $2k-8$ & 3 \\
symmetric $=$ adjoint &  \multicolumn{1}{!{\vrule width 1pt}c}{real} & $k(2k+1)$ & $k+1$ & $2k+8$ & 3 \\
\toprule 
\end{tabular}}
\caption{The constants are taken from \cite{lieart,bhardwaj} (with $c_\rho^\text{there}= 4 c_\rho^\text{here}$).}
\label{tab:ac}
\end{table}

The one-loop contribution of the anomaly polynomial is given by the sum of all these terms:
\begin{equation}
I^\text{1-loop}=I_\text{tens}+ I_\text{vec}+ I_\text{hyp}+ I_\text{hyp-bi} + I_\text{hyp-flv}\ .
\label{eq:1loop}
\end{equation}
We now need to cancel all terms involving field strengths of gauge groups from the one-loop contribution, in order to guarantee the quantum consistency of the field theory. This is done via a Green--Schwarz--West--Sagnotti (GS henceforth) mechanism \cite{green-schwarz-west,sagnotti}. However the coefficient in front of $\text{Tr}(F^4_{i})$ cannot be canceled by a GS-type mechanism; hence we need to impose that it vanish by hand. This leads to the following constraint:
\begin{equation} \label{eq:F4canc}
t_{G_i}={\epsilon}_i \alpha_{\rho_i} +\left(  \epsilon_{i\, i-1} d_{i-1}+\epsilon_{i\,i+1} d_{i+1} + \epsilon^\text{flv}_i f_i\right)\ .
\end{equation}
The rest of the one-loop gauge anomaly polynomial can be written as a product of matrices and vectors as follows:
\begin{align} \label{eq:apgm}
I^\text{1-loop}_\text{gauge $+$ mixed}=&-\frac{1}{2} \eta_{ij}\left(\frac{\text{Tr}(F_i^2)}{4}\right)\left(\frac{\text{Tr}(F_j^2)}{4}\right)-(A_{c_2})_{i}\left(\frac{\text{Tr}(F_i^2)}{4}\right)c_2(R)+\nonumber \\
&+ (A_{p_1})_{i}\left(\frac{\text{Tr}(F_i^2)}{4}\right)\frac{p_1(T)}{12}+ (A_\text{flv})_{ij}\left(\frac{\text{Tr}(F_i^2)}{4}\right)\left(\frac{\text{Tr}(F_{\text{flv}\, j}^2)}{4}\right)\ ,
\end{align} 
where $i,j=1,\ldots, N_\text{T}$ and
\begin{subequations}\label{eq:Imatrices}
\begin{align}
\eta_{ij} &:=\left(u_{G_i} - \frac{4}{3}{\epsilon}_i c_{\rho_i}\right) \delta_{ij} - 4({\epsilon_{i\,i+1}} s_{G_i} s_{G_{i+1}}\delta_{i\, i+1}+{\epsilon_{i\,i-1}} s_{G_i} s_{G_{i-1}} \delta_{i\, i-1})\ ,  \label{eq:DPfAP}\\ 
 (A_{c_2})_i &:=  h^\vee_{G_i}\ , \\
 (A_{p_1})_i &:=-h^\vee_{G_i}+{\epsilon}_i \, \Ind_{{\rho}_i}+s_{G_i}\left( {\epsilon}_i (d_{i-1}+d_{i+1}) + \epsilon^\text{flv}_i f_i\right)\ \ ,\label{eq:anconstp1} \\
(A_\text{flv})_{ij} &:=  4\, {\rm diag}\left( \epsilon^\text{flv}_i s_{G_i} s^\text{flv}_{G_{i}} \right)_{ij} \ .
\end{align}
\end{subequations}
In particular, the quadratic part (in $\Tr F_i^2$) of the one-loop gauge anomaly polynomial must have nonnegative and properly quantized coefficients $\eta_{ij}$. This is due to the Bianchi identity of the anti-self-dual two-form potential $B$, which schematically reads $dH = c \Tr (F \wedge F)$ (with $dB=H$ away from sources): That is, the instanton string in six dimensions is charged under $B$ with quantized charge $c$. Indeed one can define a Dirac pairing $\eta_{ij}$ on the six-dimensional string charge lattice \cite[Sec. 3.2]{ohmori-shimizu-tachikawa-yonekura}, $\left\langle c ,c' \right\rangle_\text{Dirac} := \eta_{ij} \,c_i \,c'_j$, that collects the charges of the instanton string of each gauge group under the $B_i$'s in the $N_\text{T}$ tensor multiplets. As one can check from \eqref{eq:DPfAP}, we can write:\footnote{\label{foot:O8+-}{However, see the discussion below \eqref{eq:etaO8+-noF} for an exception to this rule.}}
\begin{equation}\label{eq:dirac}
\eta_{ij} :=\mathbf{n}\, \delta_{ij} - \delta_{i\,i-1}- \delta_{i\,i+1} = \begin{cases} n_i & j=i \\ -1 & j=i-1,i+1 \end{cases}\ , \quad \mathbf{n}=\{n_1,n_2,\ldots, n_{N_\text{T}}\}
\end{equation}
with $i,j=1,\ldots, N_\text{T}$. The diagonal entries $n_i$ of the pairing matrix (which we collected in a vector $\mathbf{n}$) are associated with the gauge groups $G_i$ and must be integers to insure charge quantization. These integers also agree with the diagonal entries of the adjecency (or intersection) matrix of the F-theory configuration realizing the SCFT (see e.g. \cite{heckman-morrison-vafa,heckman-morrison-rudelius-vafa,delzotto-heckman-tomasiello-vafa} and in particular \cite[Sec. 3.3]{ohmori-shimizu-tachikawa-yonekura}). In the latter context each $n_i$ gives the negative self-intersection of the $i$-th compact curve in the F-theory base (the curves intersect each other at one point, hence the off-diagonal $-1$). This is the convention we use throughout the paper.

As we will see momentarily, the Dirac pairing plays a central role in the cancellation of the remaining gauge anomalies via a GS mechanism. In passing, we note that it can also be used to write down a weakly-coupled effective Lagrangian for the (bosonic part of the) tensor multiplets (when the SCFT is on its tensor branch):
\begin{equation}
\mathcal{L}_\text{eff} \supset \eta_{ij}(\partial_{\mu}\phi^i)(\partial^{\mu} \phi^j )+ \eta_{ij}\, H^i \wedge \ast H^j\ .
\end{equation}

\paragraph{GS term} The GS term can be derived via a descent mechanism involving auxiliary two-forms potentials, and its associated eight-form reads
\begin{equation}\label{eq:GS}
I_\text{GS}= \frac{1}{2} \eta_{ij}I^i I^j\ , \quad i,j=1,\ldots, N_\text{T}\ ,
\end{equation}
where the $I^i$ are defined as follows
\begin{equation} \label{eq:Ii2}
I^i= \sum_{j=1}^{N_\text{T}} \frac{1}{4}\Tr(F_j^2)+y_i c_2(R)+K_i\frac{p_1(T)}{12} +z_{i}\frac{1}{4}\Tr (F_{\text{flv}\, i}^2) \ ,
\end{equation}
and the quantities $K_i, y_i, z_{ij}$ are such that
\begin{subequations}\label{eq:YKZ}
\begin{align} 
\eta_{ij} y_j &:=(A_{c_2})_i=h^\vee_{G_i}\ ,\\
\eta_{ij} K_j &:=-(A_{p_1})_i=h^\vee_{G_i}-{\epsilon}_i \, \Ind{{\rho}_i}-s_{G_i}\left( \epsilon_{i\,i-1} d_{i-1}+\epsilon_{i\,i+1}d_{i+1} + \epsilon^\text{flv}_i f_i\right)\ ,\\
\eta_{ik} z_{kj} &:=-(A_\text{flv})_{ij}=-4  \,{\rm diag}\left(\epsilon^\text{flv}_i s_{G_i} s^\text{flv}_{G_{i}} \right)\ .
\end{align}
\end{subequations}
Adding the GS contribution to the one-loop piece of the anomaly polynomial eight-form \eqref{eq:apgm}, all the coefficients of the monomials involving the field strengths of the gauge groups vanish. The theory is completely gauge anomaly-free. (Notice that, in principle, there can be several GS contributions that cancel the gauge and mixed anomalies, and these are classified by automorphisms of the Dirac pairing. See \cite{apruzzi-heckman-rudelius} for more details.)

It now suffices to put all contributions together to obtain $\mathcal{I}$. From that, one can extract the $a$ conformal anomaly \eqref{eq:a}, as explained in section \ref{sec:aexamples}.

\subsection{The holographic limit of the exact field theory expression}
\label{sub:holoFT}
The expression \eqref{eq:a} is an exact field theory result. To perform a comparison with its gravity counterpart, we should first take the holographic limit. This limit will wash away many of its terms. When $N \to \infty$, we can safely assume all ranks of the gauge groups $G_i$ scale like $N$. As we have shown below \eqref{eq:a}, the leading ($\sim$) contribution to $a$ is then given by
\begin{equation}\label{eq:aFTholo}
a \sim \frac{192}{7} (\eta^{-1})_{ij} h^\vee_{G_i} h^\vee_{G_j}\ .
\end{equation}
This is the expression we will be comparing \eqref{eq:aholo} to. 

To compute \eqref{eq:aFTholo} we simply need to estimate the inverse of the Dirac pairing $\eta$. This is done as follows. For the case with only D6 and D8-branes, all gauge groups are $\SU(r_i)$ and, by relying on table \ref{tab:groupconst}, we conclude that $\mathbf{n}=\left\lbrace 2, 2, \ldots, 2 \right\rbrace$ from \eqref{eq:DPfAP}. Therefore the Dirac pairing \eqref{eq:dirac} is simply the Cartan matrix of $A_{N-1}$ and its inverse has already been estimated in \cite[Eq. (3.13)]{cremonesi-tomasiello}, which we reproduce below. (Notice that the left- and rightmost flavor groups engineered by $r_0,r_N$ D6-branes respectively do not have corresponding entries in the Dirac pairing among gauge groups.)

In the case with O6-planes, we have seen in section \ref{subsub:alternatingSOUSp} that we have an alternating sequence of $\SO$ and $\USp$ groups. Starting with $\SO$ ($\USp$), the string charge vector reads $\mathbf{n}=\left\lbrace 4,1,4,1 \ldots, 1,4 \right\rbrace$ ($\left\lbrace 1,4,1,4 \ldots, 1 \right\rbrace$), and the inverse of $\eta$ can be estimated recursively starting from $N_\text{T}=2,3,4,\ldots$. To write down a closed form for the latter we have actually made use of an auxiliary vector $\mathbf{v} = \left\lbrace v_i \right\rbrace_{i=1}^{N_\text{T}}:=
 \left\lbrace 1,2,1,2,\ldots\right\rbrace$ or $\left\lbrace 2,1,2,1\ldots\right\rbrace$: The first entry is 1 if the first group in is $\SO$, 2 if it is $\USp$.\footnote{We thank A.~Tomasiello for suggesting the use of $\mathbf{v}$.}

In the case with an O8-plane at $x^6=0$ between an NS5-brane and its image, we have seen in section \ref{subsub:O8pureSUk} that we have a sequence of groups starting off with $\SO$ or $\USp$ followed by a string of $\SU$'s. Therefore $\mathbf{n}=\left\lbrace 4,2,2 \ldots, 2 \right\rbrace$ or $\mathbf{n}=\left\lbrace 1,2,2 \ldots, 2 \right\rbrace$ respectively. However, given the subtlety discussed below \eqref{eq:etaO8+-noF} in the $\USp$ case, we cannot use formula \eqref{eq:dirac} to write down $\eta$ when the source is an O8$^+$-plane; rather, we must use \eqref{eq:DPfAP}. If the source is an O8$^-$, the two formulae agree and the inverse of $\eta$ is the one in \eqref{eq:eta-1O8}. If the O8$^\pm$ is stuck on a half-NS5 at $x^6=0$, all groups are $\SU$'s but the Dirac pairing is again given by \eqref{eq:eta-1O8} (due to the presence of symmetric or antisymmetric matter of the first gauge group). Finally, in case of a combined O6$^+$-O8$^-$ projection, application of formula \eqref{eq:DPfAP} produces an $\eta$ whose inverse has been written down in closed form in \eqref{eq:eta-1O8O6}. All in all we find:
\begin{subequations}
\begin{align}
&\text{regular poles ($r_0 = r_N = \alpha_0 = \alpha_N= 0$):}& &(\eta^{-1}_\text{D8})_{ij} = \frac{1}{N}\begin{cases}i(N-j) & i \leq j \\ j(N-i) & i \geq j \end{cases}\ ; \label{eq:eta-1alecrem}\\
&\text{D6 poles ($r_0, r_N \neq0$, $\alpha_0 = \alpha_N= 0$):}& &(\eta^{-1}_\text{D6})_{ij} = (\eta^{-1}_\text{D8})_{ij}\ ; \label{eq:eta-1D6}\\
&\text{O6 poles ($ r_0=  r_N = 0$, $\alpha_0, \alpha_N \neq 0$):}& &(\eta^{-1}_\text{O6})_{ij} = \frac{1}{2N}\begin{cases}i(N-j) v_iv_j & i \leq j \\ j(N-i) v_i v_j & i \geq j \end{cases}\ ; \label{eq:eta-1O6D6}\\
&\text{O8 pole at $z=0$ ($y_0 = 0$, $\alpha_0 \neq 0$):}& &(\eta^{-1}_\text{O8$^-$})_{ij} = \begin{cases}(N-j) & i \leq j \\ (N-i) & i \geq j \end{cases}\ ;  \label{eq:eta-1O8}
\end{align}
\begin{align}
&\text{O8 pole at $z=0$, O6 pole at $z=N$}& &\text{($y_0 = 0$, $\alpha_0 \neq 0$, $r_N=0$, $\alpha_N \propto y_N \neq 0$):} \nonumber \\
& & & (\eta^{-1}_\text{O6O8})_{ij} =\frac{1}{2} \begin{cases}(N-j) v_i v_j& i \leq j \\ (N-i) v_i v_j & i \geq j \end{cases}\ .  \label{eq:eta-1O8O6}
\end{align}
\end{subequations}

% fold sec (aFT)

%%%%%%%%%%%%%%%%%%%%%%%%%%%%%%%%%%%%%%%%%%%%%%%%%%

% bibliography

\bibliography{at}
\bibliographystyle{at}

\end{document}